\documentclass{aa}  

\usepackage{graphicx}
\usepackage[colorlinks,urlcolor=blue,citecolor=blue,linkcolor=blue]{hyperref}
\usepackage{adjustbox}
\usepackage{subfig}
\usepackage{natbib}
\usepackage{hyperref}

\usepackage{siunitx}

\usepackage{txfonts}
\usepackage{pifont}
\usepackage{placeins}

\def\R23{\mbox{$\rm R_{23}$}}

\def\arcsec{\hbox{$^{\prime\prime}$}}

\def\msun{M$_{_{\odot}}$}

\newcommand{\HI}{\ion{H}{I}}
\DeclareSIUnit{\mas}{mas}
\DeclareSIUnit{\dex}{dex}

\usepackage{ulem}

\begin{document}

\title{MUSE-DARK}
\subtitle{~I. Dark matter halo properties of intermediate-$z$ star-forming galaxies}
\titlerunning{Dark matter halo properties of $z\sim0.85$ SFGs}

\author{ B. I. Ciocan,
          \inst{1}
	\and 
	N. F. Bouché,
	\inst{1}
	\and
       J. Fensch, 
     \inst{1}
       \and
      W. Mercier, 
         \inst{2}
          \and 
         D. Krajnović,
         \inst{3}
         \and 
         J. Richard,
          \inst{1}
          \and
           T. Contini,
          \inst{4}
          \and 
           A. Jeanneau
          \inst{1}}
          
             \institute{Univ. Lyon1, Ens de Lyon, CNRS, Centre de Recherche Astrophysique de Lyon (CRAL), 69230 Saint-Genis-Laval, France
    \email{bianca-iulia.ciocan@univ-lyon1.fr} 
    \and
    Aix Marseille Univ, CNRS, CNES, LAM, Marseille, France
    \and 
    Leibniz-Institut für Astrophysik Potsdam (AIP), An der Sternwarte 16, 14482 Potsdam, Germany
    \and
    Institut de Recherche en Astrophysique et Planétologie (IRAP), Université de Toulouse, CNRS, UPS, CNES, 31400 Toulouse, France
    }
   \date{}

\abstract{\textit{Context:} The core–cusp problem is a key challenge to the $\Lambda$ Cold Dark Matter  ($\Lambda$CDM) model. Stellar feedback and other baryonic processes have been proposed as solutions to reconcile the tension between the predicted cuspy profiles and the observed cores of low- and intermediate-mass galaxies. However, at $z>1$, disk–halo decompositions usually focus on massive star-forming galaxies (SFGs) with $\log(M_{\star}/M_{\odot})>10$, as observations typically lack the depth and sensitivity to probe lower-mass systems, where cores are predicted to occur. \\
\textit{Aims:} In this study, we analyse the dark matter (DM) halo properties of 127 intermediate-redshift  ($0.3<z<1.5$) SFGs down to low stellar masses  ($8<\log(M_{\star}/M_{\odot})<11$), using the highest signal to noise data from the MUSE Hubble Ultra Deep Field Survey, as well as photometry from the Hubble Space Telescope and James Webb Space Telescope.  \\
\textit{Methods:} We employ both a traditional 2D line fitting algorithm and a 3D forward modelling approach to analyse the kinematics of our sample, enabling us to measure individual rotation curves extended up to $2-3$ times the effective radii. We performed a disk-halo decomposition with a 3D parametric model, which includes stellar, DM and gas components, as well as corrections for pressure support. We test our methodology using mock data cubes generated from idealised disk simulations, and we perform several cross-checks, such as comparing our 3D disk–halo decomposition with results obtained from external data, as well as 1D decompositions, finding good agreement.  We test  six  DM density profiles, including the Navarro-Frenk-White, Burkert, Einasto and the generalised $\alpha\beta\gamma$ profile of \cite{dc14} (DC14), as well as a baryon-only model using a Bayesian analysis.  \\
\textit{Results:} 
Marginalising against the unknown neutral gas content, we find that the mass-dependent DC14 DM profile, which accounts for the response of  DM to baryonic processes such as stellar feedback, performs as well or better than the  other halo models for most of the sample ($\gtrsim80\%$), and that 
baryon-only models seem to be disfavored with respect to DM models for 84\% (107/127) of the galaxies.
 We find that 89\% of the SFGs have DM fractions $(f_{\rm DM}(<R_e))$ larger than 50\%, and we extend the $f_{\rm DM}(<R_e)$-$\Sigma_{M_\star}$ relation at lower masses.  Using DC14 DM profiles,  we infer DM inner slopes $\gamma<0.5$ for 66\% of the sample, indicative of cored DM density profiles.  The stellar--halo mass  and concentration -- halo mass relations inferred from our 3D modelling agree with the theoretical expectations, albeit with larger scatter. Our results confirm the anticorrelation between the halo scale radius and DM density with a slope of  $\sim-1$, which seems to evolve with redshift. While the halo scale radii are $z$-invariant, we find tentative evidence that DM halos of  $z\sim1$ SFGs were denser (by $\sim0.3$ dex) than those in the local Universe. \\
 \textit{Conclusions:} We measured DM halo properties of intermediate-$z$ SFGs down to $10^8$ $M_{\odot}$, finding that a substantial fraction of the sample can be described by cored DM density profiles. This may point toward core formation driven by baryonic processes in the context of $\Lambda$CDM.}

\maketitle

%#######################################################################

%\newpage

\setcounter{section}{0}
%##############################################################################
\section{Introduction}

The matter content of the Universe is dominated by DM from sub-galactic to cosmological scales. The concept of DM entered mainstream research in the 1970s when observations revealed that galaxy rotation curves (RCs) remain flat at large galactocentric distances (e.g., \citealt{vdh}, \citealt{Rubin}), contrary to the expected Keplerian decline. These flat RCs could not be explained by the Newtonian gravity of the baryonic matter alone, but instead suggested the presence of an extended DM halo.

DM halos are a cornerstone of the $\Lambda$CDM cosmological model, which is a successful framework for predicting and explaining the large-scale structures of the Universe and their evolution with cosmic time. However, on scales smaller than $\sim 1$ Mpc, this model faces several challenges (see \citealt{bull17} for a review). For example,  DM-only simulations predict that DM assembles into haloes that develop steeply rising inner radial density profiles, i.e.  'cusps', in the absence of any baryonic effects. These simulated DM halos can be described by: $\rm{\rho(r) \propto r^{-\gamma}}$ with $\gamma\sim 0.8-1.4$, and are well fitted by a Navarro–Frenk–White profile (NFW; \citealt{nfw}, \citealt{Klypin}), independently of initial conditions and cosmological parameters. 
On the other hand, observations of low surface brightness galaxies have demonstrated that these systems show density profiles that are consistent with a constant-density 'core' at the centre, with $\gamma\sim 0-0.5$ (e.g. \citealt{fp}, \citealt{sal}, \citealt{debok}, \citealt{Weldrake}, \citealt{sal2}, \citealt{oh}). This disagreement between observations and simulations is known  as the core-cusp problem and has  posed a major challenge to  $\Lambda$CDM for the past two decades \citep{debok10}.  This discrepancy either hints at the inadequacy of  DM-only simulations to capture the DM dynamics on small scales, due to the absence of  phenomena connected to baryonic physics, or  the fact that a  modification of the whole $\Lambda$CDM paradigm is needed (e.g. self-interacting DM - \citealt{sidm}, axion-like fuzzy DM- \citealt{hu}).

Several baryonic mechanisms within the $\Lambda$CDM framework have been proposed to solve the core-cusp problem. For example, infalling gas clumps can transfer angular momentum to DM via dynamical friction,  resulting in a shallower central density profile (e.g. \citealt{El-Zant},  \citealt{rod}, \citealt{joh}). Alternatively, energy could be transferred to the outer halo through resonant effects induced by a central stellar bar, which can also transform cusps into cores \citep{wk}. Cores may also be created when baryons, after condensing at the centre of a halo, are suddenly expelled by feedback processes. In  many high-resolution cosmological simulations, strong stellar feedback from massive stars and supernovae was shown to drive large-scale gas outflows during repeated starburst events, leading to an overall expansion of the DM halo and the creation of  a cored DM density profile (\citealt{nav}, \citealt{rg}, \citealt{gov}, \citealt{pon}, \citealt{tey}, \citealt{br}, \citealt{free}, \citealt{Dekel-Zhao2}, \citealt{lifre}, \citealt{Jackson}, \citealt{aza}). Core formation can also be linked to active galactic nuclei (AGN) activity in high-mass galaxies and galaxy clusters  (\citealt{marit}, \citealt{Peirani}).

In the stellar feedback scenario, many studies claim that   core formation is strongly dependent on $M_{\star}/M_{\rm{halo}}$, with this mechanism  being most efficient for $ (M_{\star}/M_{\rm{halo}}) \sim 3-5\:\rm{x} \:10^{-3}$, corresponding to a stellar mass regime of $10^9<M_{\star}/M_{\odot}<10^{10}$.  In more massive  haloes, the outflows become ineffective at flattening the inner DM density, and the haloes have increasingly cuspy profiles.  Similarly, at $M_{\star}/M_{\rm{halo}}< 10^{-4}$,  there is not enough supernova energy to efficiently change the DM distribution, and the halo retains the original NFW profile (\citealt{dc14} - hereafter DC14,  \citealt{read}, \citealt{Lazar}, \citealt{aza}). Similar results were obtained from simulations that also include AGN  feedback in addition to stellar feedback ( e.g.  \citealt{Maccio}). \cite{toll} have demonstrated that the relationship between the inner DM density slope and  $M_{\star}/M_{\rm{halo}}$ holds approximately up to $z\sim2$, implying that in  all haloes, the shape of their inner density profile changes over cosmic time, as they grow in stellar and total mass. Nevertheless, the dependence of the DM inner slope on $M_{\star}/M_{\rm{halo}}$ is not universally seen across all hydrodynamical simulations (e.g., \citealt{Bose}).

RCs are fundamental tools for probing the mass distribution in SFGs, as they provide one of the few direct observables of the DM density on galactic scales \citep{Rubin2}. In the local Universe, kinematic studies have employed various dynamical tracers to characterise the DM halo properties of disk galaxies (\citealt{Kormendy}, \citealt{katz}, \citealt{alle}, \citealt{katz2}, \citealt{read3}, \citealt{kor}, \citealt{sal3}, \citealt{li19}, \citealt{li}, \citealt{2025A&A...699A.311M}). Beyond mapping the structure of local DM halos, understanding their evolution over cosmic time is crucial, making kinematic studies of higher-$z$ galaxies essential for constraining the nature of DM. 

Fortunately, over the past decade, deep Integral Field Unit (IFU) observations (e.g., \citealt{foster}, \citealt{genzz}, \citealt{Wuyts}, \citealt{udf1}, \citealt{wizz}, \citealt{girar}, \citealt{lef}) using instruments such as the Spectrograph for INtegral Field Observations in the Near Infrared (SINFONI; \citealt{eisenhau}), the Multi-Unit Spectroscopic Explorer (MUSE; \citealt{bacon10}), the K-band Multi-Object Spectrograph (KMOS; \citealt{Sharples}),  the  Near-Infrared Spectrograph (NIRSpec) (\citealt{Jakobsen}), and sub-mm observations from facilities such as the Atacama Large Millimeter Array (ALMA),  together with  
advancements in 3D modelling tools (\citealt{galpak},\citealt{3dbar}, \citealt{lee}), have opened a new avenue for probing the dynamics of high-$z$ galaxies. Indeed, studying the DM halo properties of distant SFGs at $\rm{z}\gtrsim 1$  requires very deep observations (\citealt{genz}, \citealt{genz2},\citealt{prince}, \citealt{pug}, \citealt{shach}), or stacking techniques (\citealt{lang}, \citealt{Sharma}, \citealt{til}, \citealt{Danhaive}). Many of these studies often find declining outer RCs and consequently low DM fractions, but it is important to note that they primarily focus on high-mass systems ($\log(M/M_{\odot})\gtrsim10$).  More recently, ALMA and the James Webb Space Telescope (JWST) have opened the possibility to study the properties of DM halos at $z=4-5$ (\citealt{Rizzo}, \citealt{Herrera-Camus}, \citealt{Roman}, \citealt{Lee_ALMA}). 

Some of these studies performed  disk-halo decompositions (e.g. \citealt{genz2}, \citealt{prince}, \citealt{Fraternali}, \citealt{shach}, \citealt{Lelli_highz}, \citealt{Roman}, \citealt{shach2}). In a pilot study of nine $z\sim1$ SFGs, \cite{nicolas} pushed this type of analysis on disk-halo decompositions to lower stellar masses ($8.5<\log(M/M_{\odot})<10.5$) using extremely deep MUSE observations \citep{udf1}. They found a variety of RC shapes, reminiscent of the diversity observed at $z = 0$,  DM fractions reaching $>60-95\%$ in the lower mass regime, with 50\%\ of the sample showing strong evidence for cored DM profiles. 

In this paper ("MUSE-DARK-I"), we will apply the same disk-halo decomposition technique used in \citet{nicolas} on a substantially larger sample of 127 SFGs at $z\sim1$. While \citet{nicolas} tested two DM density profiles (DC14 and NFW), here, with a larger sample of $>100$ SFGs, we aim to test more DM halo models, alongside a baryon-only model. Our galaxies span a wide range in redshift ($0.28 < z < 1.49$) and stellar mass ($7 < \log(M_{\star}/M_{\odot}) < 11$), enabling us to explore how halo properties vary with both cosmic time and galaxy mass. To our knowledge, this represents one of the largest samples of distant SFGs with disk–halo decompositions analysed to date. Indeed, current studies are limited to small samples  with 22 SFGs in \cite{pug}, 41 SFGs in \cite{genz2,prince}, 
  and 100 in \cite{shach}, most with $\log(M/M_{\odot})\gtrsim10$.  With the series of "MUSE-DARK" projects, we aim to fill this knowledge gap by performing a detailed disk-halo decomposition analysis on a statistical sample of several hundreds intermediate redshift  ($0.2<z<1.5$)  SFGs with $8<\log(M/M_{\odot})<11$ from the MUSE {\it Hubble} Ultra Deep Field (MHUDF, \citealt{udf2}), the MUSCATEL (programme 1104.A-0026; PI: Wisotzki, L.) and the lensing cluster \citep{Richard} data-sets.

This paper is the first of a series focused on the MHUDF sample and is organised as follows: in section \ref{methods}, we present the methodology used to determine the morpho-kinematics of the sample, as well as describe the 3D disk-halo decomposition. In section \ref{valid}, we validate our 3D methodology using mock data cubes generated from idealised disk simulations. Section \ref{data} describes the observations employed in this study, as well as the data selection and global properties of the sample. The main results are presented in section \ref{results}.  In section \ref{discussion}, we discuss the core formation scenario, as well as the caveats of this work, whereas in section \ref{concl}, we present our summary and conclusions.  

Throughout this paper, we use a ‘Planck 2015’ cosmology \citep{Planck} with $\Omega_M= 0.307$, $\Lambda = 0.693$, $H_{0} = 67.7\rm{\:km\:s^{-1}\: Mpc^{-1}}$. With these cosmological parameters, 1" subtends $\sim8.23$ kpc at $z=1$. We also consistently use ‘$\log$’ for the base-10 logarithm.   We assume a \cite{chabrier03}  initial mass function (IMF) for all the derived stellar masses and star formation rates (SFRs).\\

\section{Methodology}
\label{methods}

To accurately probe the DM distribution in intermediate to high-$z$ galaxies, it is essential to measure RCs at large radii, up to $10-15$ kpc, corresponding to $2-3$ times the effective radii. However, at these large galactocentric distances, the S/N per spaxel of the emission line of interest drops below unity, making velocity measurements difficult without very deep observations 
\citep{genz,genz2,nicolas,shach} or the use of stacking techniques \citep[e.g.][]{lang,til}.  Additionally,  galaxies at $z\geq0.5$ are observed with spatial resolution  (0.5" or 4 kpc in diameter)  comparable to their sizes, which span $3 - 5$ kpc typically (e.g. \citealt{ward}), meaning that these galaxies are only marginally resolved.

Fortunately,   kinematic analyses of the DM distributions in distant SFGs are now feasible thanks to advancements in 3D forward modelling tools, such as {\textsc{GalPaK$^{\rm 3D}$}} \citep{galpak}, \texttt{DYSMALpy} (\citealt{ubler3},   \citealt{prince}), and \texttt{3DBarolo} \citep{3dbar}. These tools construct 3D disk models that can be directly compared to 3D observational data   and are designed to disentangle galaxy kinematics from resolution effects by taking into account any instrumental effects (spatial or spectral resolution). ~\footnote{ Compared to {\textsc{GalPaK$^{\rm 3D}$}} and \texttt{DYSMALpy} which offer the capability to perform disk-halo decomposition of the RC directly in 3D, \texttt{3DBarolo} outputs 1D RCs, requiring a disk-halo decomposition on a few (often correlated, \citealt{posti}) 1D data points. For a comparison between these three different tools,   see \cite{lee}.}

For this study, we have chosen to use {\textsc{GalPaK$^{\rm 3D}$}}, a full 3D forward-modelling approach that fits the entire 3D cube directly rather than the measured 1D or 2D kinematics. This method is well suited for high-$z$ systems where the spatial resolution and S/N are limited,  beacuse (1)   it uses all the spatial and spectral information contained in thousands of spaxels; (2) it allows probing the low S/N regions in the outskirts of galaxies, where the S/N of the spectral line of interest drops below unity, by leveraging the collective signal of all low-S/N spaxels ~\footnote{We note that 2D methods such as Voronio binning \citep{cap1} or 1D slits (as used in \citealt{genz2, prince}) also enable probing low-S/N regions in the outskirts of galaxy disks.};  (3) it constrains the intrinsic morphological and kinematical parameters of the disk simultaneously, thereby breaking the $v_{\rm{max}}$-inclination degeneracy; and lastly (4) it computes the likelihood directly from the 3D data without losing information. This framework assumes an axisymmetric disk, hence our rather strict criteria for selecting a sample of regular, unperturbed, rotation-dominated galaxies (see section \ref{sampleselection}). We note, however,  that non-axisymmetric features or non-circular motions could bias individual parameter estimates. For more details, we refer the reader to \cite{galpak} and \cite{nicolas}.

First, we analyse the morpho-kinematics in order to select a sample of rotation-dominated SFGs with $v_{\rm{max}}/\sigma>1$ (see Sect.~\ref{sampleselection}) using the 3D morpho-kinematic modelling (Sect.~\ref{gpk1}).
Second, we perform a disk-halo decomposition (Sect.~\ref{gpk2}) using the methodology presented in \citet{nicolas}.

In both cases, for the PSF and LSF in {\textsc{ GalPaK$^{\rm 3D}$}}, we use a circular Moffat PSF, as characterised by \cite{udf2}, and  Eq. (7) and (8) of \cite{udf1} for the LSF. It is worth mentioning that in the wavelength range relevant for our emission lines, the MUSE velocity resolution is around $\sim$45-55\,km\,s$^{-1}$.

{\textsc{ GalPaK$^{\rm 3D}$} optimises the model parameters (see Sect. \ref{gpkparamssect}) using various  Markov chain Monte Carlo (MCMC) algorithms, and for this study, we use the Python version of \texttt{MultiNest} (\citealt{feroz}, \citealt{Buchner}).}  The following \texttt{pyMultiNest} configuration was used for the bulk fits: 400 live points, sampling efficiency 0.8 and evidence tolerance 0.5, which is optimised for efficient and robust posterior estimation.

\subsection{3D morpho-kimematics approach}
\label{gpk1}

 As described in \cite{galpak,nicolas1},   \textsc{GalPaK$^{\rm 3D}$} builds a 3D model of a disk with a \cite{sers} surface brightness profile, $\rm{\Sigma(r)}$,  Sérsic index, $n_{\rm{gas}}$, and effective radius, $R_{\rm{e}}$.  The disk model can be inclined to any inclination, $i$, and position angle (PA), whereas the thickness of the disk is assumed to be Gaussian, with a scale height $h_{\rm{z}} = 0.15\: \times\: R_{\rm{e}}$. 
For the disk kinematics, the model uses a parametric form for the RC, $v_{\rm{c}}(r)$, and for the dispersion profile, $\sigma(r)$. 

The velocity profile $v_{\rm{c}}(r)$ can be any functional form, and more details can be found in \cite{galpak,nicolas1}. 
 Here, as in \cite{nicolas}, we parameterised $v_{\rm{c}}(r)$ using the universal RC (URC) from \cite{persic}, which allows both rising and declining RCs.

As described in \cite{galpak,nicolas1}, the velocity dispersion profile  $\sigma(r)$ consists of the combination of a thick disk $\sigma_{\rm{thick}}$, defined as $\sigma_{\rm{thick}}(r)/v_c(r)=h_{\rm{z}}/r$,  where $h_{\rm{z}}$ is the scale height,  and a dispersion floor, $\rm{\sigma_{0}}$, added in quadrature following  \cite{Genzel06,cresci09,forsterschreiber18,wizz,ubler19}. 

 This 3D model has 13 parameters, including the [OII] doublet ratio (listed in Table \ref{gpkparams}). As discussed in \citet{nicolas}, there is generally a good agreement between the morphological parameters (S\'ersic $n$, size, $i$) obtained from [OII] MUSE data with those obtained from HST. Here, we use priors on the inclination (see Sect. \ref{gpkparamssect} for more information), while the other structural parameters are left free.  While we do not show the comparison here, we verified that the structural parameters derived from MUSE are consistent with those obtained from HST/F160W (and stellar mass maps  -see Section~\ref{Morphology} for details on the photometric data), with a scatter of $\sim0.14$~dex in $\log(R_{\mathrm{e}}/\mathrm{kpc})$ and typical differences in Sérsic index of $\Delta n \lesssim 0.5-0.6$.

\subsection{3D disk-halo decomposition}
\label{gpk2}

Following \cite{nicolas}, we adopt a 3D disk-halo decomposition from the decomposition of the gravitational acceleration $v^2/r$ into the contributions from a DM component, $v_{\rm{DM}}(r)$, a disk (stellar$+$molecular gas) component, $v_{\rm{disk}}(r)$,   a neutral gas component, $v_{\rm{\HI}}(r)$, and when the bulge-to-total ratio, $B/T>0.2$, an additional ‘bulge’ component, $v_{\rm{bulge}}(r)$, such that:
\begin{equation}
v_{\rm{c}}^2(r) = v_{\rm{DM}}^2(r) + v_{\rm{disk}}^2(r) + v_{\rm{\HI}}(r) |v_{\rm{\HI}}(r)| ( + v_{\rm{bulge}}^2(r)).
\label{eq:vdecomp}
\end{equation} 
Compared to other high-$z$ studies performing an RC decomposition (e.g. \citealt{genz},  \citealt{genz2}, \citealt{shach}), we include an unknown \HI\ neutral gas component, which we marginalise over. Also, compared to \citet{nicolas}, we use $v_{\rm{\HI}}(r) |v_{\rm{\HI}}(r)|$ to account for the possibility of a net outward force in the case of central HI mass depression \citep{Casertano}. We do not apply the same treatment to the other components, as we do not expect them to exhibit central depressions that could result in a net outward force.

As in \citet{nicolas}, \cite{weij}, \cite{burkertad} (and others) we include a correction for  pressure support $P$ (sometimes called asymmetric drift),
such that: $v_{\rm{c}}^2(r) = v_{\perp}^2(r) + v_{\rm{AD}}^2(r)$, where $v_{\rm{AD}}=-\sigma^2\frac{{\rm d}\ln P}{{\rm d}\ln r}$, with $P$ being the  pressure and  $\sigma$ the gas velocity dispersion in the radial direction.
Following   \cite{Dalcanton}, we use the correction for turbulent star-forming disks,  $v_{\rm AD}=\alpha\;\sigma^2(r/r_d)=0.92\;\sigma^2(r/r_d)$. This pressure support correction  is consistent with the correction found by \cite{Kretschmer} in the VELA  cosmological zoom-in simulations at $z = 1 - 5$, which indicate $\alpha\sim1.1$ for the galaxy mass range probed in this study.
We refer the reader to Appendix A of \cite{nicolas} for a comparison to other prescriptions for the pressure support correction.
 
As discussed in \citet{nicolas},    $v_{\rm{disk}}(r)$ is determined from the ionized gas
surface brightness profile (S\'ersic $n$)  because  the ionised gas  and stars follow similar surface brightness profiles \citep[e.g.][]{Nelson,wilm,linlin}, and consequently have similar $v(r)$~\footnote{The assumption that the ionized gas and stellar kinematics are similar is supported by observations of intermediate-$z$ SFGs \citep{guer,ubler2} in spite of the fact that the ionised gas might also be susceptible to hydrodynamical effects, such as non-circular motions, and impact of feedback.}. In practice, the shape of $v_{\rm{disk}}(r)$ is determined from the shape of the surface brightness profile, using a Multi Gaussian Expansion (MGE)  (\citealt{Monnet}, \citealt{mge}), whereas the normalisation of the ${v_{\rm{disk}}(r)}$ is given by ${M_{\star}}$.~\footnote{The MGE $v(r)$ is pre-tabulated for a grid of $n$, which is equivalent to using the \textsc{vcdisk} code (Posti et al.), but is significantly faster.}
 Given that the molecular gas mass fractions are typically 30-50\%\ in our redshift range ($z\sim1$) \citep[e.g.][]{fre2,tac}, this leads to a systematic uncertainty of 0.1-0.15~dex in $M_{\rm disk}$. In other words, the contribution of the molecular gas is effectively included in our disk component.  The fact that we do not separately model the molecular gas contribution remains a caveat of the current approach. Future inclusion of spatially resolved molecular gas maps (e.g., from ALMA) will be crucial for addressing this uncertainty.  

As discussed in \citet{nicolas}, the  contribution of the \HI\ gas might be important, especially at large radii. Given the well-known \HI\ size-Mass relation \citep{Broeils97,Martinsson,Wang16,Wang2}, the \HI\ disks are 20-90 kpc in radius, i.e. much larger than $R_{\rm e}$, and the extent of our data. Given the unknown \HI\ surface profile, one can assume  (i)  a constant $\Sigma_{\HI}$ surface mass density~\footnote{The \HI\ size-mass relation ($D_{\HI}-M_{\HI}$) with a slope of 0.5  \citep{Broeils97} indicates a uniform characteristic \HI\ surface density    \citep{Wang16}.} leading to $v(r)\propto\sqrt{\Sigma_{\HI}r}$ which qualitatively reproduces the observations of local galaxies \citep[e.g.][]{alle}; (ii)  an average \HI\ profile from the Disk Mass Survey \citep{Martinsson}~\footnote{The Martinsson et al. profile, namely $\Sigma_{\HI}\propto\exp(-(x-0.4)^2/0.36^2)$ where $x=r/r_{\HI}$,  leads to a slight \HI\ depression with $R_{\HI}$.}, which is similar to the compilation of \citet{Wang16}; or (iii)  the more recent stacked profile from 35 late-type spirals obtained by 21cm observations from the Five-hundred-meter Aperture Spherical radio Telescope (FAST)  down to 0.01 M$_\odot$~pc$^{-2}$ by \citet{Wang2}.
The circular velocity  $ v_{\rm{\HI}}(r)$ can then be found by solving the Poisson equation for a thick disk using the fitted $M_{\rm{HI}}$, following \citet{Casertano}, particularly their Equations 4–6 and their Appendix A. 
The three assumptions yielded very similar results, which agree within the uncertainties for $\sim90\%$ of the sample (see appendix \ref{app:comparison HI} for a comparison of the inferred DM inner slopes, $\gamma$, yielded by the different HI parametric models), and we used model (i) in the remainder of the paper.

Following  \citet{nicolas}, when a  bulge  is present ($B/T>0.2$, see Sect.~\ref{Morphology}) even though our sample is selected against galaxies with large bulges,  then a bulge component is added to the flux profile (using  $B/T$ as a free parameter), as well as a \cite{Hernquist} kinematic component $v_{\rm{bulge}}(r)$ to Eq. \ref{eq:vdecomp},  which has  free parameters the Sérsic index $n_{\rm{bulge}}$  (allowed to vary between  $2<n_{\rm{bulge}}$ <4),  and the bulge effective radius ($r_{\rm{bulge}}$).  In other words, we decouple the light and mass profiles, given that a bulge can be made of old stars and little to no ionised gas.  

For the DM component $v_{\rm{DM}}(r)$, we consider six different DM density profiles:
namely DC14 \citep{dc14}, NFW \citep{nfw}, Dekel-Zhao (\citealt{Dekel-Zhao}, \citealt{fre}), Burkert \citep{Burkert}, coreNFW \citep{read}, and Einasto \citep{einasto} profiles, which are detailed in Appendix~\ref{appendix:DM}. In this context, a useful parameterisation is the generalised $\alpha -\beta -\gamma$ double power-law (e.g. \citealt{Hernquist}, \citealt{zhao}):
\begin{equation}
\rho(r)=\frac{\rho_{\rm{s}}}{\left(\frac{r}{r_{\rm{s}}} \right)^{\gamma}\left(1+\left(\frac{r}{r_{\rm{s}}} \right)^{\alpha}\right)^{(\beta-\gamma)/\alpha}}, 
\end{equation}
where ${r_{\rm{s}}}$ is the scale radius, $\rm{ \rho_{\rm{s}}}$ the scale density, and $ \alpha,\: \beta, \: \gamma$ are the shape parameters of the DM density profile.~\footnote{E.g., a NFW \citep{nfw} profile has $\alpha,\beta,\gamma=(1,3,1)$, the Dekel-Zhao (\citealt{Dekel-Zhao}, \citealt{fre}) has $\alpha,\beta,\gamma=(0.5,3.5,a)$.}
$\beta$ is the outer slope,  $\gamma$ the inner slope and $\alpha$ describes the transition between the inner and outer regions. 
For DC14 (\citealt{dc14}), the values of these shape parameters depend on the stellar-to-halo mass ratio, namely $\alpha(X),\beta(X),\gamma(X)$ where $X=\log(M_{\star}/M_{\rm{halo}})$, described in Appendix~\ref{appendix:DM} (Eqs.~\ref{alpha}-\ref{gamma}).
Finally, we also consider baryon-only models setting $v_{\rm{DM}}(r)=0$.

\subsection{Model parameters and priors}
\label{gpkparamssect}
While the URC and no DM models have 12 - 13 free parameters,  the disk-halo models have between 14 - 16  free parameters (depending on the used halo model and whether we include a bulge component or not), namely: $x$, $y$, $z$, the total line flux, the inclination $i$, the Sérsic index $n_{\rm{gas}}$ for the ionised gas disk,  the major-axis position angle PA, the effective radius $R_{\rm{e}}$, the virial velocity $v_{\rm{vir}}$, the concentration $c_{\rm{vir}}$, the velocity dispersion $\sigma_0$, the HI gas density,  and the doublet ratio $r_{\rm{O2}}$ for [OII] emitters. Depending on the DM halo model,  additional parameters include $X$ for DC14 and Dekel-Zhao,  $M_{\star}$ for all the other halo models, and $\alpha_{\epsilon}$ for Einasto. For cases with a bulge component, there are 4 additional parameters:  $n_{\rm{bulge}}$, $r_{\rm{bulge}}$ and the $B/T$. , 
Table \ref{gpkparams} summarises the parameters of each model. 

We use loose flat (uninformative) priors covering a wide physical range for all these parameters, except for  the inclination, $i$, and stellar masses, $M_{\star}$, for which we use more conservative priors. More specifically, for the inclinations, we use priors based on the values obtained with \texttt{Galfit} from the HST/F160W images (\citealt{morph}, see section \ref{Morphology}), such that $i=i_{\rm{HST}}\pm5^{\circ}$. For the small sub-sample for which we add a bulge component, we use priors on for the $B/T$ parameter, such that $B/T=B/T_{\rm{mass \: maps}}\pm0.1$, for the bulge radius, $r_{\rm{bulge}}=r_{\rm{bulge}}\pm1$ (in pixel) and for $n_{\rm{bulge}}$, such that $2<n_{\rm{bulge}}<4$  (see Sect.~\ref{Massmaps} for more details). For the stellar masses, we use as priors the $M_{\star}$ values obtained from SED fitting (see section \ref{Morphology}), with  $\log(M_{\star}/M_{\odot})=\log(M_{\star}/M_{\odot})_{\rm{SED}}\pm0.15$. We note, however, that for DC14,  Dekel-Zhao and baryon-only models,  we used no priors on $M_{\star}$, as for the former two, the disk-halo degeneracy is broken through the use of ${X}$. 
We tested {\textsc{GalPaK$^{\rm 3D}$}} using both flat and Gaussian priors for all relevant parameters and found consistent results within the uncertainties. We also examined whether applying priors to additional structural parameters beyond $i$--such as $R_{\rm e}$ and $n$--affect the results, and  found agreement within the errors.

The parameter values are derived from the posterior distributions. Throughout this paper, we adopt the median of each marginalised posterior as the best-fit parameter estimate. The associated uncertainties are quoted as  95\% confidence intervals, defined by the 2.5th and 97.5th percentiles of the posterior.

Covariances between parameters were not explicitly included when computing the best-fit values, as each parameter was treated independently from its marginalised posterior. We inspected the posterior distributions of all fitted parameters across the full sample to assess potential covariances. Overall, we find no significant covariance between baryonic and DM parameters. Fewer than 10\% of the galaxies show correlations between the baryonic masses or surface densities ($M_{\rm disk}$; $M_{\rm HI}$, or $\Sigma_{\rm HI}$--depending on the adopted H I parametric model) and the virial velocity.  These covariances are generally weak and do not significantly affect the recovered marginalised parameter estimates. As expected, the concentration ($c_{\rm vir}$) shows some correlation with the virial velocity ($v_{\rm vir}$), since $c_{\rm vir}=R_{\rm vir}/r_{\rm s}$ imposes a natural coupling between these parameters.  The same applies for $\log(X)$ and $v_{vir}$ ($M_{vir}$) and for $\log(X)$ and $M_{\star}$. 

The one-dimensional posteriors are generally well-behaved and approximately Gaussian. The posterior shapes are data-driven for the majority of the sample, as only a small number of cases ($<15\%$) show posteriors approaching prior limits, typically for more weakly constrained parameters such as the halo concentration. Example corner plots for the three representative galaxies discussed in the text are shown in Appendix~\ref{corner}.

\section{Validation of the methodology}
\label{valid}

In this section, we perform a validation check of our 3D disk-halo decomposition introduced in Sect. \ref{gpk2}, by applying our 3D methodology on mock MUSE data cubes. 
We will perform several additional checks in Sect.~\ref{results}.

\subsection{Description of the simulations and creation of synthetic MUSE data-cubes}

To  validate the 3D disk-halo decomposition introduced in \ref{gpk2} and to estimate potential systematic errors, we generated mock MUSE observations from two idealised disks simulated using the Adaptive Mesh Refinement (AMR) hydro-dynamical code \texttt{RAMSES} \citep{Teyssier2002}  with specific initial conditions that control the DM profile shape using the \texttt{MAGI} code \citep{Miki2018}
as we will simulate two galaxy models, one representative of a cuspy and one of a cored DM distribution.

Specifically, the idealised disks are set in a cubic box of length $132$~kpc, and the gas cells have sizes between 1~kpc  and 8~pc depending on the level of refinement. A cell is refined if either: (i) the number of initial conditions particles in the cell is above 50; (ii) its mass, including embedded new star particles, is above $5\times10^4$~M$_\odot$; (iii) the local Jeans length is shorter than four times the cell size.
DM and initial conditions stars are modelled by $1\times 10^4$ and $8.8\times 10^4$~M$_\odot$ mass particles, respectively. Their effect on the gravitational potential is treated at a maximum refinement of 32~pc.
The numerical recipes are described in more detail in \cite{Fensch2021} and summarised here. Briefly, we use  heating and atomic cooling at solar metallicity from \cite{Courty2004}, and we allow cooling down to 100~K and prevent numerical fragmentation with a pressure floor at high densities. We model star-formation with a Schmidt law with an efficiency per free-fall time of 1\% for gas cells denser than 300 $H$/cc and cooler than $2\times10^4$~K where new stars have a mass of $4\times10^3$~M$_\odot$. We include two types of star formation feedback: type II supernovae and H\texttt{II} regions, using the same model as {\it Medium Feedback} in \cite{Fensch2021}.

We ran two disk models, one representative of a cuspy and one of a cored DM distribution, respectively ID 3 and ID 982 from \cite{nicolas}, using the shape parameters $\alpha,\beta,\gamma$ described in Table \ref{simulpara}. 
The amount of ionised gas in these galaxies is not constrained by our data, therefore, we use, for the initial conditions, a gas mass corresponding to the SFR measured in \cite{nicolas} for these two galaxies. This gas mass corresponds to a (molecular) depletion timescale of 0.7 Gyr  typical for intermediate redshift galaxies (\citealt{tac}, \citealt{Saintonge}).\footnote{In the snapshot used for mock creation, the effective total gas depletion timescale is around 0.85 Gyr for the two models.} 
We do not attempt to simulate the neutral HI gas.

We first run the simulations with an isothermal equation of state, at temperature $5\times 10^4$~K, for $\simeq~300$ Myr, to let the gas component relax in the gravitational potential of the stars. After this first phase, we let the gas cool and activate star formation and its feedback. Measurements are made 100 Myr after the activation of gas cooling and star formation, and feedback. 

\begin{table*}
\caption{Parameters used to create the initial conditions for the numerical simulations.}
\centering
\begin{adjustbox}{width=1\textwidth}
\centering
\small
\begin{tabular}{lcccccccccccc}
 Galaxy       & $\log(M_{\star} / M_\odot)$  & $\log(M_{\mathrm{gas}} / M_\odot)$  & $\log(M_{\mathrm{vir}} / M_\odot)$  & $\alpha$ & $\beta$ & $\gamma$ & c$_\mathrm{vir, -2}$ &  Disk height & Disk radius   \\

      model      &   &    & &  &  &  &  & [kpc] &  [kpc] \\

\hline 

ID3   &        10.02     &   9.62   &   11.71    &  1.49    &   2.7    &  0.82    &   11.8    &  0.84   &   5.61                           \\
ID982 &        9.54   & 9.61&    12.22   &  2.53    &   2.51&  0.23   &  7   &    0.71      &      2.85        \\
\hline
\end{tabular}
\end{adjustbox}
\label{simulpara}
\end{table*}

In order to test our ability to measure the DM profiles of these simulated disks,  we create synthetic observations to input into the  {\textsc{ GalPaK$^{\rm 3D}$}} fitting routine, in the same way as for observations. For this purpose, we create mock data cubes with resolution 85~pc $\times$ 85~pc $\times$ 22.5 km/s. The value in each gas cell is proportional to the local gas density. The synthetic cubes are then
resampled to MUSE spaxels (i.e. 0.2" $\times$  0.2" $\times$  55 km/s), are convolved with the MUSE PSF and LSF, and noise is added to obtain a similar S/N to the observations (namely S/N $\sim20$). 

\subsection{3D Disk-Halo Decomposition on Mock MUSE Data Cubes} 

\begin{figure*}
   \centering
    \includegraphics[width=1\textwidth,angle=0,clip=true]{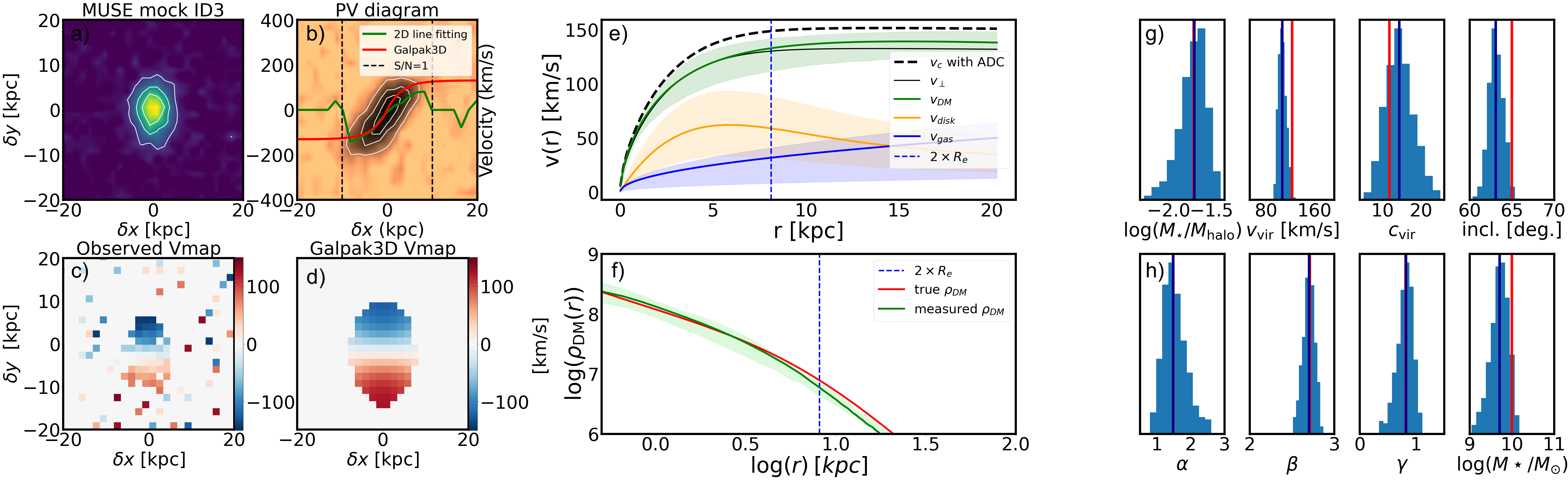}
   \centering   
 \caption{ Results from the 3D disk-halo decomposition applied to the mock MUSE cube of the simulated galaxy ID3. Panel (a) shows the mock MUSE white-light image with flux intensity contours overlaid. Panel (b) shows the position velocity diagram, with the observed velocity profiles as measured with  {\textsc{ GalPaK$^{\rm 3D}$}}  and \texttt{MPDAF} \citep{mpdaf} overlaid in red and green, respectively. The dotted black line shows the region beyond which the S/N of the emission line falls below unity, whereas the white contours show the flux intensity. Panels (c) and (d) show the observed velocity field obtained using a traditional 2D line fitting code and the modelled velocity field obtained from 3D disk-halo decomposition with  {\textsc{ GalPaK$^{\rm 3D}$}}. Panel  (e) shows the contribution of the different components (stars-orange, gas-blue, DM-green curve) to the RC (dot-dashed curve; corrected for pressure support)  and  panel (f) compares the measured DM density profile in green to the true DM density profile in red. The light shaded regions in these 2 panels show the 95\% confidence interval.  Panels (g) display the posterior distributions (in blue) for a subset of the parameters we fit for: $\log(X) = \log(M_{\star}/M_{\rm{halo}})$, the virial velocity, the concentration, and the disk inclination. Panels  (h) show the posterior distributions for parameters derived from the fitted ones, including the DM density profile shape parameters $\alpha$, $\beta$, and $\gamma$ (computed using equations \ref{alpha}, \ref{beta}, and \ref{gamma}, respectively), as well as the stellar mass $\log(M_{\star}/M_{\odot})$. The recovered values are shown as the dark-blue lines, while the values used as inputs for the simulation are shown by the red lines.}
\label{mock}
\end{figure*}

\begin{figure*}
   \centering
    \includegraphics[width=1\textwidth,angle=0,clip=true]{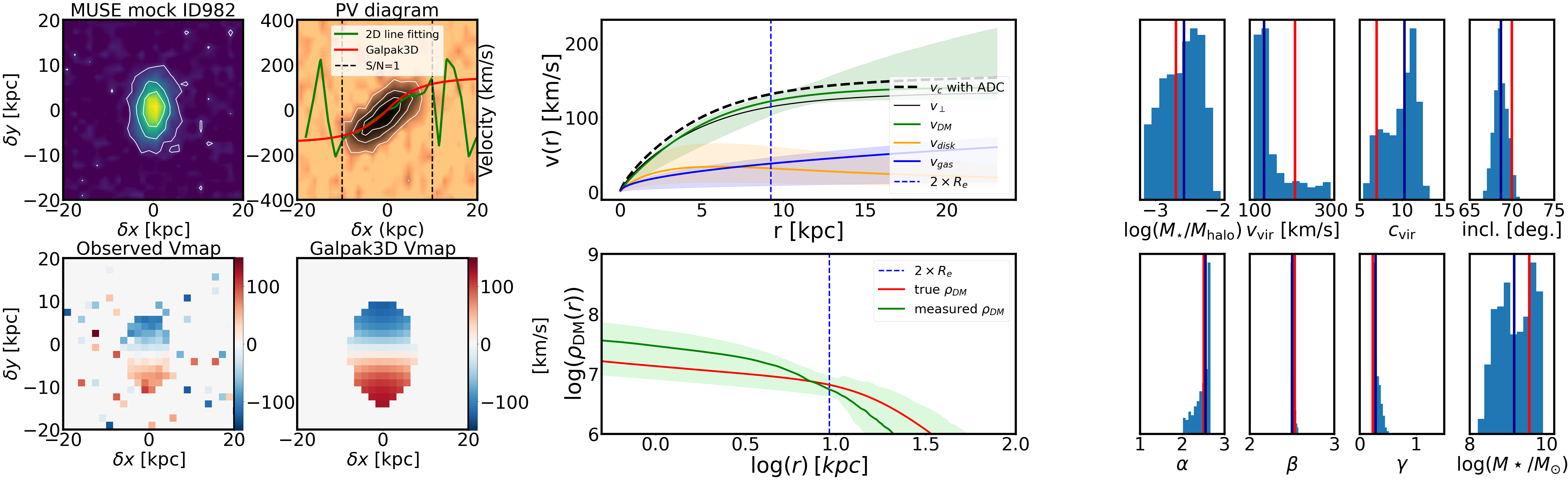}
   \centering   
    \caption{Same as Fig. \ref{mock} but for the mock MUSE cube of the simulated galaxy ID982.} 
\label{mock2}
\end{figure*}

We tested the 3D disk-halo decomposition (section \ref{gpk2}) on the mock MUSE data cubes described above using the DC14 halo model (and no priors in our modelling). Figures \ref{mock} and \ref{mock2} present the results obtained with {\textsc{GalPaK$^{\rm 3D}$}} for simulated galaxies ID3 and ID982, respectively. 

Panels c) and  d) of these figures show that the velocity fields modelled with our 3D disk-halo decomposition align well with the observed ones. Similarly, the modelled and observed velocity profiles (red and green curves, respectively, in panels (b) showing the position-velocity diagrams) closely match in regions where the $S/N > 1$. As illustrated, {\textsc{GalPaK$^{\rm 3D}$}} allows us to probe low-S/N regions in the galaxy outskirts, enabling more robust constraints on the outer-disk RC, compared to traditional 2D line fitting algorithms. Panels f) of  Figures \ref{mock} and \ref{mock2} further demonstrate that our 3D disk-halo decomposition reliably recovers the true density profile within the uncertainties for both ID3 and ID982. Additionally, the recovered values of the fitted parameters ($\log X$, $v_{\rm vir}$, $c_{\rm vir}$, and $i$ - panels g) and the derived parameters ($\alpha$, $\beta$, $\gamma$, and $M_{\star}$, panels f), indicated by the dark-blue lines, all agree with the simulation input values, shown as the red solid lines, within 1$\sigma$. The most challenging parameter to recover is the halo concentration, $c_{\rm{vir}}$, which depends on estimating the virial radius, $r_{\rm{vir}}$, which lies well beyond the radial extent of the observed RC. 

Overall, these results indicate that our 3D kinematic modelling framework provides reasonable estimates of galaxy physical parameters, even when kinematic coverage is limited.

\section{Data: intermediate-$z$ SFGs}
\label{data}
\subsection{MUSE Hubble Ultra Deep Field}
\label{mhud}

For this study, we exploit 3D spectroscopic observations from the MHUDF (\citealt{udf1}, \citealt{udf2}) to investigate the DM halo properties of intermediate-$z$ SFGs which consists in three data sets :  (1) the $\rm{3x3\:arcmin^2}$ mosaic of nine MUSE fields at a 10 hr depth (MOSAIC);  (2) a single $\rm{1x1\:arcmin^2}$ 31 hr depth field (UDF10), as well as (3) the 140 hr  MUSE eXtremely Deep Field (MXDF). 

We made use of the publicly available catalogues and advanced data products, which can all be obtained from the  AMUSED web interface \footnote{ \url{https://amused.univ-lyon1.fr}.}. In particular, we used the source files which contain generic information related to the source  (e.g. identifier, celestial coordinates, PSF model and FWHM, etc),   images (e.g., white-light, narrow bands, HST images and segmentation maps, etc), and the MUSE data-cubes centred at the source location
which we used for our kinematic analysis.

To prepare the data for the kinematic analysis, we truncated the sub-cubes in wavelength (with a width of 30\,\AA), centred on the emission line of interest using the spectroscopic redshifts, and subtracted the underlying continuum. No additional spectral/wavelength masking was applied.

We note that for some IFU sub-cubes, masks were applied during the kinematic fitting. Specifically, segmentation maps created with DS9 \citep{ds9} were used in crowded regions where multiple galaxies at similar redshift appeared within the same sub-cube ($\sim13\%$ of the sample). Pixels outside the target galaxy were set to NaN and excluded from the fits. No additional weighting was applied.

\begin{figure*}
\sidecaption
    \subfloat{{\includegraphics[width=12cm]{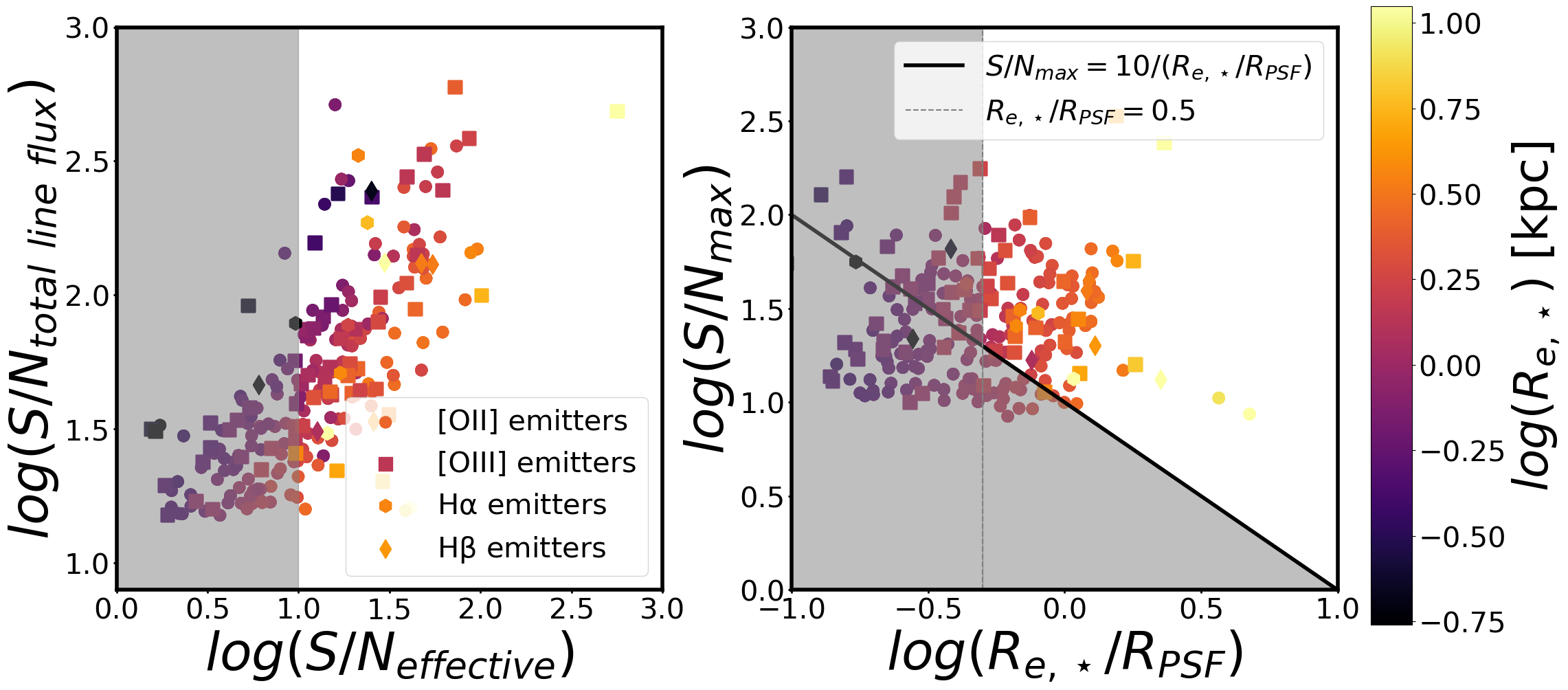} }}%   
        \caption{\normalsize Left:  Effective  S/N for the brightest emission line using equation \ref{s}, as a function of the S/N of the total line flux from the integrated spectrum. Right:: Ratio between stellar half-light radius and the PSF radius as a function of the S/N in the brightest spaxel. The grey shaded area in both panels marks the exclusion region, i.e, all galaxies with $\rm{S/N_{eff}<10}$ and $R_{\rm{e}}/R_{PSF}<0.5$ are removed from the sample.  The data points are colour-coded according to their half-light radii.  The circles, stars, diamonds and squares represent the [OII], $\rm{H\alpha}$, $\rm{H\beta}$ and [OIII] emitters, respectively.  }
\label{snr}
\end{figure*}

\subsection{Sample selection}
\label{sampleselection}
We used the   MHUDFs main catalogue to select SFGs with $z$<1.5 for our kinematic study,  
which have an effective signal to noise $\rm{S/N_{eff}}> 10$ for the brightest emission line  (either $\rm{[O\textsc{ii}]}\:\lambda 3727$, $\rm{H\beta}$,  $\rm{[O\textsc{iii}]}\:\lambda 5007$,  $\rm{H\alpha}$). The effective S/N accounts for the fact that the surface brightness of the galaxy alone is not sufficient to determine the accuracy in the fitted disk structural and kinematic parameters, as the compactness of the galaxy with respect to the PSF also plays an important role (\citealt{galpak}).  The effective S/N is defined as follows:  
\begin{equation}
\rm{S/N_{eff}}= (R_{\rm{e}}/R_{\rm{PSF}})\cdot \rm{S/N_{max}}, \label{s}
\end{equation}
 where $R_{\rm{e}}$ stands for the stellar half-light radius, $R_{\rm{PSF}}=\rm{FWHM}/2$ is the radius of the MUSE PSF, and $\rm{S/N_{max}}$ is the  S/N of the emission line of interest in the brightest spaxel of the MUSE data cube centred on the galaxy. 
We also impose  $R_{\rm{e}}/R_{\rm{PSF}}>0.5$ for our sample selection, to further avoid unresolved galaxies.
 Additionally, we only select systems which have an inclination  $i>30^{\circ}$ for our kinematic study. 
  
Figure \ref{snr} displays on the left the $\rm{S/N_{eff}}$,  as a function of the $\rm{S/N}$ of the total line flux for the [OII]$\lambda3726,\:\:\lambda3729$, $\rm{H\alpha}$, $\rm{H\beta}$ and [OIII]$\lambda5007$ emitters from the MHUDF sample and on the right  $R_{\rm{e}}/R_{\rm{PSF}}$ as a function of the S/N of the brightest spaxel.  From the whole MHUDF sample, 183 galaxies met the aforementioned selection criteria, namely  $\rm{S/N_{eff}}> 10$,    $R_{\rm{e}}/R_{\rm{PSF}}>0.5$, and $i>30^{\circ}$. 

Moreover, we require our galaxies to be axisymmetric disks- i.e. unperturbed - in order to accurately model their morpho-kinematics and infer their DM halo properties.  Therefore, we exclude merging systems from our sample. Additionally, we flag  galaxies, which show signs of gravitational interactions such as tidal arms; clumpy SF regions; as well as large residuals in the  {\textsc{ GalPaK$^{\rm 3D}$}} fits, and plot them with different symbols in the subsequent plots (the stars in all the Figs. denote the 'perturbed' galaxies). This was done by visually inspecting the galaxies using  the  JWST and  HST images, as well as investigating the catalogues from \cite{vent}, who analysed the merger fraction in MUSE deep fields.
This leads to a  sample of 173 galaxies. \\

Finally, for the disk-halo decomposition described in section \ref{gpk2}, we require the galaxies to be rotationally supported, therefore, we exclude  dispersion-dominated systems from our analysis. 
 Using the URC described in Sect.~\ref{gpk1} to estimate the kinematics and shown in Fig.~\ref{gpkplots}, we excluded galaxies
 with $v_{\rm{max}}/\sigma<1$, where $\sigma =\sigma(2R_{\rm{e}})$.  The overall selection criterion results in a final sample of 127 intermediate-$z$ main sequence galaxies that  have $\rm{S/N_{eff}} > 10$, $R_{\rm{e}}/R_{\rm{PSF}}$>0.5,  inclinations $i > 30^{\circ}$, and $v_{\rm{max}}/\sigma > 1$.

\subsection{Ancillary Data}
\label{Morphology}

We also exploit the ancillary data available in the MHUDF area, namely the  HST \citep{Dickinson2003,  Giavalisco2004, Grogin2011, Koekemoer2011} and  JWST  \citep{Bunker2023, Eisenstein2023, Eisenstein2023b, RiekeDoiJADES, Rieke2023, dEugenio2024, WilliamsDoiJEMS, Williams2023} images. 

\subsubsection*{Morphology}
For the stellar disk structural parameters, we use the F160W morphological analysis performed by \cite{morph} with the GALFIT tool \citep{galfit}, providing the half-light radii ($\rm{Re}$), Sersic index ($n$), axis ratios (b/a), and position angles (PA). We used the morphological parameters derived from the F160W band, as this is the reddest HST filter available and therefore best traces the underlying stellar mass distribution. The inclinations from this catalogue were used as priors in our dynamical modelling.

As already mentioned, the gas disk structural parameters yielded by  {\textsc{ GalPaK$^{\rm 3D}$}} from the MUSE data are in good agreement with the ones derived for the stellar component from photometric observations for the vast majority of the SFGs analysed in this study. Similar results were found by \cite{contini16} and \cite{nicolas} for a subsample of the MHUDF galaxies.

\subsubsection*{SED}

To obtain the stellar masses and SFRs,  SED fitting was performed using the MUSE spectroscopic $z$ and  11 HST broadband photometry measurements from the  \cite{Rafelski} catalogue.  The SED fits were carried out with the \texttt{Magphys} \citep{Cunha} software on the HST photometry using a \cite{chabrier03} IMF, keeping the redshift fixed to the spectroscopic one,  as well as with the \texttt{ Prospector} SED fitting code \citep{j}, observing good agreement between the results offered by the two different tools, with a scatter for $\log(M\star/M\odot)$ and SFRs in the order of $\sim0.11$ dex for our sample. 

 We note that in this study, we used the \texttt{Magphys} derived  SFRs and $M_{\star}$, which assumed exponentially declining star formation histories (SFHs) with superimposed bursts. The SFRs correspond to the average star formation over the last 0.1 Gyr, and as such are not very sensitive to short, intense bursts of star formation. 

\subsubsection*{Mass maps}
\label{Massmaps}

We used all HST and JWST images available with a \SI{30}{\mas} pixel scale in the footprint of the MHUDF to produce high-resolution resolved stellar mass maps using the pixel-per-pixel SED fitting library \texttt{pixSED}\footnote{\url{https://wilfriedmercier.github.io/pixSED/}}.
The details are presented in Appendix~\ref{Appendix:mass}, and an example is shown in Fig.~\ref{fig:example_mass_morpho}.

As a first application, we used the mass maps for the whole sample to independently estimate the stellar disk component’s contribution to the disk-halo decomposition, employing the MGE method of \cite{mge}  (see section \ref{res:mge:mass} for more details).

As a second application, we used the stellar mass maps to estimate the bulge-to-total ratio (B/T) without relying on single-band observations, nor on the assumption of a constant mass-to-light ratio throughout the galaxies. To do so, we fitted the resolved mass maps using the \texttt{AstroPhot} package \citep{astrophot}. First, we fitted a single Sérsic profile to the mass maps to estimate the centre position, ellipticity, and position angle of the galaxies. Then, we performed a second fit with a bulge-disk decomposition using the parameters from the first run as initial values for the disk component, where the disk was set to have a Sérsic index of $n = 1$ and the (circular) bulge was allowed to vary between $n = 2$ and $n = 4$.
Furthermore, the bulge  radius was constrained to be less than \SI{3}{\arcsecond}, with an initial value of \SI{0.5}{\arcsecond}, to force it to fit the inner parts of the galaxies. 
This analysis is relevant to our disk-halo modelling. For the 15 galaxies showing B/T>0.2, we added a bulge component in the 3D disk-halo decomposition on the MUSE data, described in Sect.~\ref{gpk2}.  It is also worth noting  that the structural parameters recovered from the bulge-disk decomposition on the mass maps agree well with those from the \cite{morph} catalogue.

\subsection{Global properties of the sample}
\label{global_properties}

\begin{figure}
   \centering
    \includegraphics[width=0.5\textwidth,angle=0,clip=true]{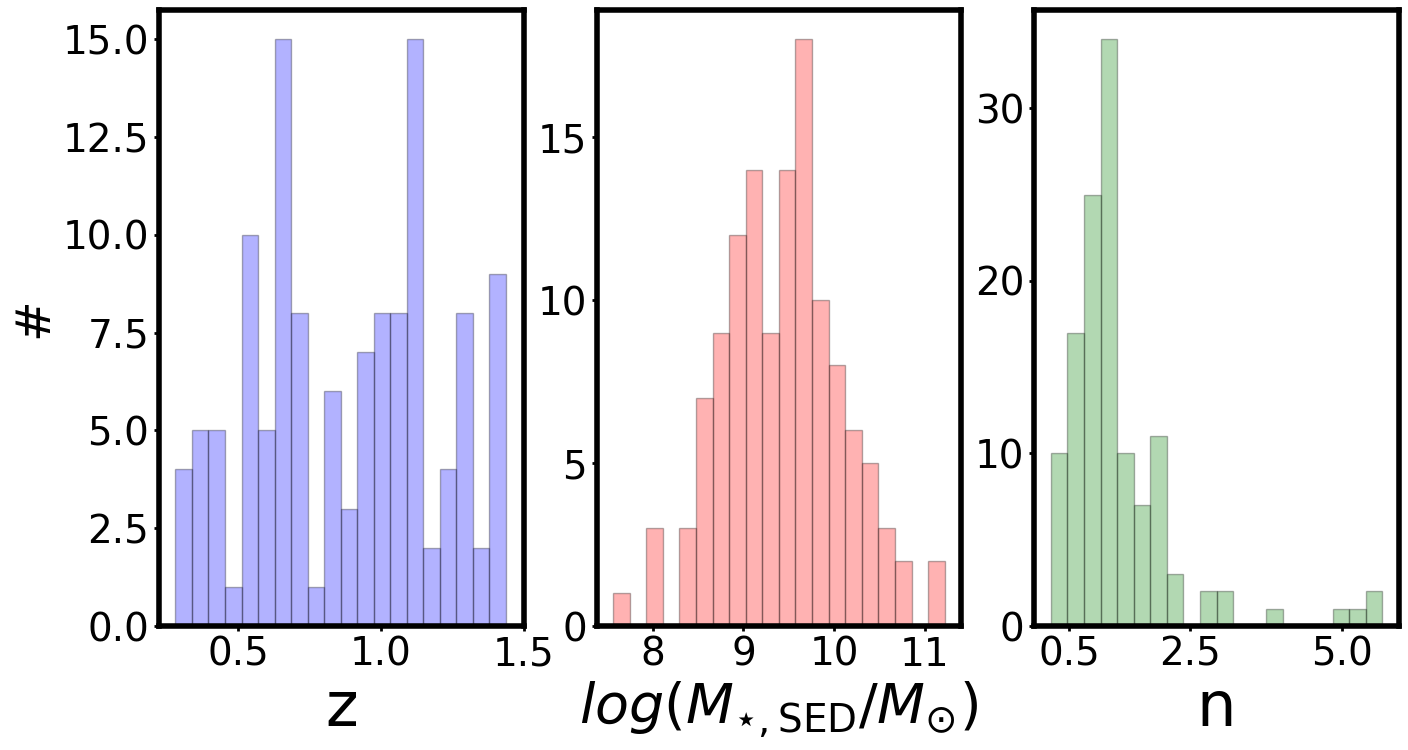}
   \centering   
    \caption{ 
    Left: Histogram showing the redshift distribution of the MHUDF  SFGs. Middle: Histogram showing the stellar mass distribution of sample \citep{udf2}. Right: Histogram showing the distribution of the Sérsic indices  (\citealt{morph}). } 
\label{histzm}
\end{figure}

\begin{figure}
   \centering
    \includegraphics[width=0.5\textwidth,angle=0,clip=true]{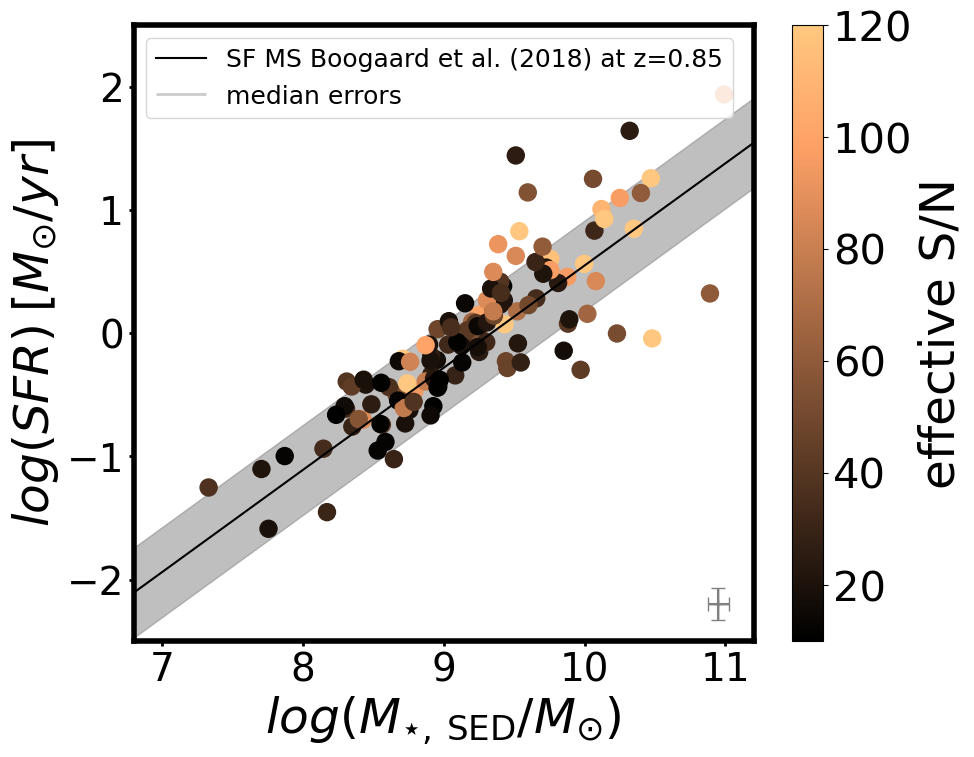}
   \centering   
    \caption{ Stellar mass - SFR relation for the MHUDF sample. The data points are colour-coded according to their effective S/N (equation  \ref{s}).      The black line shows the star-forming main sequence at $z$=0.85 (the mean redshift of our sample), as derived by \cite{bog}, whereas the grey shaded region shows the 0.44 dex scatter around the relation. The grey cross in the lower-right corner indicates the median $1\sigma$ statistical uncertainties in $M_{\star}$ and SFR  from \texttt{Magphys}.}
\label{sfr}
\end{figure}

In this section, we present the physical properties of the galaxies that met our previously described selection criteria. Figure \ref{histzm} shows the distribution of redshifts, stellar masss and Sérsic indices for our sample. The galaxies span redshifts between $0.28<z<1.49$  ($z_{\rm{mean}}$=0.85) and stellar masses $7.5<\rm{log}(M_{\star}/M_{\odot})<11$, with the distribution peaking around the mass regime where core formation is expected to be most efficient (e.g.  \citealt{dc14}, \citealt{toll}). Our sample contains disk galaxies with Sérsic indices peaking around $n=1$, whereas only a few galaxies have $n>2$. 

Figure \ref{sfr} shows the stellar mass-SFR relation for the MHUDF sample. Both stellar masses and SFRs were measured from SED fitting, as described in sections \ref{mhud}.  Most galaxies analysed in this study can be classified as star-forming main sequence (SFMS) galaxies, with only a subsample classified as starburst and/or in the process of quenching.

\section{Results}
\label{results}

In this section, we first test the six DM profiles and baryons-only model presented in Sect.~\ref{gpk2} using a Bayesian analysis to identify which best describe the data (Sect.~\ref{res:residuals}) and find that the DC14 profile is a good representation of the data. We then present the results  from the 3D disk-halo decomposition using the DC14 DM profile (Sect.~\ref{res:decomp}), and several cross-validation checks (Sect.~\ref{res:mge:mass}-\ref{res:crossvalid}). Finally, we present results on the DM fraction (Sect.~\ref{res:DMfractions}), on the DM profiles and the core-cusp fractions (Sect.~\ref{res:DMprof}), on the stellar-halo mass relation (Sect.~\ref{res:MsMh})  on the concentration-mass relation (Sect.~\ref{res:concentration}) and on concentration-density relation (Sect. ~\ref{res:concdens}).
 
\begin{figure*}%
    \centering
   \includegraphics[width=17cm]{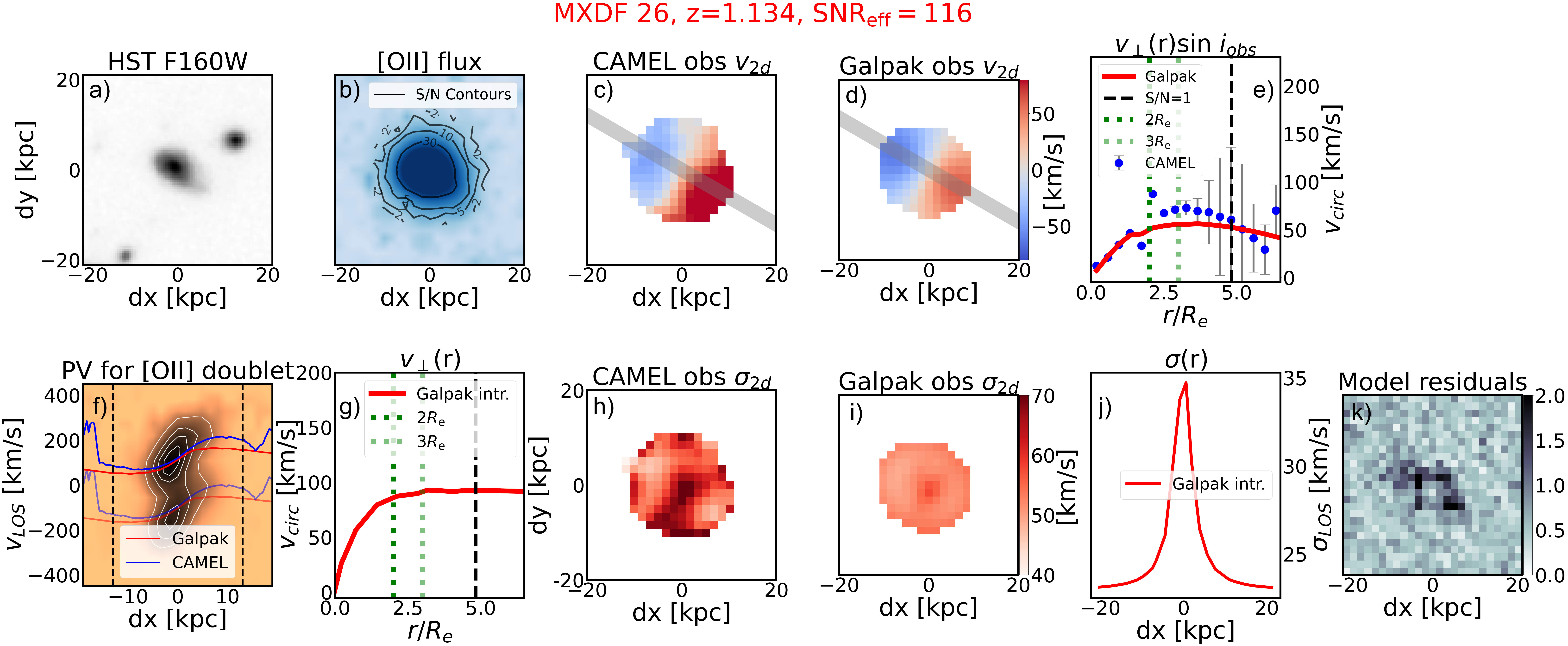} 
    \includegraphics[width=17cm]{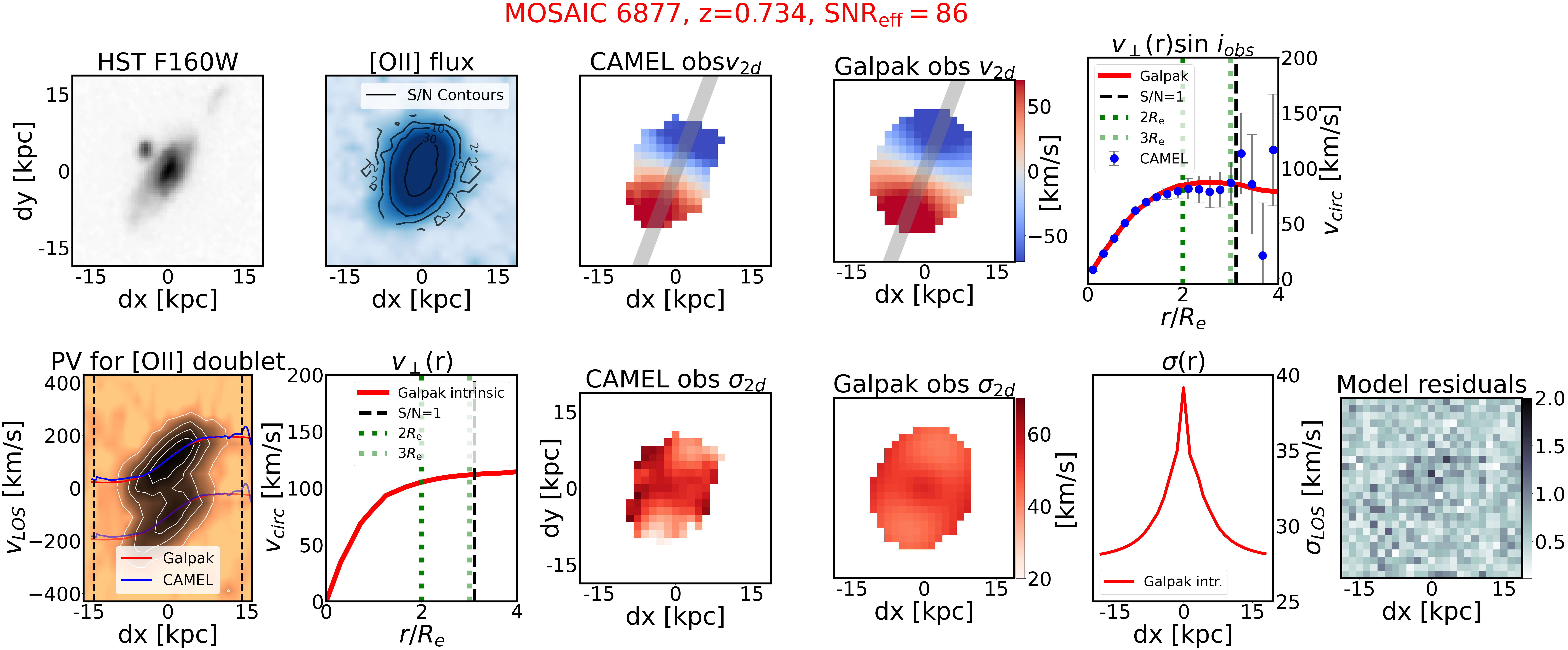}
    \includegraphics[width=17cm]{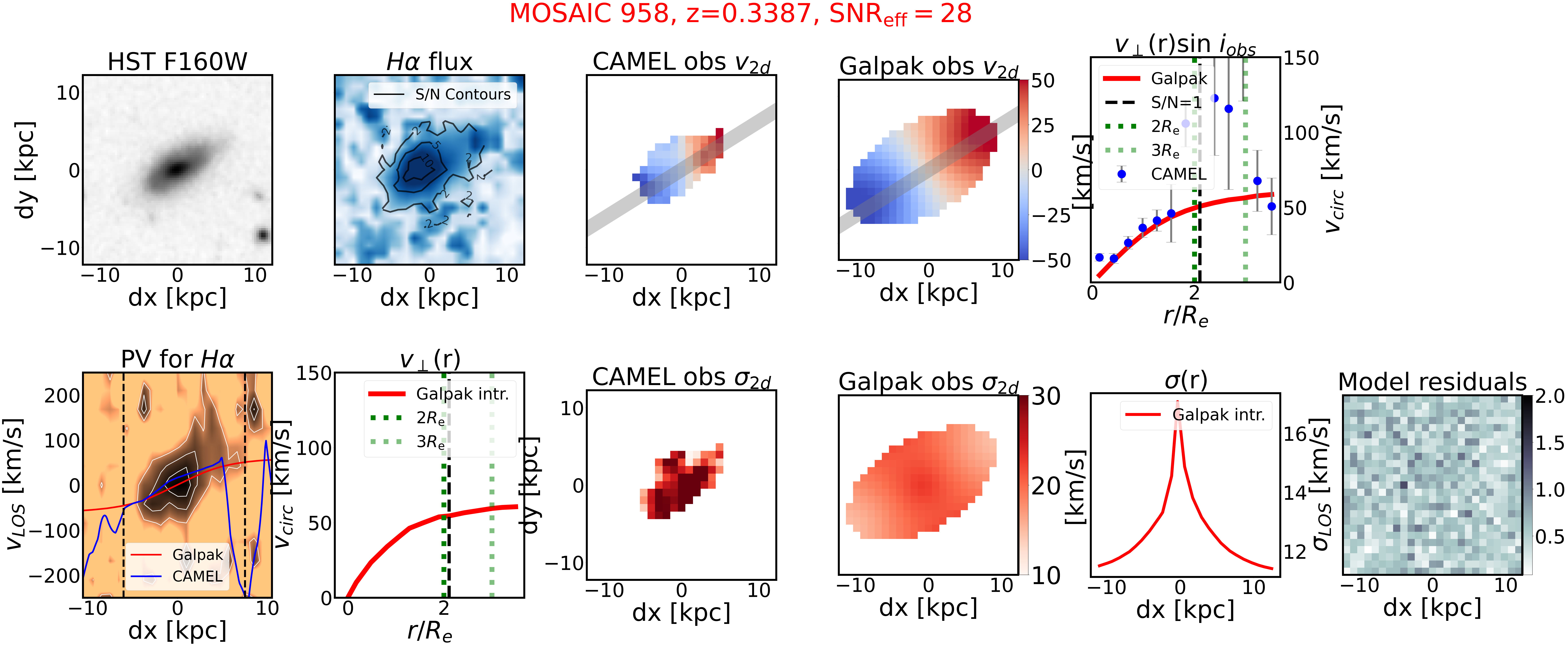} 
        \centering

   \caption{Example of morpho-kinematic maps for three galaxies, ID 26, 6877 and  958,  covering the range of S/N$\sim28-116$. For each galaxy, the 11 panels show (a) the HST F160W image, (b) the   emission-line flux map,   with the S/N contours overlaid, (c+d) the observed velocity maps in [km/s]  with the \texttt{CAMEL} \citep{camel}  and with \textsc{GalPaK$^{\rm 3D}$} (URC), with the gray line showing the major-axis,  (e) the observed velocity profile $v_{\perp}(r) \sin(i)_{\rm obs}$ extracted along the major-axis, (f) the position-velocity diagram  extracted along the major-axis, (g) the intrinsic velocity profile $v_{\perp}(r)$, i.e. corrected for inclination and instrumental effects, (h+i) the observed velocity dispersion maps in [km/s], (j) the intrinsic velocity dispersion profile in [km/s], and (k) the residuals map derived by computing the standard deviation along the wavelength axis of the normalized residual cube (see~\citealt{nicolas}). In panels (e, f) and (g), the vertical black dashed line represents the radius at which the S/N$\simeq1$, whereas the vertical dotted (light) green lines represent ($3R_{\rm{e}}$) $2R_{\rm{e}}$.}
        \centering

    \label{gpkplots}%
\end{figure*}

Before discussing the disk-halo decomposition, we show in Fig.~\ref{gpkplots} examples of the 3D morpho-kinematic fits for three SFGs (ID 26, 6877, 958) with various SNR ranging from $\sim30$ to $\sim100$ using the methodology described in Sect.~\ref{gpk1}. Additionally, the fluxes, S/N, velocities and velocity dispersions of the emission lines of interest in each spaxel of the MUSE cubes were also measured with a traditional 2D line fitting algorithm, \texttt{CAMEL}  \citep{camel}.  The position–velocity diagrams (PVs) shown in panel f) of the Fig. were extracted from the MUSE cubes, with the PV slice  taken along the kinematic major axis through the galaxy kinematic centre, using a 3-pixel-wide (0.6\arcsec) slit.

  Figure.~\ref{gpkplots}  demonstrates that owing to our deep data with high S/N,  both  \texttt{CAMEL} and {\textsc{GalPaK$^{\rm 3D}$}} enable us to probe the outer disk RCs, as illustrated in panels (e) by the (light-) green dotted lines which show ( $3R_{\rm{e}}$) $2R_{\rm{e}}$.  Nonetheless, {\textsc{GalPaK$^{\rm 3D}$}} proves more effective at tracing the outer parts of the RC, as it exploits the full  3D information from all spaxels, including those with low S/N. In contrast, the \texttt{CAMEL} measurements tend to become increasingly uncertain at large radii, where the low S/N of individual spaxels leads to noisier and sometimes biased velocity estimates.  This Fig. also shows that the disk kinematic 3D modelling reproduces the 3D data well,  in some cases with very little residuals. These fits were used to select rotation-dominated systems in Sect.~\ref{sampleselection}.

\begin{figure*}
   \centering
   \subfloat{{ \includegraphics[width=7.3cm]{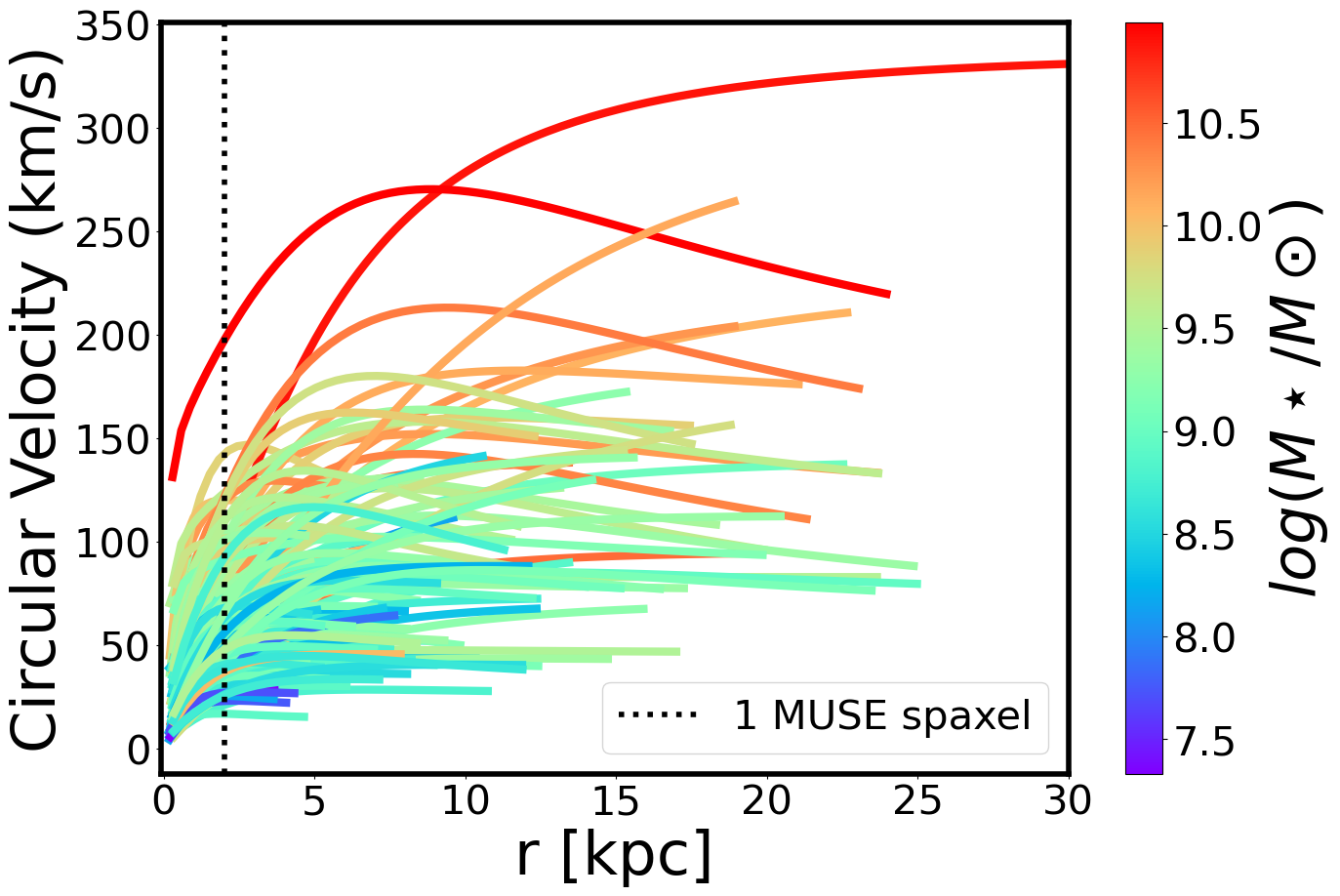}}}
    \qquad
    \subfloat{{\includegraphics[width=7.3cm]{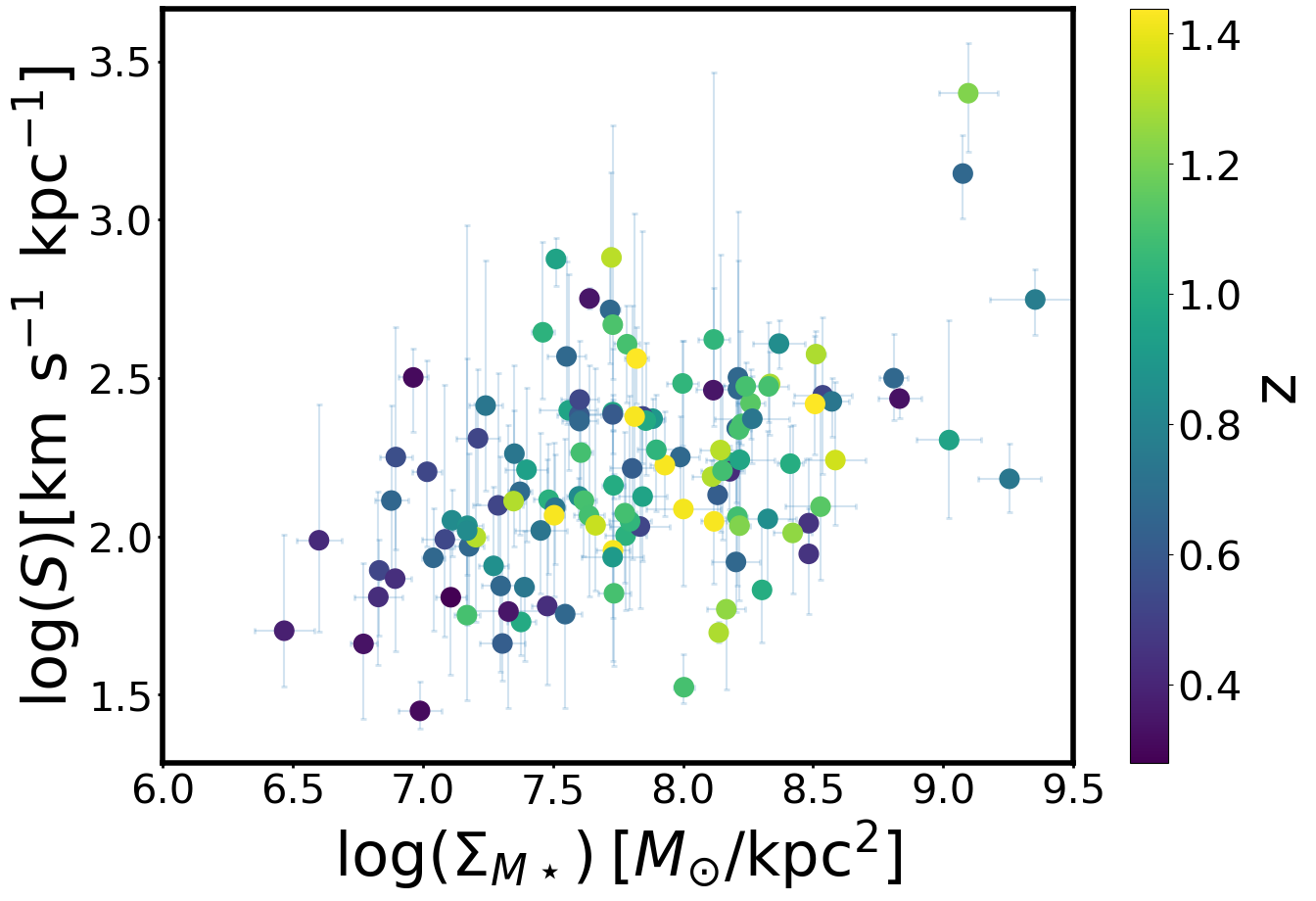} }}%   
    \caption{ Left: The RCs (using the URC model, Sect.~\ref{gpk1}) of our sample of intermediate-$z$ SFGs, color-coded by stellar mass.  The dotted black line shows the physical extent of 1 MUSE spaxel at $z\sim1$, namely 1.65 [kpc].  Right: The inner RC slopes, $S$ (corrected for inclination), as a function of the stellar mass surface density. The data points are colour-coded according to their $z$. The error bars on the y-axis show the 95\% confidence intervals from our 3D kinematic modelling, while the x-axis shows the $1\sigma$ uncertainties in $\Sigma_{M\star}$.}
\label{rcs}
\end{figure*}

Next, we show in figure~\ref{rcs}  the RCs of the full sample, estimated with the URC model (Sect.~\ref{gpk1}). The left panel shows model RCs colour-coded by stellar mass, while the right panel plots the inner RC slope, $S$, against the stellar mass surface density. Both panels reveal the same trend: galaxies with higher stellar mass surface densities—typically more massive, brighter systems—exhibit steeper inner slopes, whereas lower-mass galaxies show shallower rises. The sample spans a wide variety of RC shapes, from rising to flat and even mildly declining at large radii, broadly consistent with trends observed in the local Universe (e.g. \citealt{persic}, \citealt{katz}, \citealt{li}, \citealt{frosst}).

In the remainder of this paper, we will only present results obtained with the 3D disk-halo decomposition methodology presented in Sect.~\ref{gpk2}.

\label{res:morphokine}

\subsection{Model comparison and model selection}
\label{res:residuals} 

Using the  3D disk-halo decomposition methodology, we first identify which of the seven models --DC14, NFW, Burkert, Dekel-Zhao (DZ), Einasto,  coreNFW, baryons-only-- best describe the data  by computing the Bayes factor for each pair of models. Namely, we  use the model evidence $\log(\mathcal{Z})$ or marginal probability log$P(y\rm{|}M)$, defined as:
\begin{equation}
\log(\mathcal{Z}) = \ln \left( \int P(y|\theta, M) \, P(\theta|M) \, d\theta \right),
\end{equation}
 i.e. the integral of the posterior over the parameter space yielded by our MCMC fitting routine with \texttt{pyMultiNest} (\citealt{Buchner}).
$\log(\mathcal{Z})$ serves as a measure of how well the model explains the observed data, accounting for both goodness of fit and model complexity.  This metric is closely related to the Bayesian Information Criteria (BIC), which is BIC=$-2\ln \mathcal{\hat L}+k\ln(n)=\chi^2+k\ln(n)$ where $\mathcal{\hat L}=P(x| {\widehat {\theta }, M})$ is the maximum likelihood, $k$ the number of parameters, and $n$ the number of data points.

Following  \cite{kr}, we rescale the evidence $\log(\mathcal{Z})$ by a factor of  -2 so that it is on the same scale as the usual information criterion (BIC,  Deviance Information Criterion). We then compute the Bayes factor, defined as the ratio of the marginal probabilities of two competing models $M_1$ vs $M_2$: 
\begin{equation}
B_{12}=\frac{P(y|M_{1})}{P(y|M_{2})}=\log(\mathcal{Z}_{M_{1}})-\log(\mathcal{Z}_{M_{2}})
\end{equation}
for each pair of models (e.g. \citealt{Trotta}).  
Considering our scaling factor of -2 (see also \citealt{kr}), positive evidence against the null hypothesis,  i.e. that the two models are equivalent, occurs when the Bayes factor is $>3$,  whereas strong evidence occurs when the Bayes factor  $>20$. This corresponds to a logarithmic difference  $\rm{\Delta log(\mathcal{Z})}$ of 2 and 6, respectively.

\begin{figure*}
   \centering
    \includegraphics[width=1\textwidth,angle=0,clip=true]{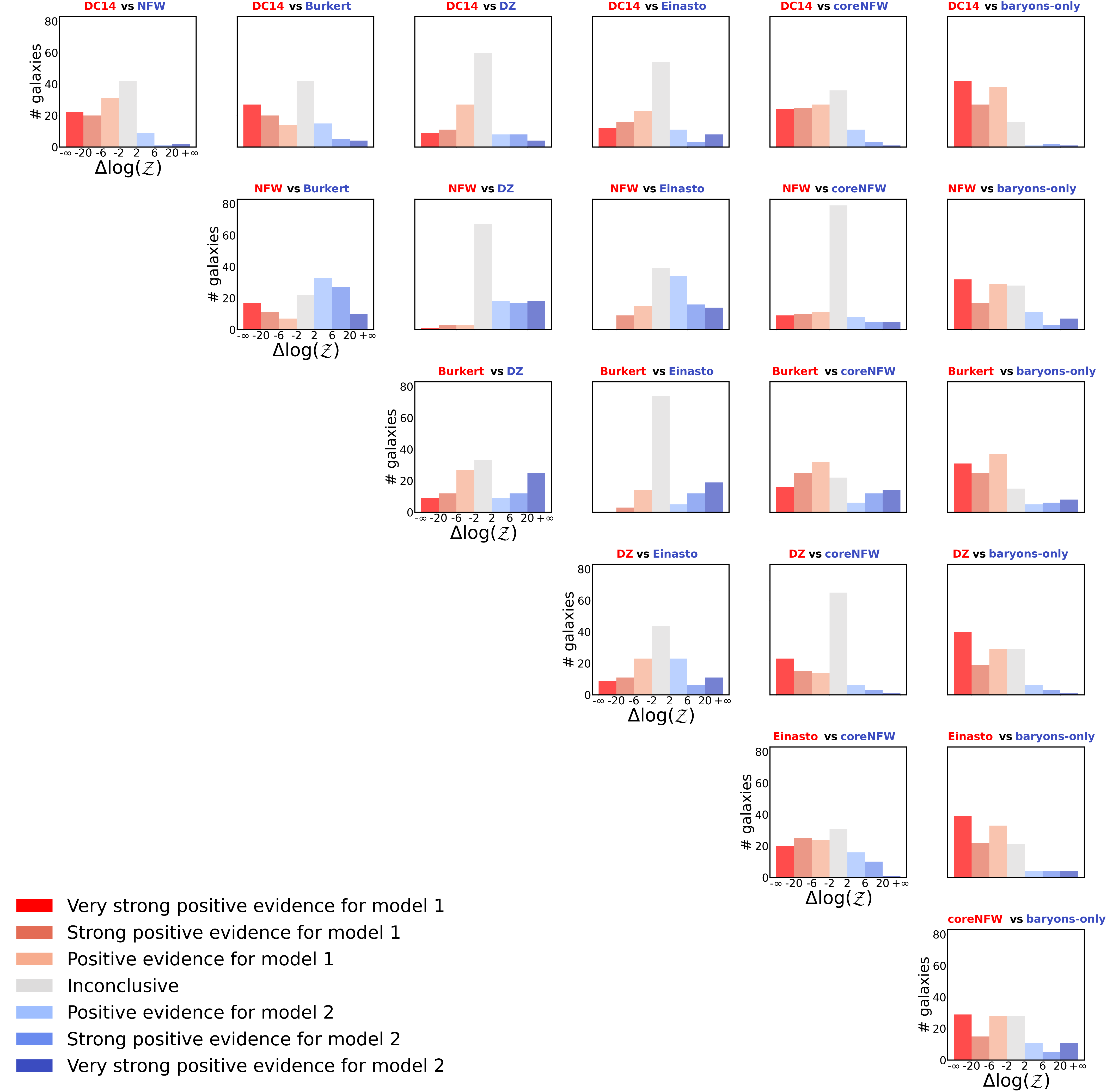} 
   \centering   
    \caption{ Histograms showing the Bayes factor for all the pairs of halo models used in this study.  The red columns show the number of galaxies that display a very strong positive evidence towards the first model, i.e. which have $-10^{3}<\rm{\Delta log(\mathcal{Z})}<-20$. The two lighter-red columns depict the number of galaxies that show strong positive evidence and positive evidence for the first model, i.e. which have  $-20<\rm{\Delta log(\mathcal{Z})}<-6$ and $-6<\rm{\Delta log(\mathcal{Z})}<-2$, respectively. The blue columns show the number of galaxies that show very strong positive evidence for the second model, i.e. $20<\rm{\Delta log(\mathcal{Z})}<10^{3}$, while the lighter-blue columns display the number of systems which show strong positive evidence and positive evidence  for the second model, namely $6<\rm{\Delta log(\mathcal{Z})}<20$ and $2<\rm{\Delta log(\mathcal{Z})}<6$, respectively. The grey columns display the number of galaxies for which both  models perform similarly, having $-2<\rm{\Delta log(\mathcal{Z})}<2$. In each panel, model 1 refers to the first model listed in the title, while model 2 refers to the second one.}
    \label{hist_baye}
\end{figure*}

Before discussing the results for the entire sample, we  present  in  Fig. \ref{figappendix:resid} from the Appendix  \ref{appenix:resid} the  residual maps  (generated from the residual cube as in Fig. \ref{gpkplots}) for each of the six DM profiles and the baryon only model  for the same three galaxies from Fig.~\ref{gpkplots} and one extra galaxy from our sample.  Alongside the maps, we report the Bayesian evidence (black text) and the Bayes factor (colour-coded as in Fig.~\ref{hist_baye}) between DC14 and the competing models, and highlight the connection between the Bayes factor and the residual structure. We show that when the Bayes factor indicates strong positive evidence  for  a particular model,  the corresponding residual map consistently exhibits lower residuals compared to the alternatives. We refer the reader to  Appendix  \ref{appenix:resid} for more details.

Regarding the entire sample, Fig.~\ref{hist_baye} shows $\Delta \log(\mathcal{Z})$ for all the DM model pairs  where  the red (blue) bars show the number of galaxies that display strong evidence for (against) the first model, while the grey bars correspond to the number of galaxies for which the two models are equivalent. 
The first panel of this figure indicates that, compared to NFW, DC14 provides fits that are equally good or better in $\sim$90\% of the sample, broadly consistent with the trends reported by \cite{katz} for the SPARC sample.

Similarly, the first row shows that, compared to Burkert, Dekel-Zhao, Einasto, coreNFW, and baryon-only models, DC14 performs at least as well or better in 81\%, 84\%, 82\%, 88\%, and 96\% of the sample, respectively (see Table~\ref{tab1} for details). Taken together, these comparisons suggest that DC14 is among the profiles that best represent the data, performing comparably to the Dekel-Zhao and Einasto models. This is in line with earlier findings, such as those of \cite{li19}, who showed that cored profiles like DC14 and Einasto can provide better fits than cuspy NFW models for the local SPARC sample. 

Fig.~\ref{hist_baye} reveals that for many galaxies ($\approx10$--50\%) in our sample, the two models under consideration are statistically equivalent, i.e. fit the data similarly well  (grey bars in Fig. \ref{hist_baye}).  In order to understand  whether the S/N of the data played a role, we repeated this analysis using only systems which have $\rm{S/N_{eff}}>50$ (44 galaxies) and found that the number of galaxies for which the two models are indistinguishable decreases below 20\% in most cases (see Table \ref{tab2} for more details). Indeed, we found that the DC14 model tends to  describe the kinematics  of the high S/N data better than the alternatives.
 
For the baryon-only model, the last column of Fig.~\ref{hist_baye} shows that it generally performs less well than models that include a DM halo. In addition, the baryon-only model yields dynamically inferred disk masses ($M_\star + M_{\rm mol}$) that are systematically higher than the stellar masses derived from SED fitting. After accounting for molecular gas using scaling relations \citep[e.g.,][]{fre2,tac}, we find that the baryon-only model still overpredicts stellar masses by $\sim$0.76 dex (for a Chabrier IMF) or $\sim$0.53 dex (for a Salpeter IMF), with offsets up to $\sim$1 dex in some cases (plots not shown). These offsets confirm that even when adopting a heavier IMF, the baryon-only model cannot reconcile the observed kinematics with the SED-inferred stellar masses. Furthermore, the typical uncertainties in SED-inferred stellar masses due to variations in modelling assumptions—such as SFH, dust, metallicity, and IMF—are estimated to be in the range of 0.2-0.3 dex at maximum (e.g. \citealt{Mobasher}), which is  below the magnitude of the observed offset in our sample. In contrast, the DC14 model yields dynamically inferred disk masses that agree well with the SED-based estimates (see Section~\ref{res:selfcheck} for details). The challenges we find in reconciling baryon-only models with both kinematic and SED-based constraints are in line with earlier results such as those of \citet{bar}, who suggested that disk galaxies are typically not maximal, especially in the mass range probed here. 

To conclude, our analysis suggests that the DC14 density profile (Eq.~\ref{dc14rhosrs}) provides, on average, a better description of the kinematics of intermediate-$z$ SFGs than the other profiles considered, with the Dekel-Zhao and Einasto models also performing comparably well, while baryon-only models are broadly disfavored with respect to models which include a DM component.
In the remainder of the paper, we will  show results obtained with the DC14 halo model. We refer the reader to the appendix (\ref{a1}, \ref{a2} and \ref{a3} ) for the results inferred using all six DM halo models.

\begin{table}
\caption{Comparison of DM halo models for the whole sample.}
\centering
\small
\begin{tabular}{lccccl}
\hline
 DM halo model  &  More likely &Inconclusive &  Less likely   \\                                                                                              
 \hline
NFW&12 &42& 73\\
Burkert &  24 &42 & 61\\
Dekel-Zhao&20 &60& 47\\
Einasto&22 &54& 51\\
coreNFW&15 &36& 76\\
baryons-only&4&16&107\\\hline
\end{tabular}
\tablefoot{Numbers indicate the count of galaxies for which each DM model is more or less likely than the reference model, DC14, based on Bayesian evidence.}
\label{tab1} 
\end{table}

\begin{table}
\caption{Comparison of DM halo models for high S/N galaxies.}
\centering
\small
\begin{tabular}{lccccl}
\hline
 DM halo model  &  More likely  & Inconclusive &  Less likely  \\                                                                                              
 \hline
NFW & 8 &5& 31\\
Burkert & 7 & 3 & 34\\
Dekel-Zhao& 15 &8& 21  \\
Einasto&12 &6& 26\\
coreNFW&10 &  2 & 32\\
baryons-only&2&1&41\\\hline
\end{tabular}
\tablefoot{Same as Table \ref{tab1}, but only considering galaxies with $\rm{S/N_{eff} > 50}$.}
\label{tab2} 
\end{table}

\subsection{3D Disk-halo decompositions}
\label{res:decomp}
\begin{figure*}
   \centering
    \includegraphics[width=1\textwidth,angle=0,clip=true]{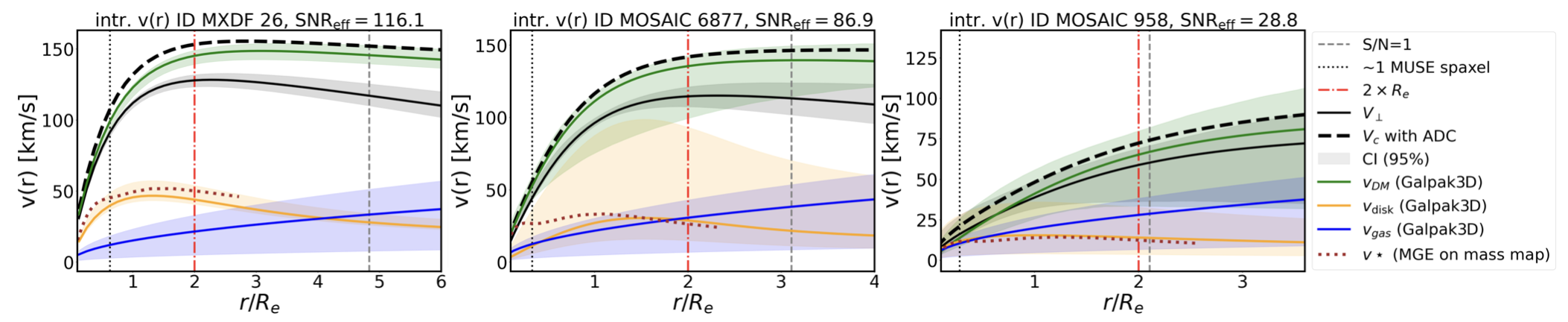}
   \centering   
     \caption{Examples of the 3D disk-halo decomposition for the three galaxies show in Fig.~\ref{gpkplots} using a DC14 DM profile (Eq. \ref{eqdc14}).  
The solid black line represents the rotational  velocity $v_{\rm{\perp}}(r)$. The thick dashed black line represents the circular velocity $v_{\rm{c}}(r)$, that is $v_{\rm{\perp}}(r)$ corrected for pressure support (Sect.~\ref{gpk2}). The green curve represents the RC of the DM component. The solid orange and blue curves represent the disk and \HI\ components, respectively.  The light shaded regions show the 95\% confidence intervals.  The dotted   brown  lines represent the stellar component obtained using the MGE modelling of the stellar mass maps (see Sect. \ref{res:mge:mass}). All velocities are ‘intrinsic’, i.e. corrected for inclination and instrumental effects, including seeing (PSF). 
  The black dotted line shows the physical extent of a MUSE spaxel, the red dot-dashed  line shows $2\times R_{\rm e}$, while the grey dashed line shows the region beyond which S/N<1.}%
    \label{RCs}%
\end{figure*}

 Figure~\ref{RCs} shows the results from the 3D disk-halo decompositions for the same three galaxies shown in Fig.~\ref{gpkplots}   
 using the DC14 DM halo profile (Eqs.~\ref{eqdc14}-\ref{gamma}). 
 All velocities are ‘intrinsic’, i.e. corrected for inclination and instrumental effects, including seeing.  This figure shows that the RCs are dominated by the DM component. Consequently, it is worth noting that, the RC data contains information on the halo shape parameters ($\alpha(X),\beta(X),\gamma(X))$, which in turn, yields a constrain on the ratio $M_{\rm disk}/M_{\rm vir}$ for DC14 DM profiles. As a result, this allows to break the traditional disk-halo degeneracy.

Before discussing the results on the DC14 profiles, we perform several cross-checks of our 3D disk-halo decomposition.
We first compare (Sect.~\ref{res:mge:mass}) our disk component to the one obtained from the stellar maps derived from HST and JWST photometry (described in Appendix~\ref{Appendix:mass}).
We then perform 1D disk-halo decomposition (Sect.~\ref{res:crosscheck}) from external data using the stellar surface brightness and several \HI\ profiles.
Finally, we perform some self-consistency checks (Sect.~\ref{res:selfcheck}-\ref{res:crossvalid}) for DC14.

\subsubsection{Stellar gravitational field from mass maps}

\label{res:mge:mass}
  
To estimate the stellar disk contribution to the disk–halo decomposition, we applied the MGE method (\citealt{Monnet}, \citealt{mge}) to the stellar mass maps derived from pixel SED fitting (Appendix~\ref{Appendix:mass}). This was done using the \texttt{mgefit} Python package of \cite{cap}, and we considered the PSF when fitting the 2D stellar maps. Each MGE model comprises concentric 2D Gaussians characterised by peak intensity, dispersion, and axial ratio or flattening~\footnote{Following  \cite{sc}, we  optimise the allowed range of axial ratios  until the fits become unacceptable, using as a lower limit the axial ratio  from   Sect.~\ref{global_properties}. 
%Convergence is achieved when the mean absolute deviation of the model for a given axial ratio pair increases by less than 10\% in the subsequent step.
}. Finally, based on the MGE parameters, we use the  \texttt{mge-vcirc} module from the \texttt{jampy} Python package of  \cite{cap2}, to calculate the gravitational acceleration ($v^2/r$) in the equatorial plane of each galaxy.  This approach models the mass distribution using an MGE representation  directly from the mass maps, under the assumption of dynamical equilibrium (thereby eliminating the need to assume a constant mass-to-light ratio as is typically required when performing MGE modelling on galaxy  images).

While the ancillary imaging used to derive the mass maps includes long-wavelength coverage that helps mitigate uncertainties due to dust attenuation, we note that stellar masses derived from SED fitting can still be subject to systematic uncertainties of up to $\sim$0.3 dex (e.g., \citealt{Mobasher}). Such uncertainties propagate into the stellar mass maps, and therefore into the inferred stellar circular velocities.  

With these limitations in mind, we show in Fig. \ref{RCs}, as the dotted brown curves, the predicted circular velocity of the stellar mass distribution, as derived from the mass maps. As this Fig. demonstrates, there is good agreement between the $v_{\star}(r)$ determined from the mass maps and the MUSE data.  For approximately $80\%$ of the galaxies in our sample, the  RCs obtained from the two methods agree within the errors.  Moreover, the half-light radii and surface brightness profiles from the mass maps are very similar to those  obtained from the ionised gas tracers.

Discrepancies in the order of $\sim 10$–20 km/s, where present, are mostly confined to the central regions, and remain within the uncertainties of the {\textsc{GalPaK$^{\rm 3D}$}} velocities (yellow shaded regions in Fig.~\ref{RCs}). These central differences likely arise from the sensitivity of the MGE method to the precise PSF modelling.  

Crucially, the agreement between the MGE-based and {\textsc{GalPaK$^{\rm 3D}$}} stellar circular velocities for the majority of the sample provides an independent validation of our disk-halo decomposition. While {\textsc{GalPaK$^{\rm 3D}$}} assumes an Sérsic  disk profile, the MGE approach does not impose a parametric form on the stellar mass distribution. The fact that both methods yield broadly consistent stellar contributions implies that the stars in our sample galaxies are indeed in (near) equilibrium disks. The results obtained in this section were used as a qualitative check against the results obtained with our 3D parametric modelling.

\subsubsection{Cross-checks with 1D disk-halo decomposition}
\label{res:crosscheck}

To independently perform the disk–halo decomposition from the observed RC, we compute the expected baryonic contribution by solving the Poisson equation for arbitrary surface density distributions (i.e. solving Equation 4 of \citealt{Casertano}) using the \texttt{vcdisk} tool~\footnote{\url{https://vcdisk.readthedocs.io/en/latest/}}, assuming a vertical structure of the disk as in Sect. ~\ref{gpk2}. For the disk component, $v_{\mathrm{disk}}(r)$, we use a Sérsic surface brightness profile with parameters derived from the HST/F160W data (\citealt{morph}, see Sect. \ref{Morphology}), assuming a constant mass-to-light ratio. For the \HI\ component, we use the radial profiles from \cite{Martinsson} or from \cite{Wang2}, where the \HI\ masses are inferred from the stellar masses of our sample using the $z\sim1$ $M_{\HI}-M_{\star}$ scaling relation of \cite{Chowdhury} (their eq. 1).

\begin{figure*}
   \centering
    \includegraphics[width=1\textwidth,angle=0,clip=true]{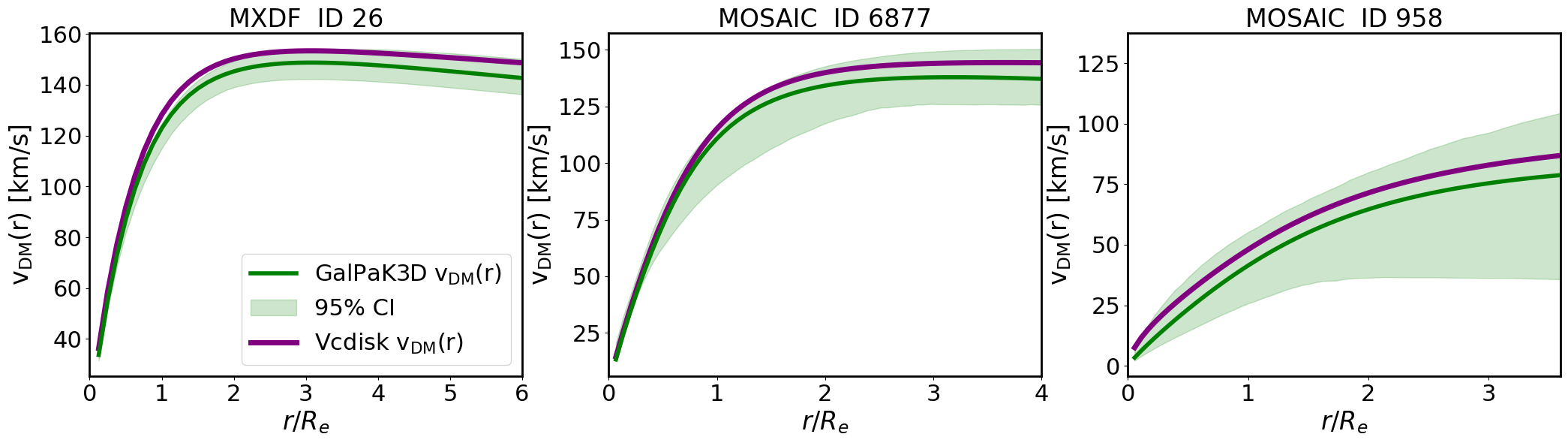}
   \centering   
\caption{Comparison of DM velocity profiles for the three galaxies shown in Fig.~\ref{RCs}. The green curves show the profiles modelled with {\textsc{GalPaK$^{\rm 3D}$}} assuming a DC14 halo, while the purple curves are obtained by subtracting the baryonic contributions obtained with \texttt{vcdisk} from the MUSE-derived $v_{\rm c}(r)$. The green shaded regions indicate the 95\% confidence intervals from the 3D disk–halo decomposition fits.}
\label{vcdisk}
\end{figure*}

The purple curves in Fig.~\ref{vcdisk} show the resulting DM velocity curves derived using Eq.~\ref{eq:vdecomp}, i.e. by subtracting the baryonic components computed with \texttt{vcdisk} from the total measured $v_{\rm c}$ from MUSE, for the same three galaxies shown in Fig.~\ref{RCs}.  For the examples shown here, we  used   the   \cite{Wang2}  \HI\ profile for the gas component; we note, however, that using the \cite{Martinsson} profile leads to similar results. The {\textsc{GalPaK$^{\rm 3D}$}}-based velocity curves, shown in green, assume a DC14 DM halo. As illustrated  in  Fig.~\ref{vcdisk}, the 1D and 3D decompositions agree within the uncertainties.  We find consistent DM contributions between the two methods for  $\sim 87\%$ of the galaxies in our sample. The discrepancy, when present, mostly occurs for the perturbed, as well as for compact, low $\rm{S/N_{eff}}$ galaxies.

\subsubsection{Self-consistency checks for DC14}
\label{res:selfcheck}

In this section, we check for consistency between the  $M_{\star}$ yielded by our 3D disk-halo decomposition and $M_\star$ derived  from SED fitting. We remind the reader that the stellar mass was a free parameter  (contained in $X$) in our disk-halo decomposition using DC14 and the disk-halo degeneracy is broken because (i) the RC shape is driven by the shape parameters which depends on  $X\equiv\log(M_\star/M_{\rm vir})$, and (ii) the RC also yields $v_{\rm vir}$ from the maximum velocity.

\begin{figure*}
   \centering
    \includegraphics[width=0.4\textwidth,angle=0,clip=true]{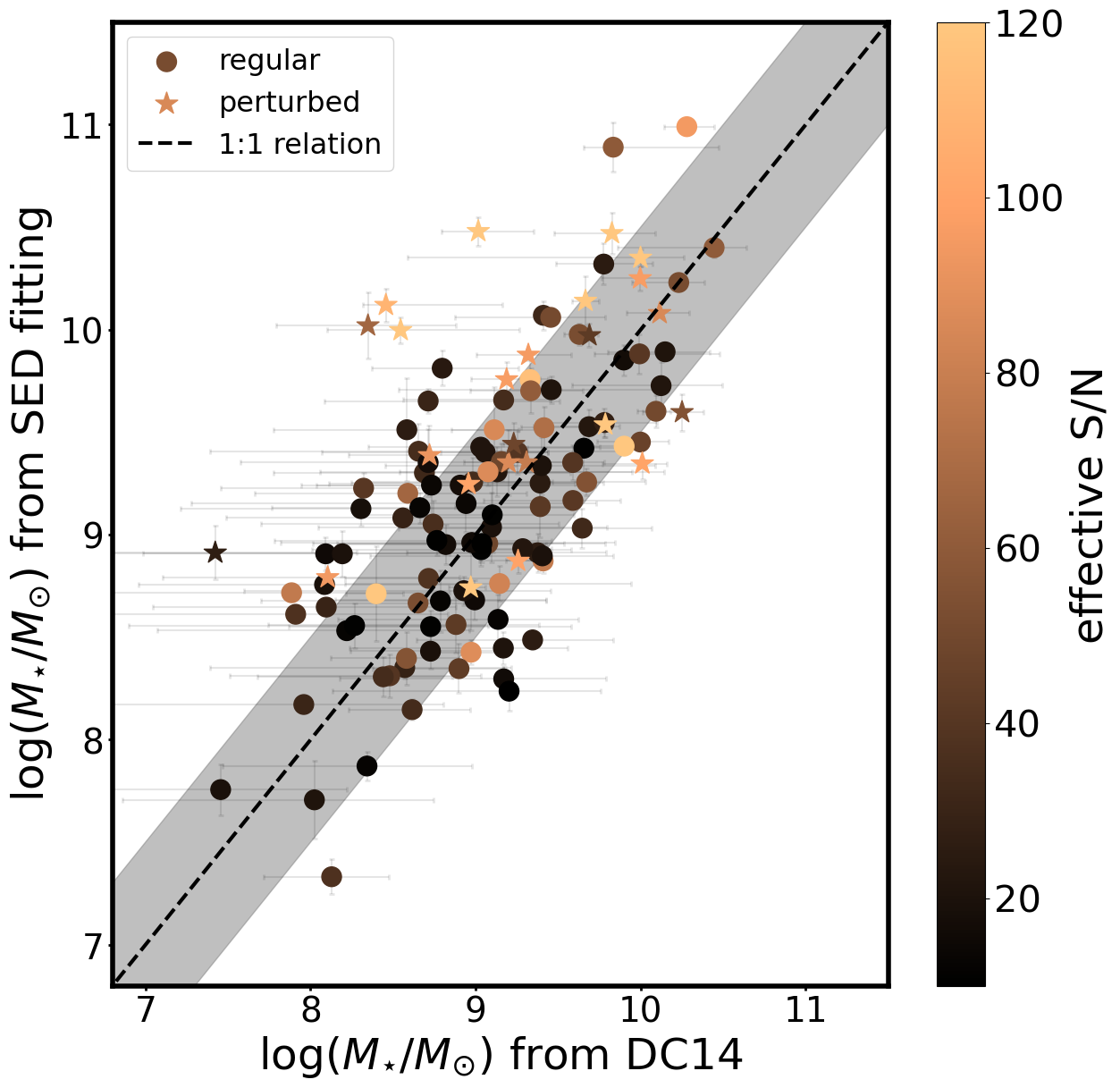}
     \includegraphics[width=0.4\textwidth,angle=0,clip=true]{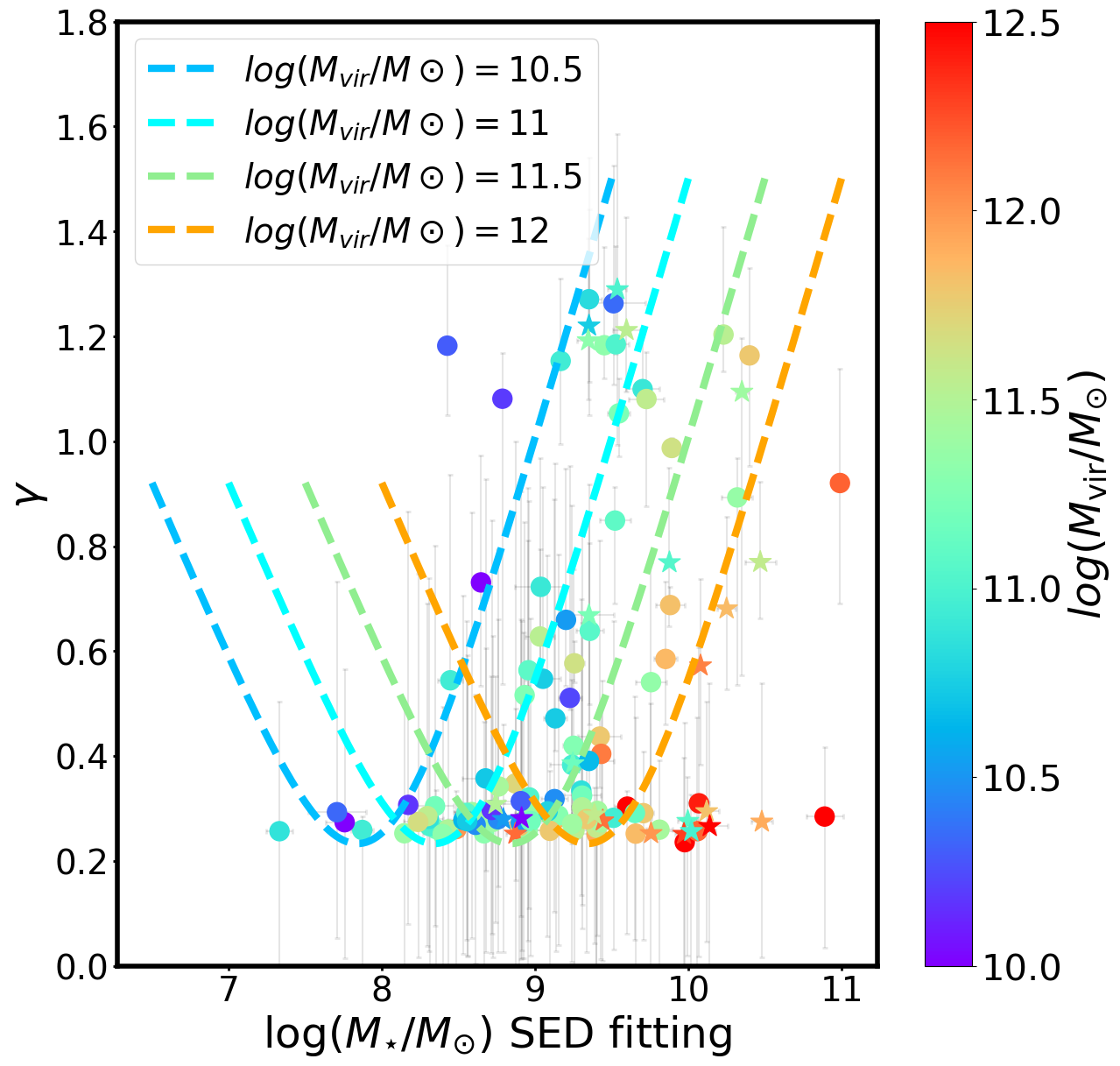}  
   \centering   
    \caption{{\it Left:} Comparison between the stellar masses inferred from our disk-halo decomposition using the DC14 halo model and the stellar masses inferred from photometric observations. The black dashed line shows the 1:1 relation, and the grey shaded region shows the 0.5 dex dispersion around the relation.   The data points are colour-coded according to their $\rm{S/N_{eff}}$.  
 {\it Right:}DM inner slope, $\gamma$, as inferred from the DC14 halo model as a function of the stellar masses derived from SED fitting. The data points are colour-coded according to their virial masses. The coloured curves depict the parametrisation of \cite{dc14} for $\gamma$ (equation \ref{gamma}), for four different virial masses, as indicated in the legend.  In both panels,  the circles show the regular MHUDF galaxies, while the stars depict the perturbed galaxies from our sample. The error bars represent the  95\% confidence intervals.
 }
\label{fig:selfcheckdc14}
\end{figure*}

Figure \ref{fig:selfcheckdc14} (left) shows the comparison between the kinematically inferred and SED-based stellar masses. The majority of galaxies lie within a $<\pm0.5$ dex region (grey shaded area) around the one-to-one relation (dashed line), indicating general agreement between the two estimates. The mean offset is modest, with dynamically inferred stellar masses being on average 0.11 dex lower than the SED-based stellar masses. The largest discrepancies between SED and kinematic-based $M_{\star}$ occur for the perturbed galaxies  as well as for compact, low $\rm{S/N_{eff}}$ galaxies. Excluding the perturbed galaxies and performing a linear fit to the data yields a slope of $\sim0.82 \pm 0.1$, which is close to the expected slope of 1 for the relation between $M_{\star}$ from SED fitting and $M_{\star}$ from the kinematic analysis. Overall, this figure demonstrates that our disk-halo decomposition produces stellar masses broadly consistent with those derived from SED fitting.

\subsubsection{Cross-validation check for DC14}
\label{res:crossvalid}

In an attempt to independently verify the DC14 framework of varying inner DM slopes with mass, we show
in  Figure \ref{fig:selfcheckdc14}(right) the inner DM slope, $\gamma$, as a function of  the stellar mass inferred from SED fitting (i.e. independent of our  {\textsc{ GalPaK$^{\rm 3D}$}} fits)  with the data points colour coded according to their virial masses obtained from the 3D kinematic modelling. This Fig.  indicates that the DM slopes ($\gamma$ values) we infer using  {\textsc{ GalPaK$^{\rm 3D}$}} agree well with the DC14 expectations for the respective masses.  % 

\subsection{DM fractions}
\label{res:DMfractions}

\begin{figure}
   \centering
    \includegraphics[width=0.5\textwidth,angle=0,clip=true]{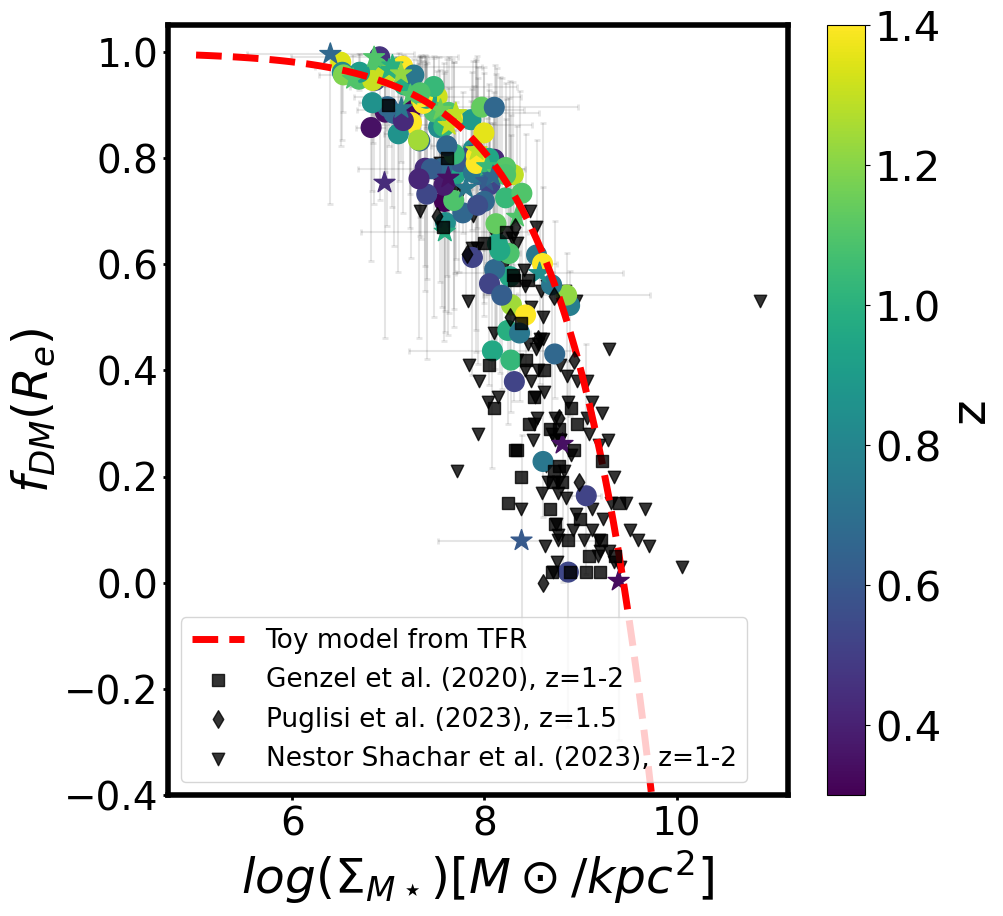}  
       \centering   
    \caption{DM fractions within $R_{\rm{e}}$ as a function of the stellar mass surface density.  The black diamonds show the DM fractions inferred by \cite{pug} for $z\sim1.5$ SFGs,  the black squares depict the $z=1-2$ sample of \cite{genz2}, while the black triangles show the RC100 sample from \cite{shach}. Our sample is colour-coded according to its redshifts.  The  circles depict the regular  galaxies, while the stars represent the perturbed systems from our sample.  The error bars represent the  95\% confidence intervals.  The red dotted line shows the toy model derived from the TFR relation (see \citealt{nicolas}, Section 5.1). }
\label{frac}
\end{figure}

In this section, we investigate the DM fraction, $f_{\rm{DM}}(<R_{\rm{e}})$, of our sample by integrating the DM and disk mass profile to $R_{\rm{e}}$.  Figure \ref{frac} shows $f_{\rm{DM}}(<R_{\rm{e}})$, as a function of the stellar mass surface density. Additionally, we also plot in this figure the DM fractions obtained by \cite{pug} for SFGs at $z\sim1.5$, as well as the DM fractions inferred by   \cite{genz2}  and \cite{shach} for SFGs  at $z$=1-2 with black symbols. We find that 89\% of our sample has $f_{\rm{DM}}(<R_{\rm{e}})> 50\%$.  

For a given stellar mass surface density, $\Sigma_{M_{\star}}$, we infer $f_{\rm{DM}}(<R_{\rm{e}})$ for the MHUDF galaxies, consistent with those reported by \cite{pug}, \cite{genz2} and \cite{shach} for their samples. However, compared to the SFGs analysed in the three aforementioned studies, our sample extends  to the lower mass regime, and, thus,  to lower mass surface densities $\log(\Sigma_{M_{\star}}) <8 \:M_{\odot} \rm{kpc}^{-2}$.  As this Fig. illustrates, the galaxies follow a relatively tight relation in the $f_{\rm DM}(<R_{\rm e})$–$\Sigma_{M_\star}$ plane, consistent with \citet{genz2, pug, shach}. This trend likely reflects the underlying Tully–Fisher relation for disks \citep[e.g.][]{ubler}, as illustrated by the simple toy model at $z\sim1$ (red dotted line;  see \citealt{nicolas}, Section 5.1).

\subsection{DM density profiles and inner slopes}
\label{res:DMprof}
\begin{figure*}%
   \centering
   \includegraphics[width=0.87\textwidth,angle=0,clip=true]{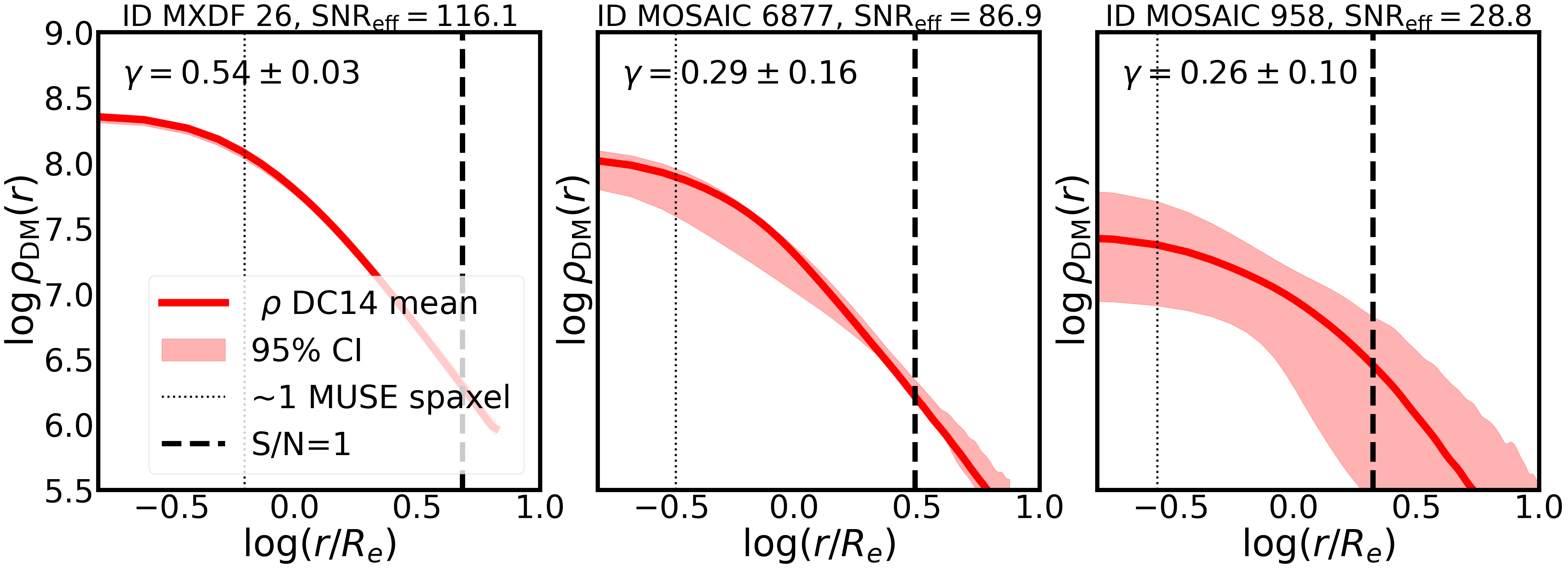}

   \caption{Example of DM density profiles for the same three galaxies shown in Fig.~\ref{RCs}. The y-axis shows the DM density, $\rho_{\rm DM}(r)$, in units of $\rm M_{\odot}/kpc^3$, and the x-axis the $\rm r/R_e$. The red curve shows the DM density profile obtained using the DC14 halo model, whereas the red-shaded region represents the 95\% confidence interval. The inferred DM inner slopes, $\gamma$, are indicated in the upper-left corner of each panel. The vertical black dashed lines represent the 0.2" spaxel scale,  indicating the lower limit of our constraints.} 
   \label{densities}%
\end{figure*}

In this section, we discuss the DM inner slopes, $\gamma$, of our sample. The  three panels of Fig. \ref{densities} show the density profiles  for the same three galaxies presented in Fig. \ref{RCs}. The inferred  $\gamma$ values are reported in the upper-right corner of each panel. From left to right, the profiles become progressively more cored, as illustrated by the overall flattening of the central density. The dashed vertical line marks the physical scale of a single MUSE spaxel. Although MUSE’s spatial resolution limits direct probing of the inner $\lesssim$1-2 kpc, the global curvature of the RC over several kpc retains information on the central DM slope: cored profiles produce a shallower rise compared to cuspy ones, and this distinction remains observable at MUSE resolution (Fig. \ref{RCs}). The robustness of our slope measurements is further supported by tests on mock MUSE cubes (Sect. \ref{valid}). These tests demonstrate that the inferred $\gamma$ values recover the true input slopes within $1\sigma$.

Globally, our analysis reveals that $\sim 66\%$ of our sample of intermediate-$z$ SFGs exhibits cored DM profiles, with $\gamma < 0.5$. 
Note, observational evidence exists for DM cores up to $z\sim2$ \citep[e.g.][]{genz, Sharma2, shach}, and possibly as far as $z\sim6$ \citep{Danhaive}.
We discuss these findings further in Sect~\ref{coreform}.

\subsection{Stellar-halo mass relation}  
\label{res:MsMh}

 The $M_{\star}-M_{\rm{vir}}$ relation in the $\rm{\Lambda CDM}$ framework is a well-known scaling relation describing the relationship between the mass of a DM halo and the stellar mass of the galaxies residing within it. This scaling relation reflects the SF efficiency, and thus the efficiency of feedback processes,  providing crucial insights into galaxy formation models (e.g. \citealt{vo}, \citealt{mos}). Several recent studies have investigated the stellar mass–halo mass relation using RCs of individual galaxies  (e.g., \citealt{read2}, \citealt{Paolo}, \citealt{Posti2}, \citealt{li}, \citealt{Romeo}, \citealt{2025arXiv251108685M}), generally finding a considerable scatter in the relation.
  
Figure \ref{msmh} from Appendix \ref{a1} shows the stellar mass--halo mass relation for our sample. We show the SED-based stellar masses vs the virial masses,  as inferred from the 3D disk-halo decomposition, using all the different halo models. The upper-left panel shows the DC14 results, which yield the tightest $M_{\star}-M_{\rm{vir}}$ relation amongst all the tested halo models. We find  that our results are qualitatively in good agreement with the relation inferred by \cite{Behroozi}, as well as the one derived by \cite{Girelli} using the (sub)halo abundance matching technique, albeit with a larger scatter.   The latter two studies quote a scatter in the $M_{\star}-M_{\rm{vir}}$ relation of about $\sim 0.2- 0.4$ dex, which is much smaller than the scatter  $\gtrsim1.5$ dex we measure for the MHUDF galaxies. Similar results were found by \cite{li} for the local SPARC sample, using DC14 without $\Lambda$CDM priors.  It is also worth noting that  our sample does not probe the high-mass end, and therefore does not reach the bend in the $M_{\star}-M_{\rm{vir}}$  relations. In Appendix \ref{a1}, we further discuss the results obtained with the other halo models.

 \subsection{Concentration-mass relation}
 \label{res:concentration}
 
A fundamental prediction of N-body $\Lambda$CDM simulations is the concentration of DM haloes, $c_{\rm vir} \equiv r_{\rm vir}/r_{\rm s}$, which encodes the assembly history of the halo. The concentration is expected to decrease with increasing halo mass and redshift, following $c_{\rm vir} \propto M_{\rm vir}^{-0.13}$ and $c_{\rm vir} \propto (1+z)^{-1}$ \citep{bull, weh}, as a consequence of the redshift evolution of the virial radius $r_{\rm vir}$. The $c_{\rm{vir}}$–$M_{\rm{vir}}$ relation has been tested using RCs in the local Universe \citep[e.g.,][]{alle, katz, li}, but remains poorly constrained at higher redshifts for galaxy-scale halos \citep[e.g.,][]{nicolas, Sharma2}, and has been primarily probed at the high-mass end using galaxy clusters \citep[e.g.,][]{biv}.

\begin{figure*}%
    \centering
    \subfloat{{\includegraphics[width=8cm]{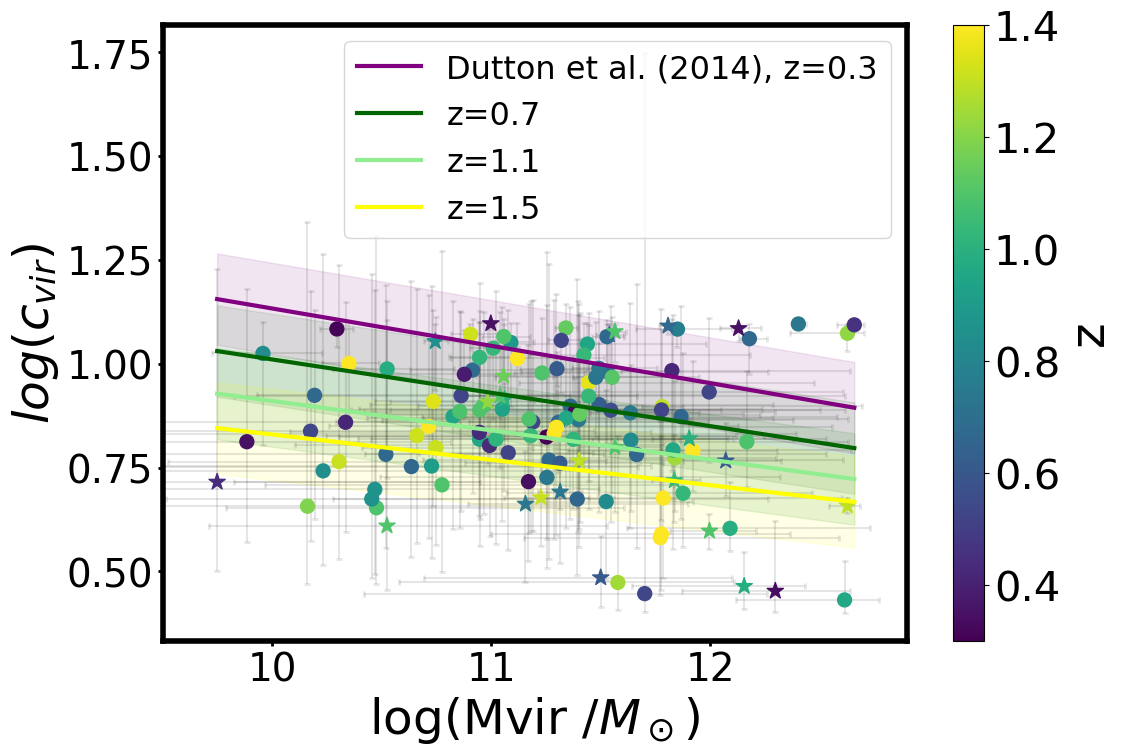} }}%
    \qquad
    \subfloat{{\includegraphics[width=8cm]{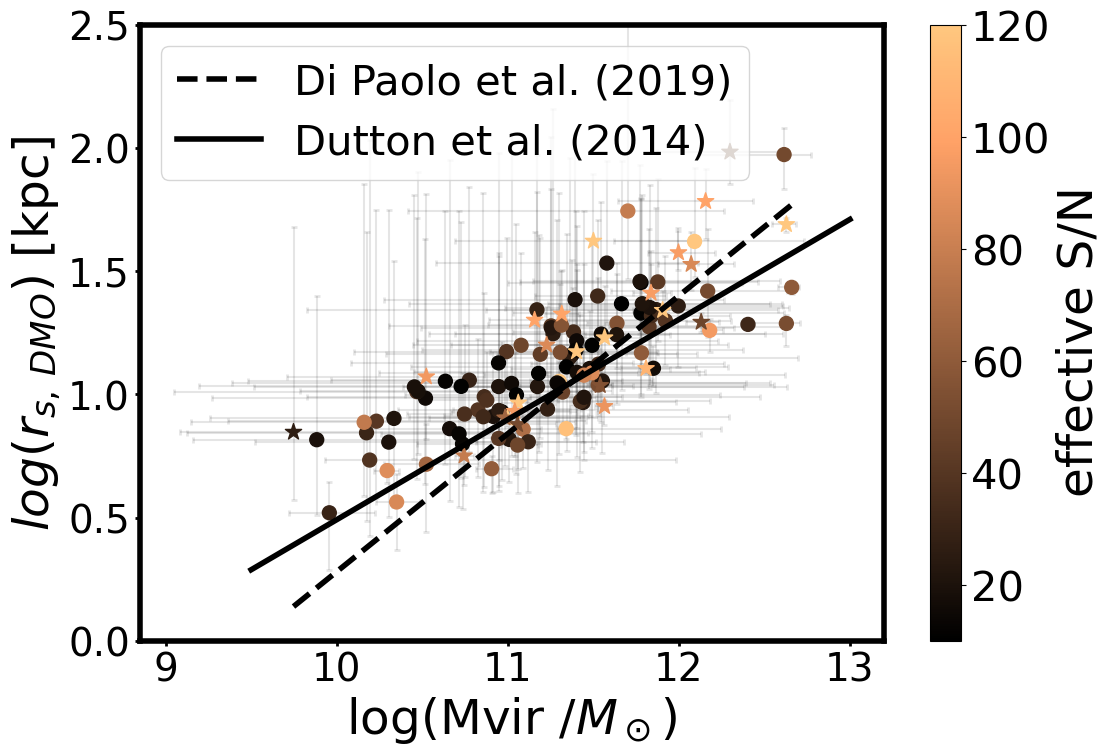} }}%
    \caption{Left: Concentration--halo mass relation for the MHUDF sample. The solid coloured lines show the $\rm{c_{\rm{vir}}-M_{\rm{vir}}}$ relation from \cite{dutton} for  0.3<$z$<1.5, while the shaded regions show the 0.11 dex scatter around the relations. The data points are colour-coded according to their redshifts.  Right:  $\rm{r_{\rm{s, DMO}}=r_{\rm{vir}}/c_{\rm{vir}}}$ in [kpc] as a function of the halo masses.  The dotted black line displays the observed scaling relation for $z= 0$ low surface brightness systems inferred by \cite{Paolo}, while the black solid line shows the predictions from \cite{dutton}.  The data points are colour-coded according to their effective S/N. In both panels, the  circles show the regular MHUDF galaxies, while the stars depict the perturbed galaxies.  The error bars represent the  95\% confidence intervals.}%
    \label{conc}%
\end{figure*}

  Figure \ref{conc} (left) shows  the concentration--halo mass relation for our sample. The solid coloured lines and shaded regions show the $c_{\rm{vir}}-M_{\rm{vir}}$ relation and scatter of 0.11 dex, respectively, from \cite{dutton} for simulated halos with  $0.3<z<1.5$.   The MHUDF galaxies show significant scatter around the expected scaling relations.  This is expected, as even DM–only simulations produce considerable intrinsic scatter \citep[e.g.,][]{cor}, and baryonic processes are known to further increase the scatter \citep[e.g.,][]{sorini}.  We quantify the agreement with the theoretical relation by accounting for both measurement uncertainties and the intrinsic 0.11 dex scatter in the $c_{\rm{vir}}$–$M_{\rm{vir}}$ relation. Approximately 70\% of the MHUDF galaxies lie within 1$\sigma$ of the \citet{dutton} relation at their respective redshifts, a level of agreement broadly consistent to that observed in the local $z=0$ SPARC sample \citep{li}, when no $\Lambda CDM$ priors are used in the modelling. For completeness, we show in Fig.  \ref{cmh} the $c_{\rm{vir}}-M_{\rm{vir}}$  relation inferred from all six different halo models described in Appendix~\ref{appendix:DM}.   

Figure \ref{conc} (right)  shows the scaling relation between the halo mass, $M_{\rm{vir}}$, and the scale radius, $\rm{r_{\rm{s, DMO}}=r_{\rm{vir}}/c_{\rm{vir}}}$, which is a redshift-independent parameter  (e.g. \citealt{bull}, \citealt{dutton}).  The sample follows a tighter sequence in the  $r_{\rm{s, DMO}}-M_{\rm{vir}}$  than in the $c_{\rm{vir}}-M_{\rm{vir}}$ plane.  However, we observe a small systematic offset: our sample tends to have larger $\rm{r_{\rm{s, DMO}}}$ values at fixed $M_{\rm{vir}}$ than predicted by simulations or inferred observationally for the low surface brightness sample of  \cite{Paolo}.  This offset likely arises from the uncertainties in estimating $r_{\rm{vir}}$, which requires extrapolation beyond the radial extent of the observed RCs. Since $r_{\rm{vir}}$ is not directly measured but inferred by fitting halo models over a limited kinematic range, this extrapolation introduces significant model-dependent uncertainties. As a consequence, derived parameters such as $c_{\rm{vir}}$ and $r_{\rm{s, \rm{DMO}}}$ are less tightly constrained than for local galaxies, where extended HI rotation curves provide direct measurements out to large radii.

\subsection{Concentration-density relation}
\label{res:concdens}
\begin{figure*}
   \centering
    \includegraphics[width=1\textwidth,angle=0,clip=true]{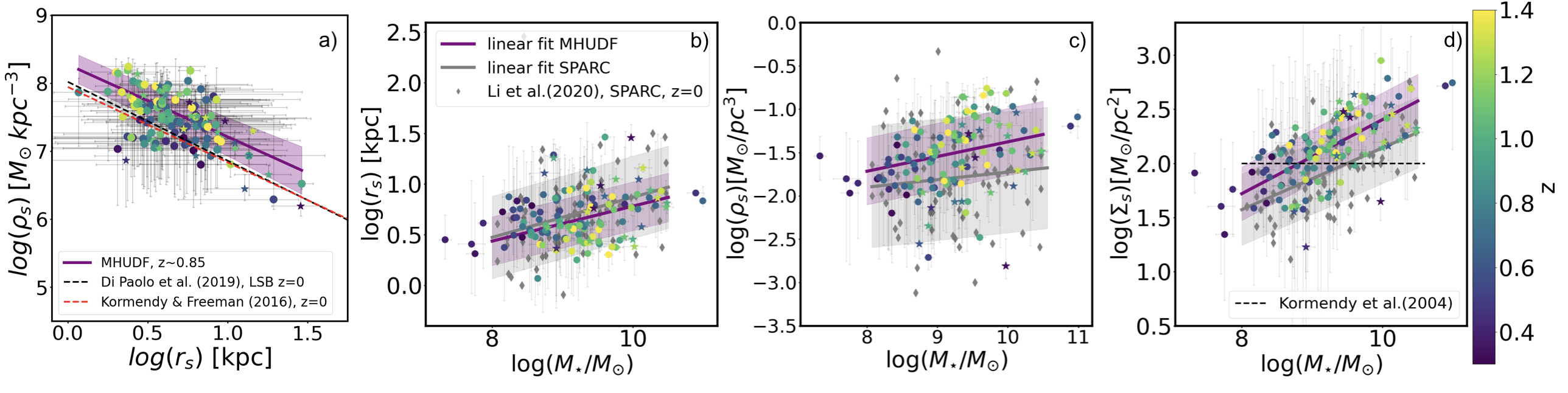} 
       \centering   
    \caption{Panel a) shows the halo scale radius-DM density relation for the MHUDF sample. The black dashed line shows the relation inferred by \cite{Paolo} for $z=0$ low surface brightness galaxies, while the red dashed  line shows the relation derived by \cite{Kormendy} for $z=0$ SFGs.  The purple line shows a power-law fit to  our mass complete sample, with $\rho_{\rm{s}} \propto r_{\rm{s}}^{-1.032 \pm 0.192}\cdot (1+z)^{ 0.541 \pm0.311}$, whereas the purple-shaded region shows the dispersion around the relation.  Panel b) shows the halo scale radius-stellar mass relation. Panel c) shows the DM density-stellar mass relation. Panel d) shows the DM surface density ($\rm{\Sigma_{\rm{s}} \propto \rho_{\rm{s}} \cdot r_s}$)-stellar mass relation.  In panels b), c), d), the grey diamonds  represent the mass-matched SPARC sample at $z=0$ from \cite{li}.  The purple and grey lines represent linear fits to the MHUDF and SPARC galaxies, respectively. The shaded regions show the dispersion around the relations. The black dashed line in panel d) shows the constant DM surface density inferred by \cite{Kormendy}.
In all the panels, our sample is colour-coded according to its $z$. The circles show the regular MHUDF galaxies, while the stars depict the perturbed systems. The error bars represent the  95\% confidence intervals.
 }
\label{dc14rhosrs}
\end{figure*}

The $c_{\rm{vir}}-M_{\rm{vir}}$ or $r_{\rm{s}}-M_{\rm{vir}}$ relations can be recast to the $r_{\rm{s}}-\rho_{\rm{s}}$ relation.
Figure \ref{dc14rhosrs} shows  this relation for the MHUDF galaxies in panel a). We observationally confirm the well-known anticorrelation  between $r_{\rm s}$ and $\rho_{\rm s}$  with a slope of $\approx-1$ (e.g. \citealt{sal}, \citealt{Spano}, \citealt{Kormendy}), consistent with expectations from hierarchical clustering of DM halos \citep[e.g.][]{GR}.  Using \texttt{HyperFit} \citep{hyperfit}, we fit a power-law model of the form $\rho_{\rm s} \propto r_{\rm s}^x (1+z)^y$, accounting for the errors. Before exploring a possible redshift evolution, we assessed the completeness of our sample. While the MHUDF survey is complete down to $\log(M_\star/M_\odot)=8.61$ for $0.2<z<1.5$, our stricter S/N cuts imply completeness only above $\log(M_\star/M_\odot)=8.8$. We therefore adopt this mass limit, excluding 22 galaxies from our fit. Restricting to this complete sample, we find $\rho_{\rm s} \propto r_{\rm s}^{-1.032\pm 0.192}(1+z)^{0.541\pm0.290}$ (purple line in Fig. \ref{dc14rhosrs}, panel a). For comparison, \cite{Djorgovski} showed that the density of DM halos, $\rho$, and the size, $r$,  follow $\rho \propto r^{-3(3+n)/(5+n)}$.   The $r_{\rm{s}}$ slope of $\sim -1.03$ that we infer yields $n\sim-1.92$. This value for $n$ is very close to $n\sim-2$ expected in $\Lambda$CDM   for halo masses of $\sim 10^{12}\:M_{\odot}$ (e.g. \citealt{Shapiro}).  The redshift term from the fit suggests a mild increase of $\rho_{\rm s}$ with $z$, consistent with higher-$z$ galaxies lying above the local relation (dashed red and black lines in panel a). While this points towards denser halos at earlier epochs, a larger sample is required to confidently confirm this trend.

Next, we show in panels  b), c), d) the halo structural parameters vs the stellar masses for our intermediate-$z$  (coloured data-points) and the SPARC $z=0$ galaxies (grey diamonds, \citealt{li}). Panel  b)  shows $r_{\rm s}$–$M_\star$ relation, panel c)  the $\rho_{\rm s}$–$M_\star$ relation, and panel d) the halo surface density $\Sigma_{\rm s}(=\rho_{\rm s}r_{\rm s})$–$M_\star$ relation. Consistent with \citet{li19}, we find that $r_{\rm s}$ and $\Sigma_{\rm s}$ increase with $M_\star$, while $\rho_{\rm s}$ remains roughly constant. This implies that more massive galaxies inhabit progressively larger halos, with  the mass and size growing in tandem to maintain a nearly constant density. The observed increase of $\Sigma_{\rm s}$ with $M_\star$ agrees with \citet{zhou} (SPARC) and \citet{Kaneda} (theory). In contrast, \citet{Donato} and \citet{Kormendy} found $\Sigma_{\rm DM}$ to be constant with stellar mass. These discrepancies likely stem from differing halo models and fitting methods. While \citet{Kormendy} used maximum-disc fits with pseudo-isothermal halos, where $\rho_0$ is the central density and $r_c$ the core radius,  we use $\rho_{\rm s}$ and $r_{\rm s}$ from DC14 profile. For example, \citet{li19} showed that adopting Kormendy’s assumptions for SPARC indeed yields a constant $\Sigma_{\rm DM}$, but argued that the maximum-disc method is unsuitable for low- and intermediate-mass galaxies, as it artificially boosts the stellar contribution at the expense of the halo.

From direct comparison of linear fits in panels b), c), d) (MHUDF -purple line; SPARC -grey line),  we find that our intermediate-$z$ galaxies have $r_{\rm s}$ consistent with the local sample. This observational result is in line  with DM-only simulations showing $r_{\rm s}$ to be nearly redshift-invariant \citep[e.g.][]{bull,dutton}.  In contrast, $\rho_{\rm s}$ appears $\sim0.3$ dex higher at intermediate $z$ than locally, suggesting denser halos at earlier epochs, consistent with the mild redshift dependence inferred from the fit in panel~a). The tentative redshift dependence of $\rho_{s}$  is in qualitative agreement with theoretical expectations  (\citealt[e.g.][]{zhao2003,zhao2009}, see  Fig. 20 of the latter paper). Such a trend arises naturally from the physics of hierarchical structure formation in an expanding Universe. At higher $z$, the mean background density is higher, and since the characteristic density  is related to the density of the Universe at the time of halo collapse, halos forming at earlier epochs tend to have higher densities.    
  
\section{Discussion}
\label{discussion}

\subsection{Core formation}
\label{coreform}
In Sect.~\ref{res:DMprof}, we found that $\sim66\%$ of the intermediate-$z$ SFGs from our sample have cored DM density profiles.  Within $\Lambda$CDM, baryon--DM interactions in the inner halo--driven by AGN feedback \citep{Dekel-Zhao2}, infalling gas clumps \citep{El-Zant, rod, joh}, central stellar bars \citep{wk}, or stellar feedback \citep{rg, gov, tey, br, dc14,toll,Bose, fre, Jackson,aza}--have been proposed as core-forming mechanisms.
Most of these studies  find that feedback can induce core formation, with the efficiency of this process depending on the stellar-to-halo mass ratio (e.g. \citealt{dc14}; \citealt{toll}; \citealt{Lazar}; \citealt{aza}),
while \citet{Bose} find no such correlations  in the APOSTLE and AURIGA simulations.
If stellar feedback-induced  core-formation is  currently  the favourite mechanism, very few authors have reported direct
observational evidence linking cores to SFHs/SFRs (e.g. \citealt{read3,nicolas}).

\begin{figure*}
   \centering
   \subfloat{{ \includegraphics[width=5.4cm]{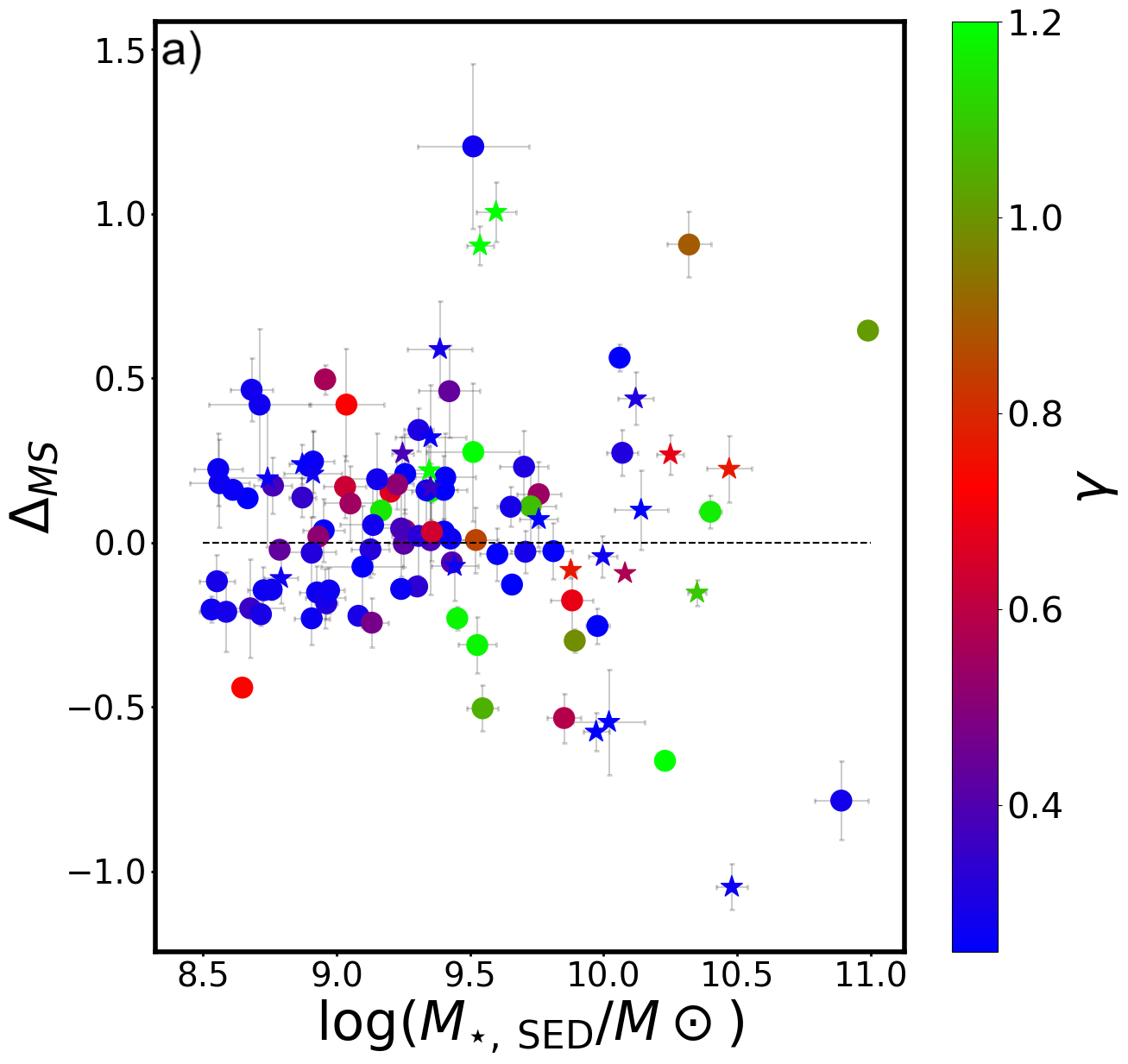}}}
    \qquad
    \subfloat{{\includegraphics[width=5.7cm]{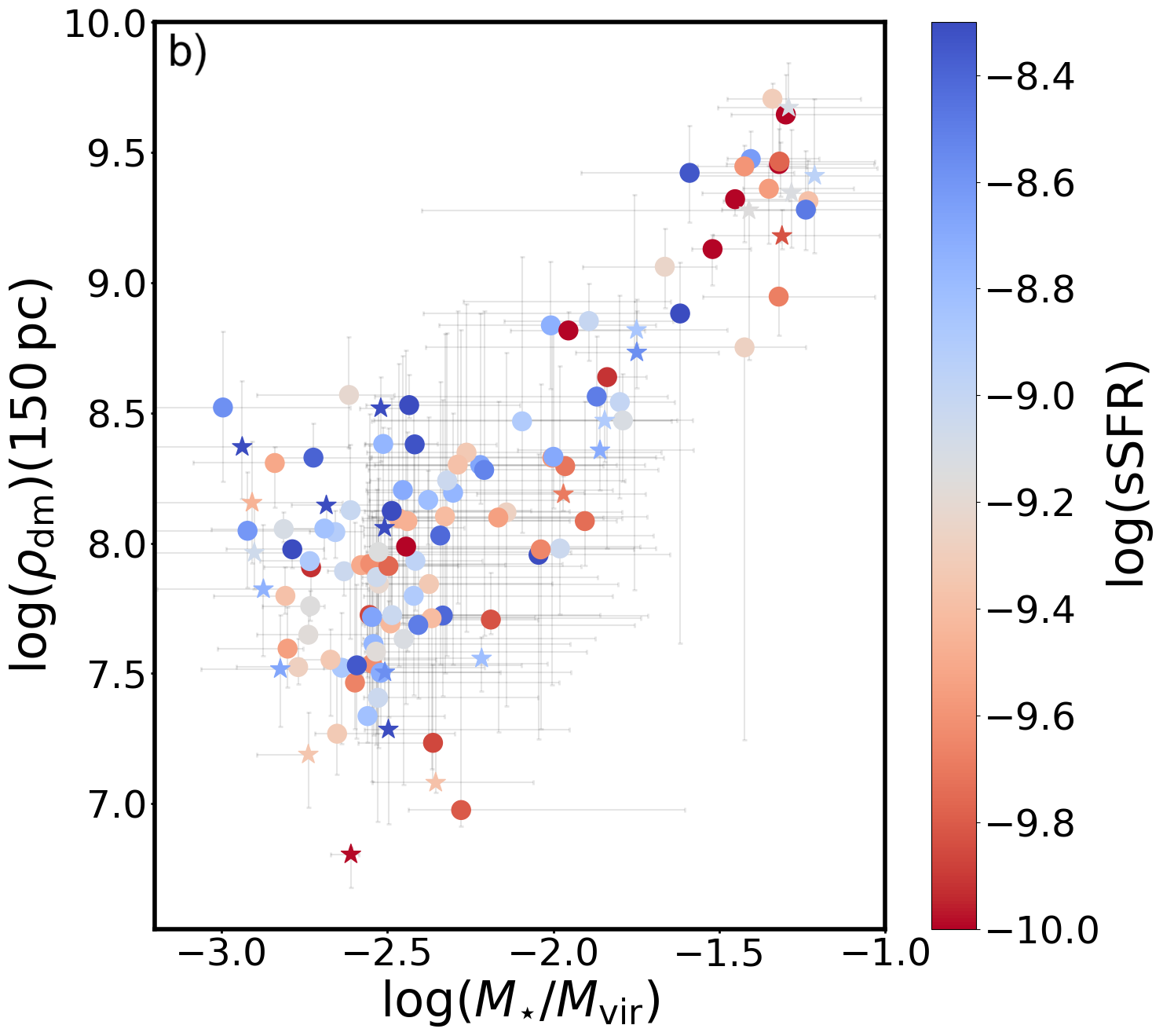} }}%   
      \qquad
    \subfloat{{\includegraphics[width=5.4cm]{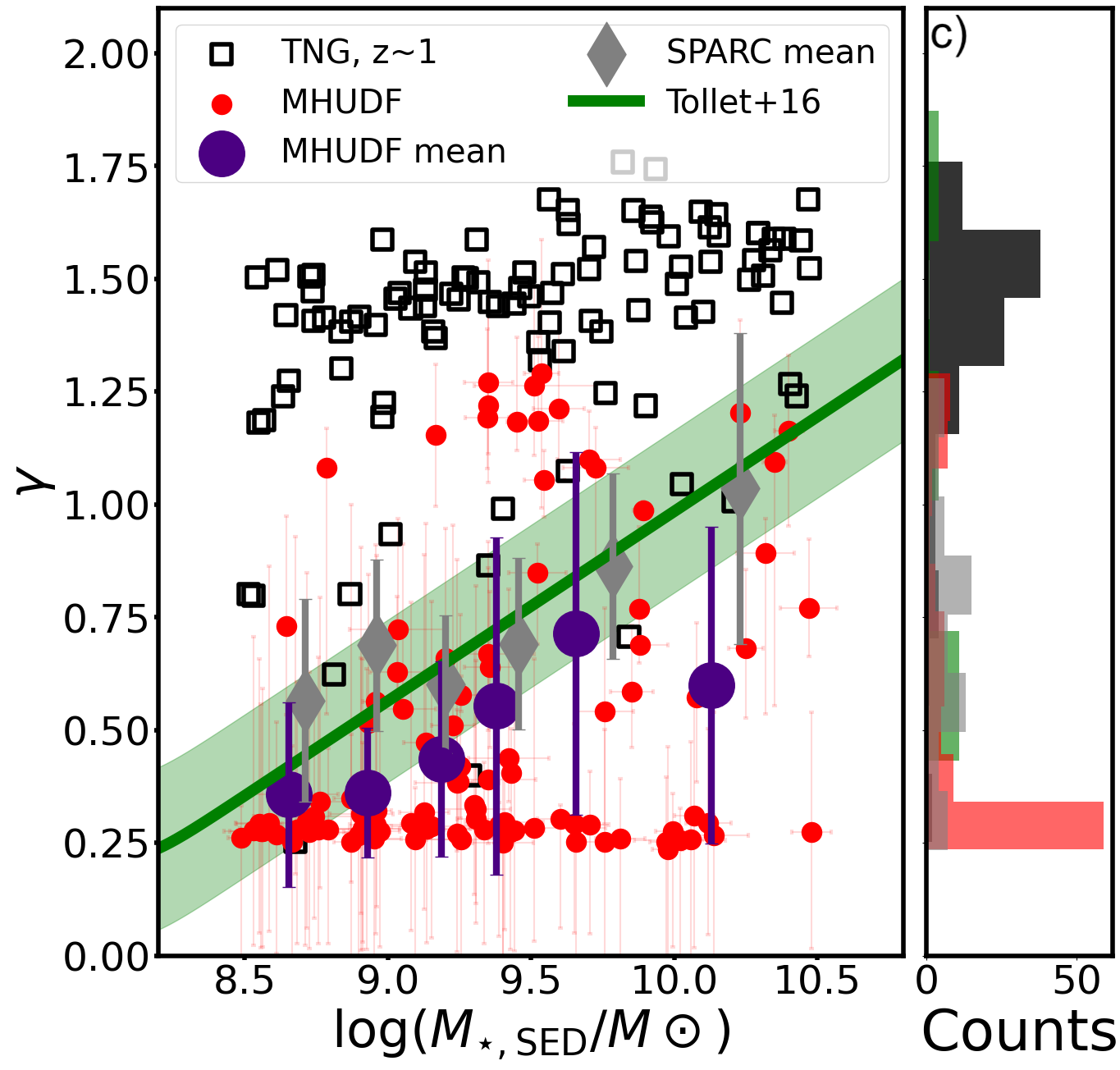} }}% 
    \caption{Panel (a): Stellar masses of the entire MHUDF sample (with $\log(M_{\star}/M_{\odot})>8.5$) as a function of the offset from the SFMS  ($\rm{\Delta MS}$) from \cite{bog}. The data points are colour-coded according to the  DM  inner slope, $\gamma$.  The error bars show the $1\sigma$ statistical uncertainties in $M_{\star}$ and SFR  from \texttt{Magphys}.  Panel (b): DM density at 150 pc as a function of $\log(M_{\star}/M_{\rm{vir}})$. The data points are colour-coded according to their sSFRs. The error bars represent the  95\% confidence intervals from our disk-halo decomposition.  In both panels, the circles show the regular MHUDF galaxies, while the stars depict the perturbed systems from our sample.  Panel (c): $\gamma$ as a function of the stellar masses for the mass-matched MHUDF sample (red circles) and the TNG galaxies  (black open squares,  from \citealt{Pill}). The large indigo data points show the mean values for $\gamma$ of the MHUDF sample in 6 mass bins, while the grey diamonds show the mean values for the mass-matched SPARC sample at $z=0$ (using the DC14 fits without $\Lambda$CDM priors from \citealt{li}). The green curve shows the predictions from \cite{toll} at $z=0$, whereas the green shaded region shows the 0.18 dex scatter around the relation. The histogram shows the distribution of the  $\gamma$ values for the mass-matched MHUDF (in red), SPARC (in grey), TNG  (in black) and NIHAO (in green) samples.}
\label{dms}
\end{figure*}

Motivated by these studies, we investigate whether the presence of DM cores in our sample  correlates with the current SF activity, quantified either by the offset from the 
SFMS or by the specific SFR  ($\mathrm{sSFR} = \mathrm{SFR}/M_\star$). 
Panel a) of Fig.~\ref{dms} shows the MHUDF galaxies with  $\log(M_\star/M_\odot) > 8.5$ in the $\Delta_{\rm MS}$--$M_\star$ plane, colour-coded by their 
inner DM slopes. We find no evidence for a correlation between $\gamma$ and $\Delta_{\rm MS}$:  both cored and cuspy systems exhibit a wide range of SFRs. 
Panel b) of Fig.~\ref{dms} shows the DM density within 150\,pc as a function of the  stellar-to-halo mass ratio, with the data points colour-coded by their sSFRs. 
Again, we find no correlation between $\rho_{\rm DM}(150\,\mathrm{pc})$ and the sSFR. 
Overall, our results indicate no link between the central DM density and the current level of  SF activity in these intermediate-$z$ galaxies.

We note, however, that this lack of a correlation with current, integrated SF activity is not unexpected, since (i) models predict that core formation is driven by repeated, rapid fluctuations in the gravitational potential, triggered by bursts of SF and gas outflows, rather than by the instantaneous SFR (e.g.  \citealt{pon}) and (ii) the integrated SFRs used here are not by themselves sensitive to bursty SFHs, limiting their diagnostic power.  Testing for the feedback-induced core-formation scenario requires full SFHs, as in \citet{read3}, which we defer to future work.

On the other hand, $\gamma$ seems to be strongly dependent on the galaxy mass. Indeed, Figure~\ref{dms}(a) already indicates that $\gamma$ is  correlated with $M_\star$, and Figure~\ref{dms}(c) shows the $\gamma-M_\star$ relation for the MHUDF galaxies.

To place our results in a broader context,  Fig.~\ref{dms}(c)
compares the MHUDF sample (red circles and red histogram; indigo circles showing the mean $\gamma$ values in six mass bins), to the $z=0$ SPARC galaxies (grey diamonds representing the mean $\gamma$ values in six mass bins  and grey histogram; using the DC14 fits from \citealt{li}), and the $z\sim1$ TNG50 SFGs (black open squares and black histogram; \citealt{Pill}).  The green curve shows the $z=0$ predictions from the NIHAO hydrodynamical simulation \citep{toll}, while the green histogram shows the distribution of the gamma values of individual NIHAO SFGs. To ensure a meaningful comparison, all  data sets are restricted to the same stellar-mass range, $8.5 < \log(M_\star/M_\odot) < 10.5$.

As this panel illustrates, the TNG50 SFGs are almost exclusively cuspy,
suggesting that TNG50 may not fully capture baryonic core-formation processes, at least for the mass and $z$ range probed in this work. In contrast, hydrodynamical simulations such as NIHAO predict a strong mass dependence of the inner slope that is in broad agreement with our observational results and those from  \cite{li}.  Note, the MHUDF galaxies with $\gamma\sim 0.25$ across a broad $M\star$ range can be explained by the intrinsic scatter of $\sim 0.25$ dex in the stellar-to-halo mass relation \citep{Behroozi} and the lack of scatter in the $\gamma-\log(X)$ relations used from DC14.

\subsection{Model limitations and caveats}

Performing a disk-halo decomposition at intermediate $z$ when the resolution/beam is comparable to the galaxy radius is challenging, but performing 3D forward modelling over several thousands of spaxels is possible provided that (i) the PSF and LSF are both stable and well characterised, and (ii) sufficient S/N is available. For illustration purposes, a galaxy with a Gaussian profile convolved by a Gaussian beam yields a Gaussian whose FWHM is $\sqrt{\hbox{FWHM}_{\rm int}^2+\hbox{FWHM}_{\rm beam}^2}$. Provided there is sufficient S/N, one can recover the galaxy intrinsic morphological parameters $\hbox{FWHM}_{\rm int}$ as discussed in \cite{galpak,nicolas1} and in \cite{Refregier12} in the context of weak lensing. The same principle applies to 3D data.

Rotation curve decompositions in lower mass galaxies $M_\star<10^{10}$ \msun{} are somewhat easier given that the DM fraction increases in galaxies with lower mass/surface-brightnesses. Hence, the RCs are more dominated by the DM component than in high-mass galaxies. This implies that information on the DM profile is contained in the shape/curvature of the RCs. For instance, a DM core of a few kpc in $\rho(r)$ is felt over several kpc in the RCs, given that the rotation velocity is driven by the cumulative mass profile $v(r)^2\propto G M(<r)/r$ (note the transition radius $r_s$ is $>10$ kpc).

Performing a disk–halo decomposition at intermediate redshifts is certainly challenging. In Sect. \ref{valid}, we tested our 3D disk–halo decomposition method on mock data cubes generated from realistic simulations with disk and DM properties similar to our galaxies. We also carried out additional tests and self-consistency checks (Sect. \ref{res:decomp}), which indicate that our methodology remains robust despite assumptions and approximations that may not hold for all the galaxies in our sample. We also examined whether the adopted baryonic parameterisation and priors introduce systematic biases in the inferred halo properties, and find that their impact is generally small. Nonetheless, some limitations remain.

First, we assume purely circular motions in axisymmetric disks, neglecting non-circular components.  Non-circular motions --such as inflows, outflows, or bars--can bias RC measurements and mimic a more cored distribution \citep[e.g.][]{Oman2019,Marasco}. Based on our MGE modelling of the mass maps and JWST imaging, we find no strong evidence for barred galaxies in our sample, although we cannot fully rule out the presence of weak non-circular motions. 
These remain difficult to model even in the local Universe \citep[e.g.][]{Di_Teodoro}, and their treatment becomes even more challenging at higher redshifts \citep[e.g.][]{Genzel_noncirc}.  We defer the analysis of non-circular motion to another paper.

Second, due to the lack of resolved \HI\ data at intermediate $z$, we adopted a constant surface density or the mean surface density profiles derived from observations of local SFGs  for the \HI\ contribution. Contrary to the vast majority of the  kinematic studies at $z>0$, we include and marginalise over an \HI\ component and show that different plausible assumptions (a constant $\Sigma_{\HI}$, or the profiles proposed by \citealt{Martinsson,Wang2}) yield consistent results  (see Appendix \ref{app:comparison HI}). However, the true diversity of \HI\ surface density profiles at $z\sim1$ may deviate from the mean, which may affect the accuracy of the inferred mass distribution. Using \texttt{vcdisk}, we estimate such deviations to be $\sim 10$–15 km/s, well within our velocity uncertainties. We therefore conclude that our results are robust against reasonable variations in \HI\ profiles. Looking ahead, facilities such as the Square Kilometre Array (SKA) will deliver spatially resolved \HI\ measurements at $z>0$, allowing a more accurate characterisation of atomic gas density profiles and, in turn, more robust disk–halo decompositions of intermediate-$z$ SFGs, in line with what is currently achievable only in the local Universe.

\section{Summary and conclusion}
\label{concl}

In this paper, we analyse the DM halo properties of a large sample of 127 SFGs, spanning a wide redshift   ($0.28<z<1.49$)  and stellar mass ($8<\log(M_{\star}/M_{\odot})<11$) range. To do so, we use deep MUSE IFU observations from the MHUDF survey \citep{udf2}. We perform a disk--halo decomposition using a 3D forward-modelling approach that accounts for stellar, gas, DM (and, where applicable, bulge) components, including corrections for pressure support.  For the DM component, we test six different density profiles, namely: (1) the DC14 profile \citep{dc14}, linking the DM profile shape to the stellar mass--halo mass ratio; (2) NFW \citep{nfw}; (3) Burkert \citep{Burkert}; (4) Dekel-Zhao \citep{fre}; (5) Einasto \citep{einasto}; and (6) coreNFW \citep{read}, as well as baryon-only models. The robustness of our methodology is assessed through application to mock MUSE data generated from idealised disk simulations and through various consistency checks. Our main findings are summarised below:
\begin{enumerate}
\item[-] The 3D approach allows us to constrain RCs to $2-3\:R_{\rm{e}}$ in individual SFGs, revealing DM dominated RCs, and a diversity in shapes, similar to what is observed in the local Universe (Fig. \ref{RCs}).
  \item[-] Compared to all the tested DM profiles and baryon-only models, DC14 generally performs as well or better across the majority of the sample (Fig. \ref{hist_baye}).
\item[-] Several internal consistency checks support the reliability of the 3D decompositions. These include independent derivations of the stellar gravitational potential from mass maps and a 1D RC decomposition, both of which yield results  consistent, within the errors, with the 3D approach for the vast majority of galaxies.
\item[-] Performing several self-consistency checks for the best fitting model, DC14, we found that the kinematically  and  SED-inferred $M_{\star}$ agree within the errors, and that the inferred DM inner slopes, $\gamma$, are consistent with the DC14 expectations (Fig. \ref{fig:selfcheckdc14}).
\item[-]  89\% of the sample has $f_{\rm{DM}}(<R_{\rm{e}})$  larger than 50\%, while only  the most massive galaxies are baryon dominated.  Globally, the $f_{\rm{DM}}(<R_{\rm{e}})-\Sigma_{M_{\star}}$  relation is similar
to the z = 0 relation (e.g. \citealt{cd}), and follows from the TFR (Fig \ref{frac}).
\item[-] 66\% of the SFGs are consistent with cored DM density profiles, with $\gamma$< 0.5. However, no correlation between core/cusp formation and the current SF activity (averaged over 0.1 Gyrs and, thus, not very sensitive to bursts) of the sample can be observed (Fig. \ref{dms}).
\item[-] The stellar mass--halo mass relation of the sample follows the \cite{Behroozi} and  \cite{Girelli} relations, but with a larger scatter (Fig. \ref{msmh}). 
\item[-] The concentration--halo mass relation agrees qualitatively with the   \cite{dutton} relation inferred from DM-only simulations, but with a larger scatter (Fig. \ref{conc} -left-hand side). The sample follows a tighter sequence in the halo mass--halo scale radius plane, in accordance with simulations   \citep{dutton} and observations \citep{Paolo},  but with a slight offset towards larger $r_{\rm{s,\: DMO}}$ values at fixed halo mass (Fig. \ref{conc} -right-hand side).  
\item[-]  We observe an anti-correlation between halo scale radius and DM density (Fig.~\ref{dc14rhosrs}, panel a), with a slope of $\sim-1$. The MHUDF galaxies are offset above the $z=0$ relations from \cite{Kormendy} and \cite{Paolo}, suggestive of a tentative evolution in $\rho_{\rm{s}}$ with redshift.
\item[-] The halo scale radius and the DM surface density increase with $M_{\star}$, while the DM  density remains relatively constant across the mass range explored (Fig.~\ref{dc14rhosrs}, panels b,c and d).
\item[-]The DM densities of $z \sim0.85$  SFGs are on average $\sim0.3$ dex higher than those of  z=0 galaxies, while the halo scale radii are consistent with those of local galaxies (Fig.~\ref{dc14rhosrs}).
\end{enumerate}
Taken together, these findings provide support for the presence of cored DM distributions in a significant fraction of  intermediate-$z$ SFGs, and offer empirical constraints on several DM halo scaling relations across a redshift and mass regime that remains relatively under-explored. This study also highlights the power of utilising 3D disk-halo decomposition on deep IFU data. In future studies, we will expand this analysis to larger samples and explore alternative DM models, including self-interacting DM \citep{sidm} and fuzzy DM \citep{hu}.

\section{Data Availability}
\label{avail}
While the MUSE data cubes used in this study are publicly available through the AMUSED web interface, we release all catalogues and data products from our disk–halo decomposition on the DARK website  \footnote{\url{https://dark.univ-lyon1.fr/data-releases/}}. This public release includes the best-fit parameters for all models used in the disk-halo decomposition together with their 95\% CI. For reference, Appendix~\ref{a3} presents a short version of the catalogue (Table~\ref{dmbestfit}), listing the best-fit parameters for the three galaxies (ID26, 6877, 958) shown in Fig.~\ref{gpkplots}. Results are provided for all DM halo models considered: (1) DC14 profile \citep{dc14}; (2) NFW \citep{nfw}; (3) Burkert \citep{Burkert}; (4) Dekel–Zhao \citep{fre}; (5) Einasto \citep{einasto}; and (6) coreNFW \citep{read}.
In addition, the data release will include the full photometry catalogue containing the structural parameters of the stellar disk, stellar masses, and star-formation rates, as well as the mass- , 2D flux-, velocity- maps, along with the chains, corner plots and residual maps for the entire galaxy sample. These data products will enable the community to reproduce our analysis and perform further comparisons.

\begin{acknowledgements}
We would like to express our deep gratitude to the anonymous referee for providing constructive comments and help in improving the manuscript. 
We would like to thank Jonathan Freundlich for the useful discussions and suggestions on improving the paper.  Many thanks to Maxime Cherry for fruitful discussions and help with technical issues. B.C.,  N.B.  and J.F. acknowledge support from the  ANR DARK grant (ANR-22-CE31-0006).
This work is primarily based on observations collected at the European Southern Observatory under ESO programs 094.A-0289(B), 095.A-0010(A), 096.A-0045(A), 096.A-0045(B) and 1101.A-0127. This research is also based in part on observations made with the NASA/ESA Hubble Space Telescope obtained from the Space Telescope Science Institute, which is operated by the Association of Universities for Research in Astronomy, Inc., under NASA contract NAS 5–26555. These observations are associated with programs AR-13252, 12060, 12061, 12062. This work is based in part on observations made with the NASA/ESA/CSA James Webb Space Telescope. The data were obtained from the Mikulski Archive for Space Telescopes at the Space Telescope Science Institute, which is operated by the Association of Universities for Research in Astronomy, Inc., under NASA contract NAS 5-03127 for JWST. These observations are associated with programs  1180, 1181, 1210, 1286, 1895, 1963, 3215 and 1963.
The data described here may be obtained from
\url{doi:10.17909/8tdj-8n28} and \url{ doi:10.17909/fsc4-dt61}.
This research made use of the following open source software:  {\textsc{ GalPaK$^{\rm 3D}$}} \citep{galpak},  \texttt{Astropy} \citep{astropy}, \texttt{numpy} \citep{numpy}, \texttt{matplotlib} \citep{plot}, \texttt{MPDAF} \citep{mpdaf}, \texttt{CAMEL} \citep{camel} and \texttt{HyperFit} \citep{hyperfit}.
\end{acknowledgements}

%###########################################################################

\bibliographystyle{aa}
\bibliography{references_fin}

\clearpage 

\begin{appendix}

\section{DM models}
\label{appendix:DM}
For the disk–halo decomposition, we considered six DM halo models, as well as a baryon-only model, described in detail below:

(1) DC14: \cite{dc14} introduce a mass-dependent density profile to describe the distribution of DM on galactic scales, which takes into account the response of DM to baryonic processes,  and can thus represent both cored and cuspy profiles. 
Within this context, the DM density profile is modelled as a generalised $\alpha -\beta -\gamma$ double power-law (e.g. \citealt{Hernquist}, \citealt{zhao}):
\begin{equation}
\rho_{\rm{DC14}}(r)=\frac{\rho_{\rm{s}}}{\left(\frac{r}{r_{\rm{s}}} \right)^{\gamma}\left(1+\left(\frac{r}{r_{\rm{s}}} \right)^{\alpha}\right)^{(\beta-\gamma)/\alpha}} \label{eqdc14}
\end{equation}
where ${r_{\rm{s}}}$ is the scale radius, $\rm{ \rho_{\rm{s}}}$ the scale density, and $ \alpha,\: \beta, \: \gamma$ are the shape parameters of the DM density profile. $\beta$ is the outer slope,  $\gamma$ the inner slope and $\alpha$ describes the transition between the inner and outer regions. The values of these shape parameters depend on the stellar-to-halo mass ratio, $X=\log(M_{\star}/M_{\rm{halo}})$ (\citealt{dc14}, \citealt{toll}), as:
\begin{eqnarray}
\alpha &=& 2.94 - \log[(10^{X+2.33})^{-1.08} + (10^{X+2.33})^{2.29}] , \label{alpha}\\
\beta &=& 4.23 + 1.34\;X + 0.26\;X^{2},  \label{beta}\\
\gamma &=& -0.06 + \log[(10^{X+2.56})^{-0.68} + 10^{X+2.56}]. \label{gamma}
\end{eqnarray}
 The parameter $X$  determines the shape of the DM halo profile and its associated ${v_{\rm{DM}}(r)}$. We restrict $X$ to $[-3.0, -1.2]$ to guarantee a solution in the upper branch of the core-cusp vs $X$ parameter space (see Fig. 1 of \citealt{dc14}).\\
The density  $\rm{ \rho_{\rm{s}}}$ is set by the halo virial velocity ${v_{\rm{vir}}}$ (or equivalently halo mass $M_{\rm{vir}}$), and following \cite{dc14},  $r_{-2}$  can be converted to ${r_{\rm{s}}}$ as:
\begin{equation}
r_{-2}=\left(\frac{2-\gamma}{\beta-2}\right)^{1/\alpha}r_{\rm{s}},
\end{equation}
where ${r_{-2}}$ is the radius at which the logarithmic DM slope is equal to -2. \\
The concentration $c_{\rm{vir}}$ is defined as:
\begin{equation}
 c_{\rm{vir},-2} = \frac{r_{\rm{vir}}}{r_{-2}},
\end{equation}
 where $r_{\rm{vir}}$ is the  virial radius of the halo. We note that the halo concentration parameter $c_{\rm{vir}}$ can be corrected to a DM-only  halo following e.g. the prescriptions of  \cite{katz} as: 
 \begin{equation}
 c_{\rm{vir, DMO}} = \frac{ c_{\rm{vir},-2} }{1+  e^{0.00001[3.4(X + 4.5)]}} \label{dmocorr},
\end{equation}

(2) NFW:  N-body DMO simulations predict cuspy DM halo profiles, which can be well described by the NFW profile \citep{nfw}. This profile can be considered a special case of the generalised $\alpha -\beta -\gamma$ profile with  $\alpha, \beta, \gamma =(1, 3, 1)$. The DM density profile is given by:
\begin{equation}
 \rho_{\rm{NFW}}(r)=\frac{\rho_{\rm{s}}}{\left(\frac{r}{r_{\rm{s}}}\right)\left(1+\left(\frac{r}{r_{\rm{s}}}\right)\right)^{2}}.
\end{equation}
which goes as $\rho \propto r^{-1}$ at small radii and ${\rho \propto r^{-3}}$ at large radii.  

(3) Dekel-Zhao: \cite{fre} introduce a halo profile with two shape parameters subject to baryonic effects, which is a special case of the  generalised $\alpha -\beta -\gamma$ profile with  $\alpha, \beta, \gamma =(0.5, 3.5, \gamma)$. This profile has a variable inner slope, $\gamma$, and concentration parameter $c_{\rm{vir}}$, and the profile parameters are correlated with the stellar-to-halo mass ratio, thereby being appropriate for describing both cusped and cored density profiles. The DM density profile is given by:
\begin{equation}
\rho_{\rm{Dekel-Zhao}}(r)=\frac{\rho_{\rm{s}}}{\left(\frac{r}{r_{\rm{s}}} \right)^{\gamma}\left(1+\left(\frac{r}{r_{\rm{s}}} \right)^{1/2}\right)^{2(3.5-\gamma)}},
\end{equation}
with $\rho_{\rm{s}} =(1-\gamma/3)\overline{\rho_{\rm{c}}}$, while $\overline{\rho_{\rm{c}}} =c_{\rm vir}^{3}\mu\overline{\rho_{\rm{vir}}}$,   $\mu=c_{\rm{vir}}^{\gamma-3}(1+c_{\rm{vir}}^{1/2})^{2(3-\gamma)}$ and $\overline{\rho_{\rm{vir}}}=3M_{\rm{vir}}/4\pi r_{\rm{vir}}^{3}$, and two shape parameters $\gamma$, and the concentration parameter $c_{\rm{vir}}$. 
The variable inner slope, $\gamma$, defined in  \cite{fre} as the absolute value of the logarithmic slope at 1\% of the virial radius (denoted as $s_1$), depends on the stellar mass- halo mass relation as follows:
\begin{equation}
s_{1}(x)=\frac{1.25}{1+\left(\frac{X}{1.30 \cdot10^{-3}}\right)^{2.98}}+0.37 \log \left(1+\left( \frac{X}{1.30 \cdot10^{-3}}\right)^{2.98} \right)
\end{equation}
with $X=M_{\star}/M_{\rm{vir}}$. 
The inner slope corresponds to the NFW slope for $\log(M_{\star}/M_{\rm{vir}} ) < -4$, to flatter inner density profiles for  $\log(M_{\star}/M_{\rm{vir}} )$ between -3.5 and -2 and to steeper than NFW inner density profiles for $\log(M_{\star}/M_{\rm{vir}} ) > -2$.

(4) Burkert:  \cite{Burkert} introduced a halo profile characterised by a cored structure, i.e. with a finite density towards the centre, as opposed to the diverging density seen in the NFW profile. The density profile is given by:
\begin{equation}
 \rho_{\rm{Burkert}}(r)=\frac{\rho_{\rm{s}}}{\left(1+\frac{r}{r_{\rm{s}}}\right)\left(1+\left(\frac{r}{r_{\rm{s}}}\right)^{2}\right)}.
\end{equation}

(5) coreNFW: \cite{read} studied the evolution of isolated dwarf galaxies using high-resolution hydrodynamic simulations, and showed that repeated bursts of star formation can transform cusps into cores. A general fitting function for the evolved DM profile, based on the NFW profile, was introduced:
\begin{equation}
M_{\rm{cNFW}}(<r)=M_{\rm{NFW}}(<r)f^{n}(r)
\end{equation}
where
\begin{equation}
f^{n}(r) = \tanh(r/r_{\rm{c}})^n
\end{equation}
This function suppresses the central cusp below a core radius, $r_{\rm{c}}$, defined as $r_{\rm{c}}=\eta R_{1/2}$, with $\eta=1.75$ and $ R_{1/2}$ being the half mass radius. 
The parameter  $n$ controls the shallowness of the core, where $n = 0$ corresponds to  a cusp and $n = 1$ to a core. This parameter is tied to the star formation time scale $t_{\rm{SF}}$, set to be $ \sim 5.8$ Gyr for $z\sim1$ SFGs.
The resulting density profile for  coreNFW is given by:
\begin{equation}
\rho_{\rm{cNFW}}=f^{n}(r) \rho_{\rm{NFW}} + \frac{nf^{n-1}(r)(1-f^2(r))}{4\pi r^2 r_c}M_{\rm{NFW}} \\
\end{equation}

(6) Einasto: Using high-resolution N-body simulations, \cite{einasto} demonstrated  that the DM halos are better described by the Einasto profile \citep{ein}, which has the following density profile:
\begin{equation}
\rho_{\rm{Einasto}}(r)=\rho_{\rm{s}} e^{-\frac{2}{\alpha_{\epsilon}}\left(\left(\frac{r}{r_s}\right)^{\alpha_{\epsilon}}-1\right)}
\end{equation}
where $\alpha_{\epsilon}$ is a shape parameter describing the rate at which the logarithmic slope decreases towards the centre.  When $\alpha_{\epsilon}>0$, the profile has a finite density.  For example, \cite{dutton} have shown that $\alpha_{\epsilon}$ is dependent on the halo mass; however, we chose to leave this parameter free in our modelling and impose flat priors such that $\alpha_{\epsilon}\in[0.1,2]$. 

(7) Baryons-only: We also tested baryon-only models, performing the RC decomposition by assuming maximal disks, i.e. without any contribution from a  DM halo.  We thereby set $v_{\rm{DM}}(r)^2=0$ in equation \ref{eq:vdecomp}.

\label{para_tab}
Table \ref{gpkparams} presents the free parameters used in our models, which are detailed in Section \ref{gpkparamssect}.

\begin{table*}[ht]
 \centering
 \caption{Parameters used in the {\textsc{GalPaK$^{\rm 3D}$}} modelling, as described in Sections \ref{gpk1} and \ref{gpk2}..}
    \begin{tabular}{|l|c|c|c|c|c|c|c|c|c|}
        \hline
        Parameters common to all models & URC & DC14 & NFW & Dekel-Zhao & Burkert & coreNFW & Einasto & baryons-only & Bulge \\
        \hline
        $x$, $y$, $z$ & \ding{51} & \ding{51} & \ding{51} & \ding{51} & \ding{51} & \ding{51} & \ding{51} & \ding{51} &  \\
        Total Line Flux & \ding{51} & \ding{51} & \ding{51} & \ding{51} & \ding{51} & \ding{51} & \ding{51} & \ding{51} &  \\
        Inclination ($i$) & \ding{51} & \ding{51} & \ding{51} & \ding{51} & \ding{51} & \ding{51} & \ding{51} & \ding{51}  &  \\
        Sérsic Index ($n_{\rm{gas}})$ & \ding{51} & \ding{51} & \ding{51} & \ding{51} & \ding{51} & \ding{51} & \ding{51} & \ding{51} &  \\
        Major-Axis PA & \ding{51} & \ding{51} & \ding{51} & \ding{51} & \ding{51} & \ding{51} & \ding{51} & \ding{51} &  \\
        Half-Light Radius ($R_{\rm{e}}$) & \ding{51} & \ding{51} & \ding{51} & \ding{51} & \ding{51} & \ding{51} & \ding{51} & \ding{51} &  \\
        Turnover radius ($r_{\rm{t}})$ & \ding{51} & \ding{55} & \ding{55} & \ding{55} & \ding{55} & \ding{55} & \ding{55} & \ding{55}  &  \\
        Outer slope of $v(\rm{r})$ ($\beta$ )& \ding{51} & \ding{55} & \ding{55} & \ding{55} & \ding{55} & \ding{55} & \ding{55} & \ding{55} &  \\
        Maximum Velocity ($v_{\rm{max}}$) & \ding{51} & \ding{55} & \ding{55} & \ding{55} & \ding{55} & \ding{55} & \ding{55} & \ding{55} &  \\
        Virial Velocity ($v_{\rm{vir}}$) & \ding{55} & \ding{51} & \ding{51} & \ding{51} & \ding{51} & \ding{51} & \ding{51} & \ding{51}  &  \\
        Concentration ($c_{\rm{vir}}$) & \ding{55} & \ding{51} & \ding{51} & \ding{51} & \ding{51} & \ding{51} & \ding{51} & \ding{55} &  \\
        Velocity Dispersion ($\sigma_0$) & \ding{51} & \ding{51} & \ding{51} & \ding{51} & \ding{51} & \ding{51} & \ding{51} & \ding{51} &  \\
        HI density & \ding{55} & \ding{51} & \ding{51} & \ding{51} & \ding{51} & \ding{51} & \ding{51} & \ding{51} &  \\
      
        Doublet Ratio ($r_{\rm{O2}})$ & \ding{51}/\ding{55} & \ding{51}/\ding{55} & \ding{51}/\ding{55} & \ding{51}/\ding{55} & \ding{51}/\ding{55} & \ding{51}/\ding{55} & \ding{51}/\ding{55} & \ding{51}/\ding{55} &  \\
        $X$ & \ding{55} & \ding{51} & \ding{55} & \ding{51} & \ding{55} & \ding{55} & \ding{55} & \ding{55} &  \\
        Stellar Mass ($M_{\star}$) & \ding{55} & \ding{55} & \ding{51} & \ding{55} & \ding{51} & \ding{51} & \ding{51} & \ding{51} &  \\
        $\alpha_{\epsilon}$ - Einasto & \ding{55} & \ding{55} & \ding{55} & \ding{55} & \ding{55} & \ding{55} & \ding{51} & \ding{55} &  \\
        \hline
        Bulge Sérsic Index ($n_{\rm{bulge}}$) &  &  &  &  &  &  &  &  & \ding{51} \\
        
        Bulge Radius ($r_{\rm{bulge}}$) &  &  &  &  &  &  &  &  & \ding{51} \\
        Bulge-to-Total Ratio ($B/T$) &  &  &  &  &  &  &  &  & \ding{51} \\
        \hline
    \end{tabular}
 % \tablefoot{The parameters  used in  our 3D forward modelling, as described in Sections \ref{gpk1} and \ref{gpk2}. Checkmarks (\ding{51}) indicate parameters used for a given model;  crosses (\ding{55}) indicate parameters not used for a given model. }
       \label{gpkparams}
\end{table*}

\section{Impact of Different HI Parametrisations}
\label{app:comparison HI}

\begin{figure*}
\sidecaption
\includegraphics[width=12cm]{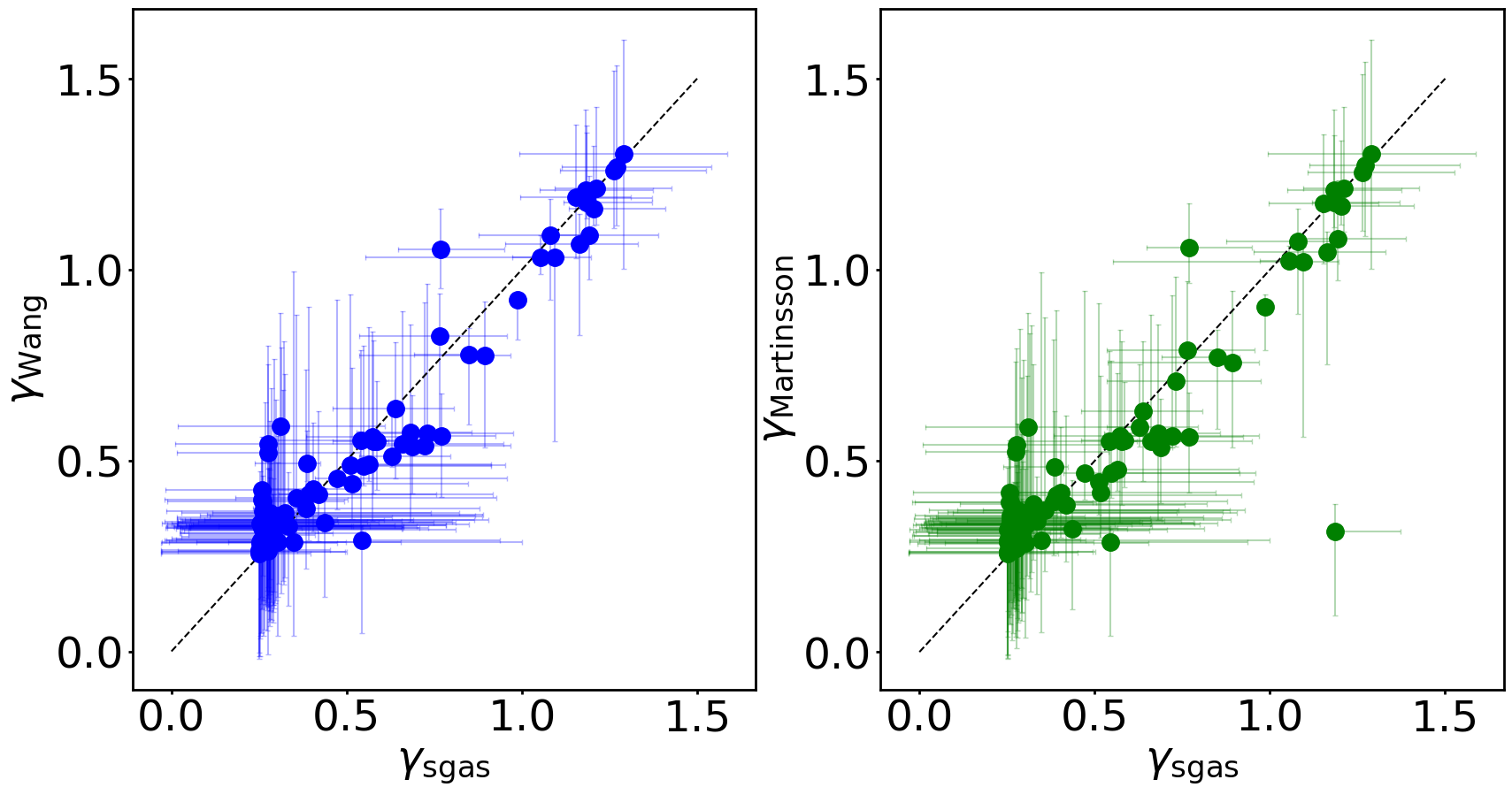}
\caption{\normalsize	 Comparison of the inner DM slope $\gamma$ obtained using different HI surface density models in the disk-halo decomposition:  constant HI surface mass density vs.\ \citet{Martinsson} profile (left) and vs.\ \citet{Wang2} profile (right). Error bars indicate  95\% confidence intervals from the MCMC posterior. The dashed 1:1 line marks perfect agreement.}
\label{fig:gamma_comparison_HI}
\end{figure*}

To assess the impact of the HI surface density model used in our disk-halo decomposition (Section~\ref{gpk2}) on the recovered DM halo parameters, we compare in Figure~\ref{fig:gamma_comparison_HI} the inferred inner DM slope $\gamma$ obtained with three different assumptions for the HI distribution. Our  model (i) assumes a constant HI surface mass density ($\Sigma_{\rm HI}\sim$ct.), yielding $v_{\rm gas}(r) \propto \sqrt{\Sigma_{\rm HI} r}$, and results in a slope $\gamma_{\rm sgas}$. We contrast this with results obtained using the models (ii) and (iii) assuming radially declining profiles from \citet{Martinsson} ($\gamma_{\rm Martinsson}$, left panel) and \citet{Wang2} ($\gamma_{\rm Wang}$, right panel), respectively. All decompositions assume the DC14 profile for the DM halo.

We find good agreement within the uncertainties for 90\% of the sample, indicating that our conclusions are robust against reasonable variations in the assumed HI profile. The uncertainties on $\gamma$ incorporate the full posterior distribution from the MCMC sampling and therefore account for the propagated uncertainty in the gas contribution.

\section{Example Posterior Distributions}
\label{corner}
\begin{figure*}[!t]
\sidecaption
\includegraphics[width=12cm]{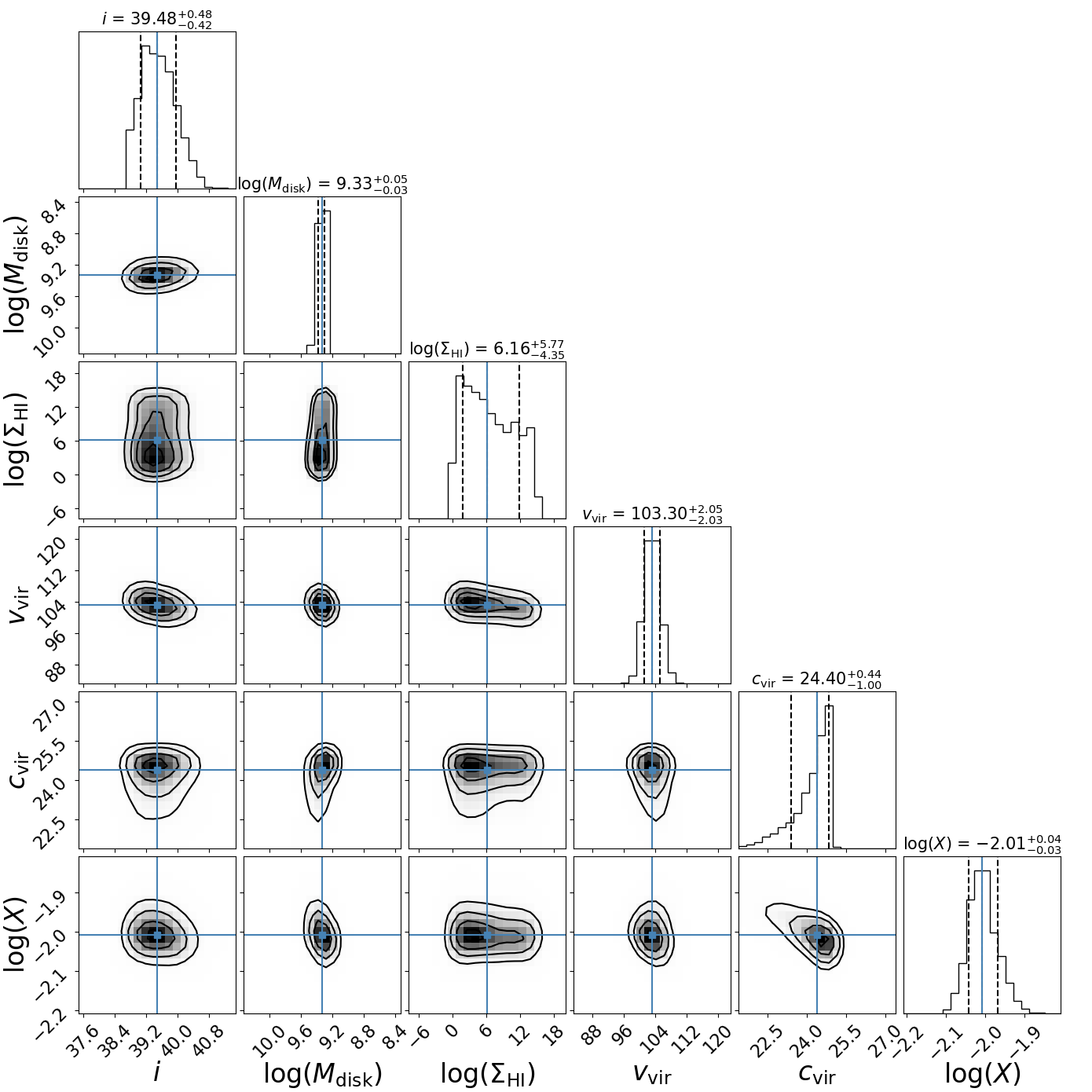}
\caption{\normalsize Corner plot for galaxy MXDF 26 showing the posterior distributions of the model parameters for the DC14 fits. The diagonal panels display the one-dimensional marginalised distributions for each parameter.  The values of the best-fit parameters and associated errors (68\% confidence intervals) are shown as the blue and black dashed lines, respectively. The off-diagonal panels show the two-dimensional joint distributions, highlighting correlations between parameters. Contours in the two-dimensional plots represent the $1\sigma$, $2\sigma$ and $3\sigma$ confidence regions. }
\label{corner1}
\end{figure*}

\begin{figure*}[!t]
\sidecaption
\includegraphics[width=12cm]{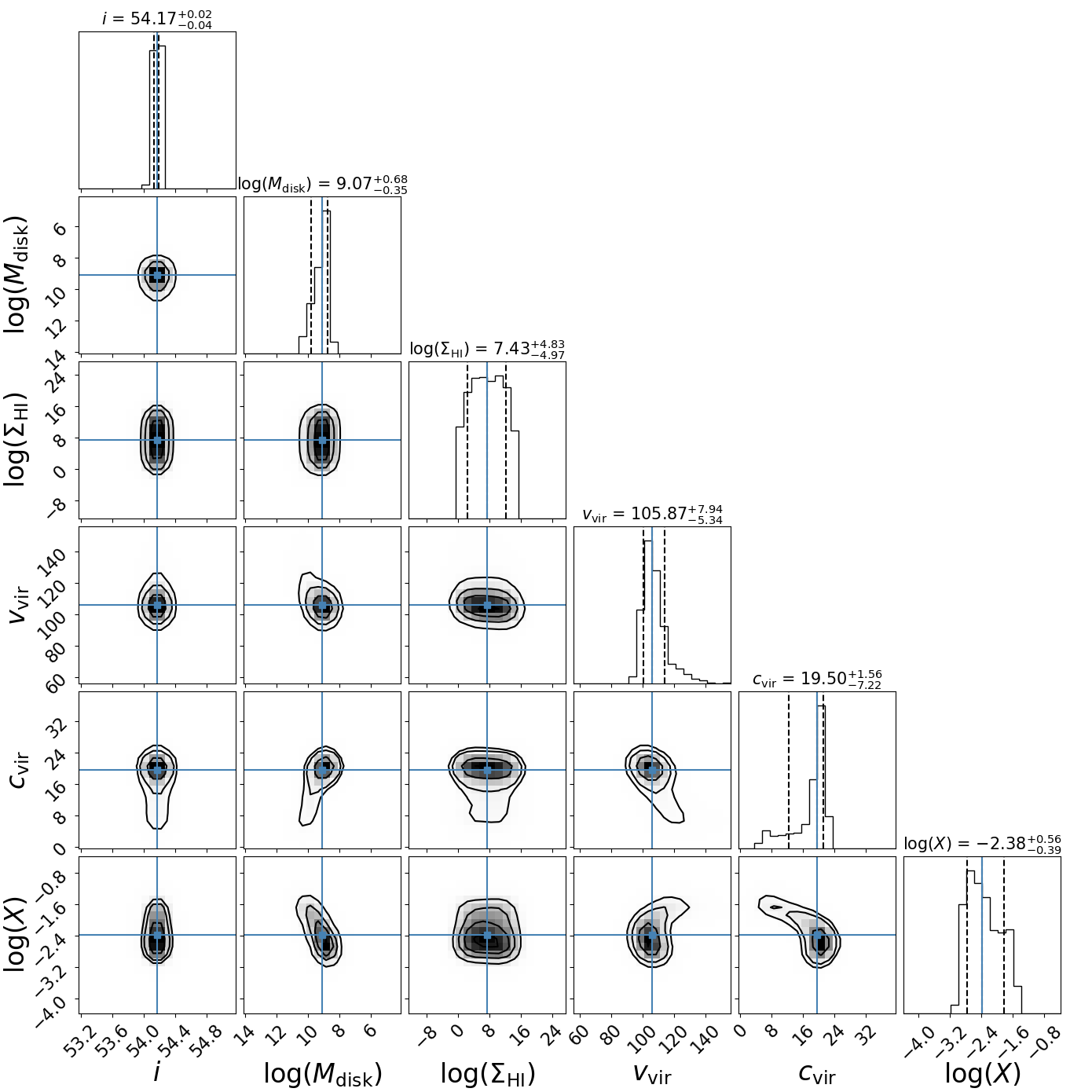}
\caption{\normalsize Same as \ref{corner1} but for galaxy MOSAIC 6877. }
\label{corner2}
\end{figure*}

\begin{figure*}[!t]
\sidecaption
\includegraphics[width=12cm]{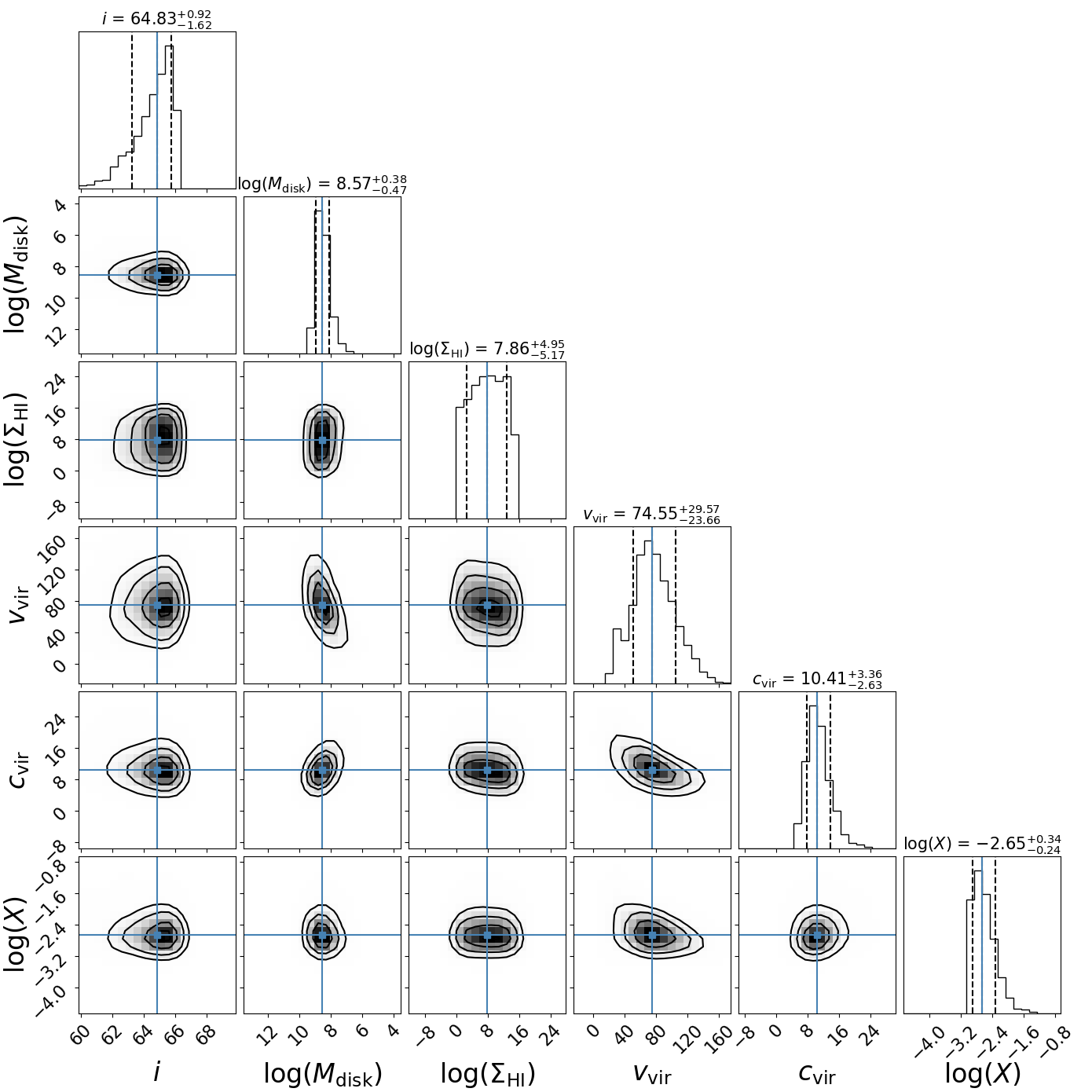}
\caption{\normalsize Same as \ref{corner1} but for galaxy MOSAIC 958.}
\label{corner3}
\end{figure*}

Figures \ref{corner1}, \ref{corner2}, and \ref{corner3} show corner plots of the posterior distributions for the three representative galaxies shown throughout the main text-- IDs 26,  6877, and  958  (from Fig.~\ref{gpkplots})-- based on the DC14 fits. The diagonal panels display the marginalised one-dimensional distributions for each parameter, while the off-diagonal panels show two-dimensional scatter plots of all sample pairs. These examples illustrate the typical behaviour discussed in the main text: most parameters exhibit well-behaved, approximately Gaussian posteriors with minimal degeneracies, which are data-driven and not prior-dominated.  We note that the concentration ($c_{\rm vir}$) shows some correlation with the virial velocity ($v_{\rm vir}$), as expected from $c_{\rm vir}=R_{\rm vir}/r_{\rm s}$. Similar correlations are present for $\log(X)$ and $v_{\rm vir}$ (or $M_{\rm vir}$) and for $\log(X)$ and $M_{\star}$, noting that $M_{\star}$ is derived from $\log(X)$, rather than independently fitted.

\section{Mass maps}
\label{Appendix:mass}

To create resolved stellar mass maps, we used the HST/ACS F435W, F606W, F775W, and F850LP bands from  the Great Observatories Origins Deep Survey South (GOODS-S, \citealt{Dickinson2003, Giavalisco2004}) and we added the F814W from the Cosmic Assembly Near-IR Deep Extragalactic Legacy Survey (CANDELS, \citealt{Grogin2011, Koekemoer2011}). 
For JWST, we used the NIRCAM F090W, F115W, F150W, F200W, F277W, F335M, F356W, F410M, and F444W bands from the JWST Advanced Deep Extragalactic Survey DR3 (JADES, \citealt{Bunker2023, Eisenstein2023, Eisenstein2023b, RiekeDoiJADES, Rieke2023, dEugenio2024}). We also complemented the observations with additional JWST/NIRCam medium bands from the JWST Extragalactic Medium-band Survey (JEMS, \citealt{WilliamsDoiJEMS, Williams2023}) as provided as part of JADES since its DR2. This includes the F182M, F210M, F430M, 460M, and 480M bands. We did not consider the HST/WFC3 F105W, F125W, F140W, and F160W bands from CANDELS since the images are only available at a coarser pixel scale of either \SI{40}{\mas} or \SI{60}{\mas}. In the end, we used five HST/ACS bands in the optical and 14 JWST/NIRCAM bands in the NIR (from roughly \SI{400}{\nano\meter} to \SI{5}{\micro\meter}) to reconstruct spatially resolved maps of the stellar mass of the galaxies. 

For each galaxy, we extracted $200 \times 200$ pixels cutouts and variance maps in all aforementioned bands and converted them to the same flux unit. For JWST, variance maps were taken as the square of the error map that includes all sources of uncertainties, whereas, for HST, the variance maps were taken as the inverse of the weight maps \citep{Casertano2000} and, by construction, do not include Poisson noise. Thus, we added in quadrature the contribution of Poisson noise to the HST variance maps following the prescription given in Sect.\,8.1 of \cite{MercierPhD}. 
Then, we convolved each cutout to the spatial resolution of the F480M band\footnote{The PSF FWHM was estimated to be around \SI{0.16}{\arcsecond}. As a comparison, the best resolution is achieved in the F115W band with a PSF FWHM of roughly \SI{0.03}{\arcsecond}.} using a matching kernel for the image and the square of the kernel for the variance map (see Sect.\,8.1 and Appendix D of \citealt{MercierPhD}).
The HST PSFs were produced using the \texttt{Photutils} \citep{photutils} implementation of the ePSF empirical reconstruction method \citep{Anderson2000, Anderson2016} using a maximum of 30 iterations and a sigma clipping of five. To reconstruct the best-fit ePSF model, we used a set of five nearby stars taken from the MXDF analysis that are located in the vicinity of the MHUDF. For JWST, we created PSF models for the JADES observations using \texttt{WebbPSF} \citep{Perrin2012, Perrin2014}, following \cite{RiekeDoiJADES}.  We did not estimate PSF variations across the field of view since the MHUDF is a relatively small field. After producing all the PSF models, we created for each band the matching kernel to F480M using the \texttt{PyPHER} software \citep{pypher} and convolved the images and the variance maps in this band using this kernel. The F480M PSF was also adopted in our MGE modelling (Sect. \ref{res:mge:mass}) to ensure consistency with the PSF-matched data and resulting mass maps.

We used the pixel-per-pixel SED fitting library \texttt{pixSED} to produce spatially resolved stellar mass maps from the PSF-matched HST and JWST images of the galaxies. An account of how the library works can be found in Sect.\,8 of \cite{MercierPhD}. Effectively, it acts as a wrapper that takes as input the redshift along with multi-band images and variance maps of a galaxy and produces resolved maps of physical parameters (e.g. stellar mass or SFR) using already existing SED fitting codes as backend. In this analysis, we used the SED fitting code \texttt{Cigale} \citep{Cigale}. Each pixel was fitted separately assuming a Chabrier IMF \citep{chabrier03}, \cite{BC03} single stellar populations, a delayed exponential SFH with a final burst/quench episode \citep{Ciesla2018, Ciesla2021}, and a \cite{CF00} attenuation law. Nebular emission was also included in the fit, using the default  \texttt{Cigale} parameters. Specifically, we adopted an ionisation parameter of $\log(U) = -2$, assumed no Ly$\alpha$ photons escape, and used a default line width of  $300$ km/s.  Furthermore, we added in quadrature a 10\% uncertainty on the fluxes, which is standard practice in SED fitting, with this treatment accounting for both calibration and model uncertainties (e.g. \citealt{cw}). To speed up the fitting process, we masked beforehand pixels that belong to nearby galaxies using the segmentation maps provided as part of JADES, and we kept the background pixels to estimate the uncertainties on the stellar mass at the pixel level when fitting pure background fluctuations. 

\begin{figure*}
    \sidecaption
     \includegraphics[width=12cm]{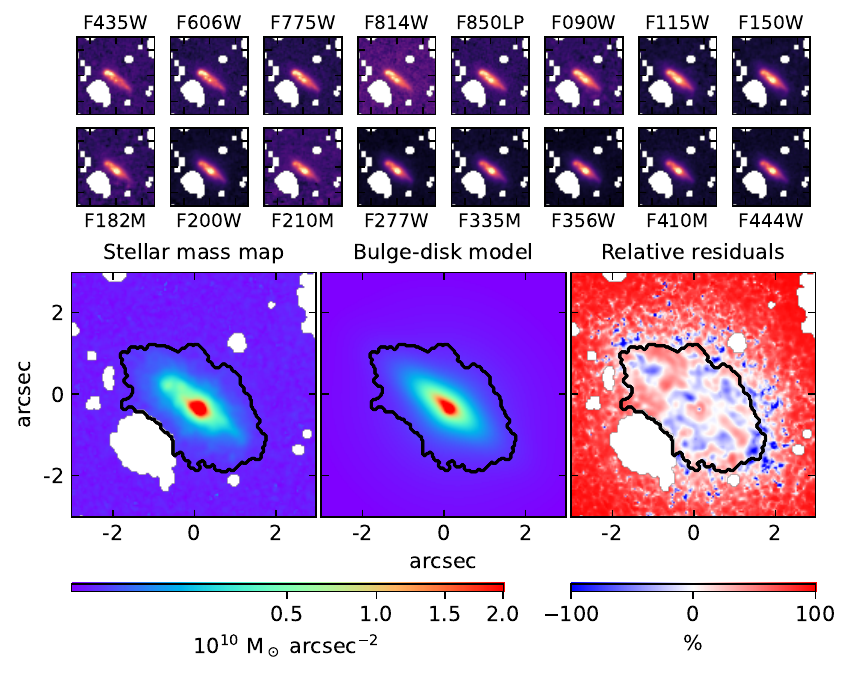}
    \caption{\normalsize Example of a stellar mass map and its associated bulge-disk decomposition for galaxy 939, which has a B/T=0.22. The top two rows show cutouts of the galaxy in the different HST and JWST bands that were used for the pixel-per-pixel SED fitting (arbitrary unit).  
  The bottom row shows the reconstructed stellar mass map (left), the best-fit bulge-disk decomposition obtained on the mass map (middle), and the residuals (right). The black line denotes the limits of the galaxy as determined by the segmentation map provided as part of the JADES survey.}
    \label{fig:example_mass_morpho}
\end{figure*}

We show a typical example for galaxy  939 at $z \approx 1.3$ in Fig.\,\ref{fig:example_mass_morpho}. 
When integrating the stellar mass map within the area defined by the JADES segmentation map, we find a total stellar mass equal to $\log(M_{\star}/M_{\odot})=10.01$, similar to what was obtained in \cite{udf2} with \texttt{Magphys}, namely $\log(M_{\star}/M_{\odot})=10.14$. It is worth mentioning that we observe good agreement between the $M\star$ obtained from the different SED fitting routines outlined in section \ref{Morphology} and in this section. 

\section{Model residuals and Bayes factor}
\label{appenix:resid}
\begin{figure*}
    \centering
       \includegraphics[width=18cm]{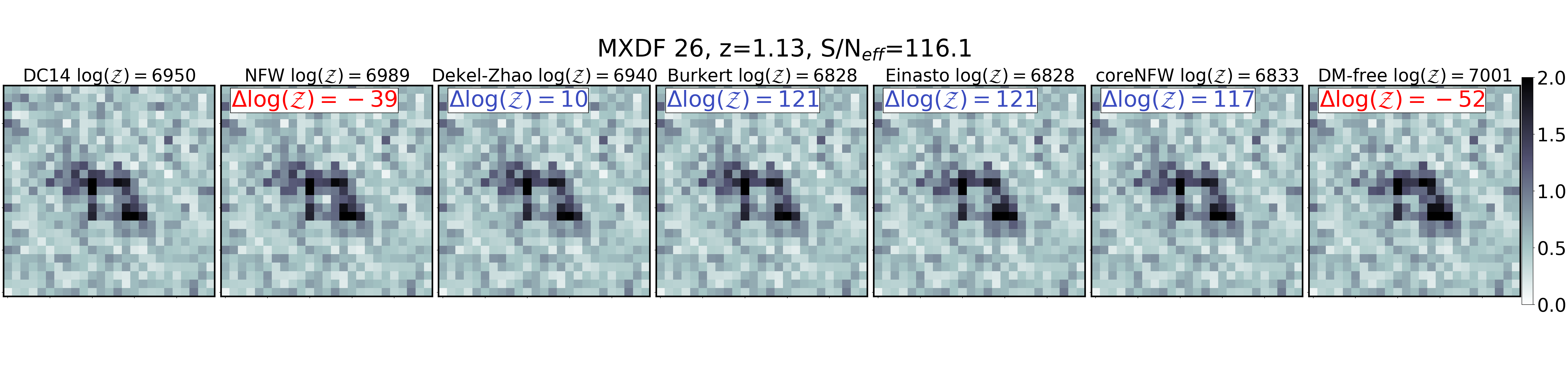}
   \includegraphics[width=18cm]{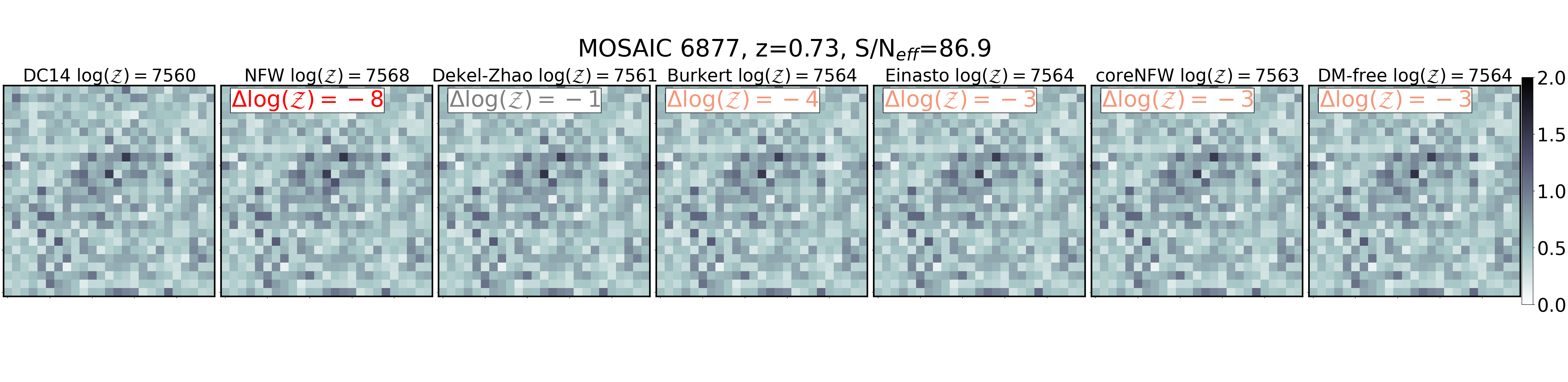}
    \includegraphics[width=18cm]{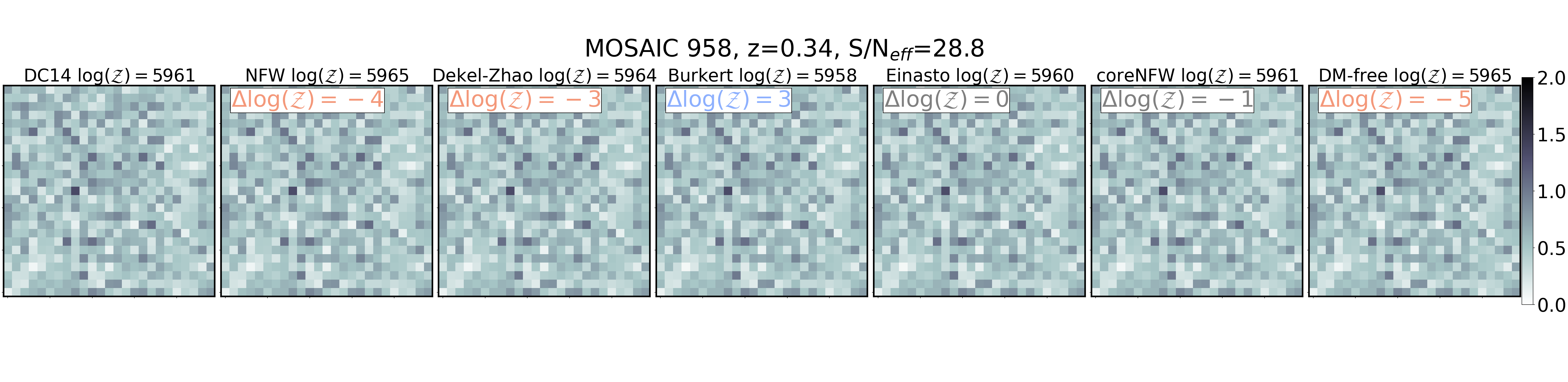}
  \includegraphics[width=18cm]{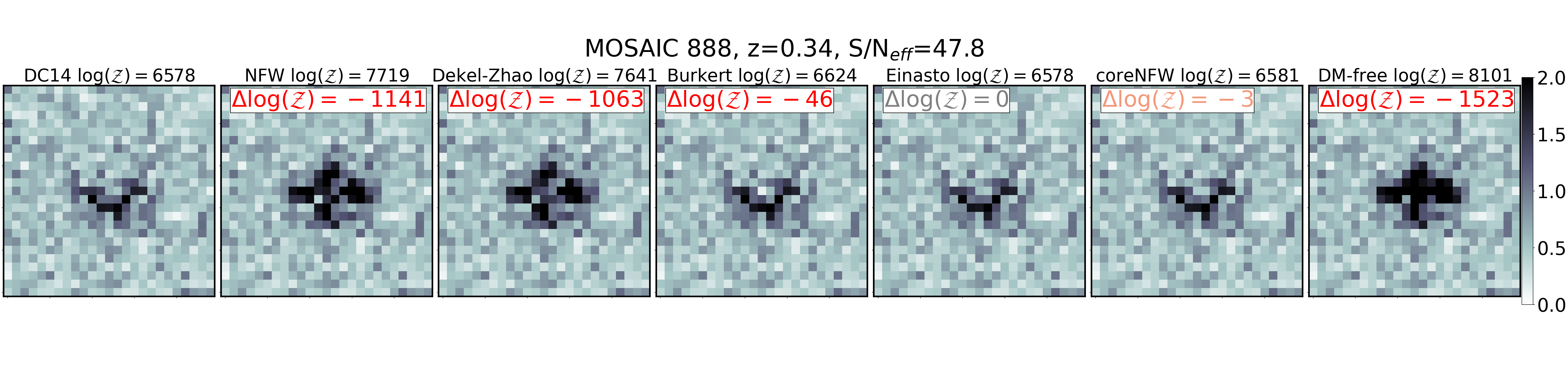}

    \caption{Residual maps derived by computing the standard deviation along the wavelength axis of the residual cube, shown for all six halo models and the baryon-only model for the same galaxies shown in Fig.~\ref{gpkplots} and one more representative galaxy from our sample, for which we find mainly strong positive evidence for the DC14 model over the alternatives (except for Einasto and coreNFW). The titles of each panel show the $\log(\mathcal{{Z}})$ inferred from our MCMC fitting routine for the respective model. The Bayes Factor computed for each model against  DC14  is shown in each panel, colour-coded as in Fig. \ref{hist_baye}.}
    \label{figappendix:resid}%
\end{figure*}

Here we explore the connection between the residual maps and the Bayes factor, which is used to select the best-fitting model for our sample in Sect.~\ref{res:residuals}. Fig. \ref{figappendix:resid} shows  the residual maps  (generated from the residual cube as in Fig. \ref{gpkplots}) for each of the six DM profiles--DC14, NFW, Burkert, Dekel-Zhao, Einasto,  coreNFW-- and the baryon only model for  the same three galaxies from Fig.~\ref{gpkplots} and one more representative galaxy from our sample.  On the maps, we report the Bayesian evidence (black text) and the Bayes factor (colour-coded as in Fig.~\ref{hist_baye}) between DC14 and the competing models.
These examples illustrate the link between the Bayes factor and the structure of the residuals: whenever the Bayes factor indicates strong evidence in favour of one model, i.e.  $\Delta\log(\mathcal{Z}) \ll -6$, the corresponding residual map of the best-fit model consistently displays lower residuals than those of the competing models. For instance, for galaxy MOSAIC 888, we infer a Bayes factor of $\Delta\log(\mathcal{Z}) = -1141$  between the DC14 and NFW profiles, and  $\Delta\log(\mathcal{Z}) = - 1523$ between the DC14 and the baryon-only models, indicating a very strong positive evidence for DC14.  The figure shows that the baryon-only and NFW models are strongly disfavored, as reflected in their pronounced residuals. However, when we infer a Bayes factor with a small value between two competing models ($-6<\Delta\log(\mathcal{Z})<6$ - inconclusive/mild positive/mild negative evidence - see middle two panels of Fig. \ref{figappendix:resid}), the residual maps look fairly similar.

 \twocolumn[
\vspace{0.5cm}
]

\section{Stellar mass halo mass relation for all halo models}
\label{a1}
\begin{figure*} 
    \centering
    \includegraphics[width=0.8\textwidth,angle=0,clip=true]{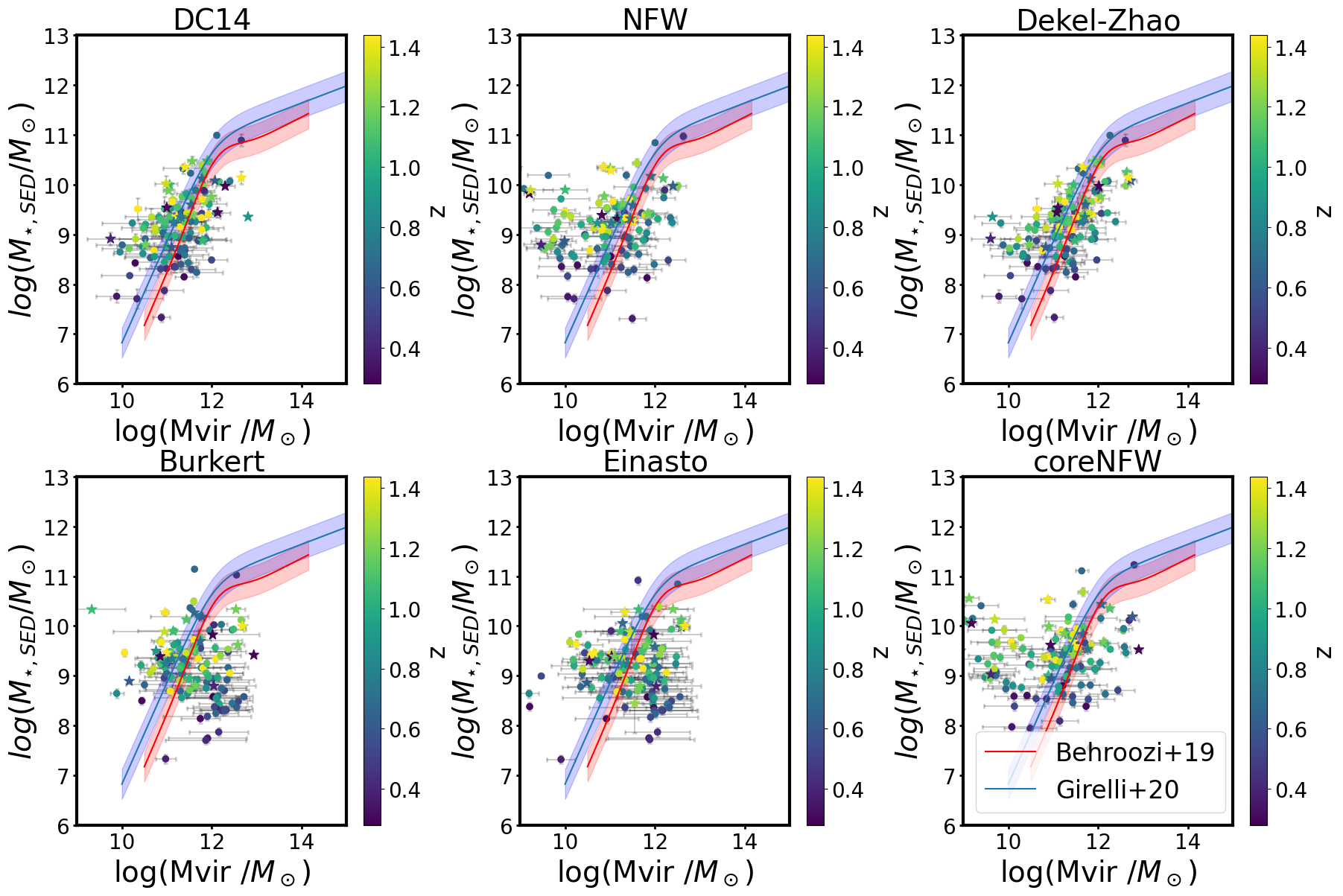}
   \centering   
    \caption{ Stellar mass -- halo mass relation for the MHUDF sample using the six halo models. The solid red curve shows the relation from \cite{Behroozi} for the mean redshift of our sample, $z$=0.85, while the red-shaded region shows the scatter of the relation. The solid blue curve shows the relation inferred by \cite{Girelli} for 0.8<$z$<1.1, whereas the blue-shaded region shows the scatter of the relation. The circles depict the regular galaxies, whereas the stars represent the perturbed galaxies from our sample. The data points are colour-coded according to their $z$. To improve the readability, the error bars show the 65\% CI ($\sim1\sigma$ symmetric errors).}
\label{msmh}
\end{figure*}

 Figure  \ref{msmh} displays  the $M_{\star}-M_{\rm{vir}}$ relation, where $M_{\star}$ is inferred from SED-fitting, and  $M_{\rm{vir}}$ from the disk-halo decomposition using  all six different halo models. As discussed in the main text, this Fig. demonstrates that the DC14 model yields the tightest $M_{\star}-M_{\rm{vir}}$, followed by Dekel-Zhao,  in qualitative agreement with the predictions from  \cite{Behroozi} and  \cite{Girelli}, albeit with larger scatter. The  $M_{\star}-M_{\rm{vir}}$ relations obtained using NFW, Burkert, Einasto and coreNFW display a much larger scatter with respect to the expected relation. Additionally, when using the other models, some galaxies exhibit $M_{\rm{vir}} < M_\star$ (for instance, as seen in the second panel from the top, where we have used the NFW profile). Upon investigating these outliers, we discovered that they are galaxies that exhibit a strong preference for cored DM distributions. Consequently, the NFW fits produce unphysical results for these cases.

 \section{Concentration halo mass relation for all halo models}
 \label{a2}
\begin{figure*}  
     \centering
    \includegraphics[width=0.8\textwidth,angle=0,clip=true]{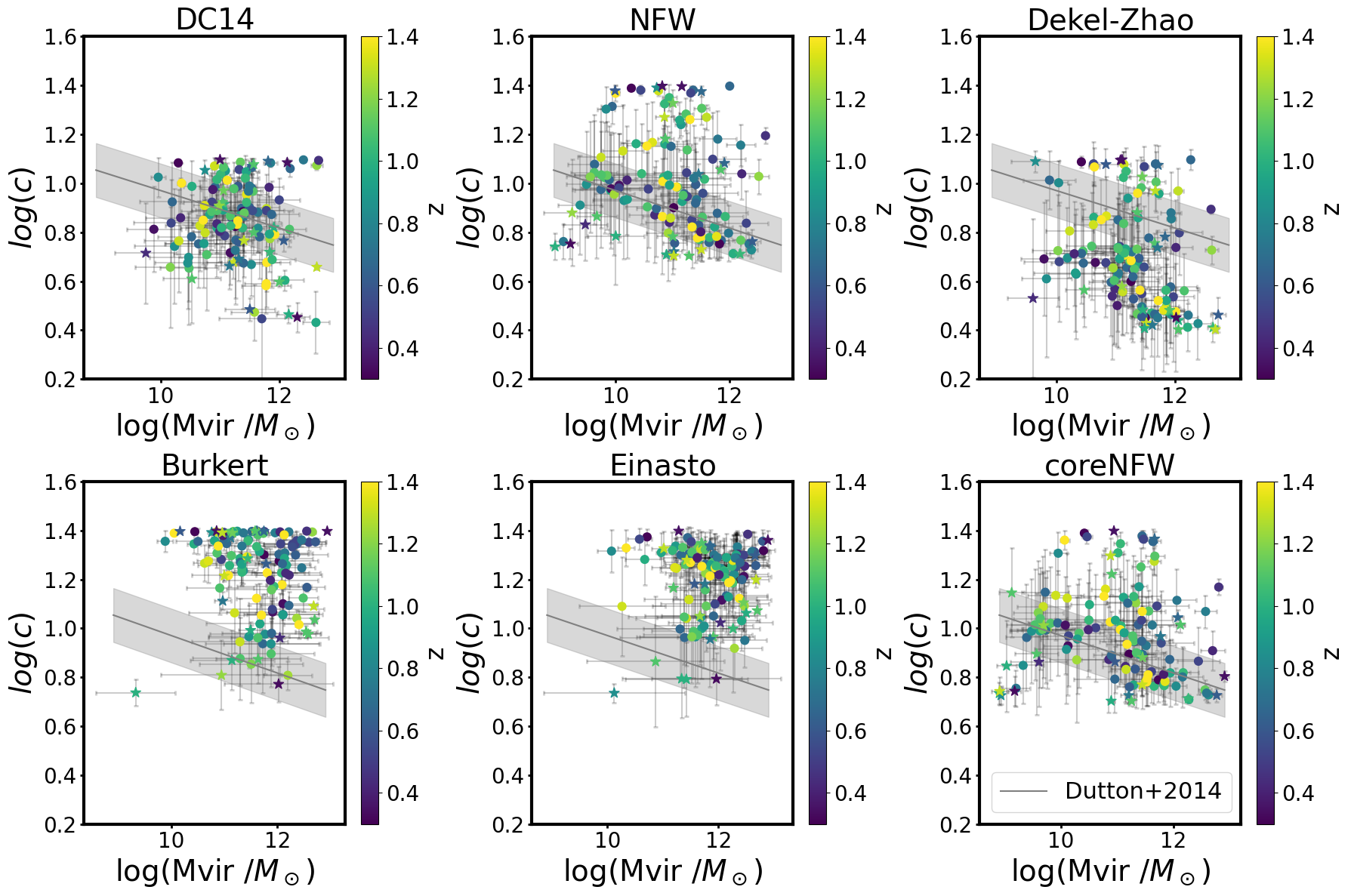}
   \centering   
    \caption{ Concentration -- halo mass relation for our sample using all six halo models. The solid grey line shows the $c_{\rm{vir}}-M_{\rm{vir}}$ relation from \cite{dutton} for the mean  $z$ of our sample, $z$=0.85,  while the shaded grey region shows the scatter of the relation. The circles depict the regular galaxies, whereas the stars represent the perturbed galaxies from our sample. The data points are colour-coded according to their $z$.  To improve the readability, the error bars show the 65\% CI ($\sim1\sigma$ symmetric errors).}
\label{cmh}
\end{figure*}

Figure  \ref{cmh} displays the $c_{\rm{vir}}-M_{\rm{vir}}$ relation inferred from all six different halo models.  This figure demonstrates that the DC14 model yields the tightest $c_{\rm{vir}}-M_{\rm{vir}}$, while the parameters inferred from the other models display a much larger scatter with respect to the simulated relation \citep{dutton}, yielding by far higher concentration values than expected. 

\section{Best fit parameters}
\label{a3}
In Table \ref{dmbestfit}, we present the best fit parameters for the same three galaxies (ID 26, 6877, 958) shown in Fig. \ref{gpkplots} using all the differen DM models, namely (1) the DC14 profile \citep{dc14}; (2) NFW \citep{nfw}; (3) Burkert \citep{Burkert}; (4) Dekel-Zhao \citep{fre}; (5) Einasto \citep{einasto}; and (6) coreNFW \citep{read}. We provide the MUSE IDs, indicate the respective DM model and list best fit parameters such as the virial velocity, $V_{\rm vir}$,  in [km/s], the virial mass in log-scale, $\log M_{\rm vir}$ [$M_\odot$], the concentration,  $c_{\rm vir}$, and the scale radius, $r_{\rm s}$, in [kpc]. For every parameter, we also report its  95\% confidence interval, computed from the 2.5 and 97.5 percentile bounds of the posterior distribution. A full catalogue can be made available upon request or can be found on the DARK webpage.
%\FloatBarrier

\begin{table*}[!htbp]
\centering
\caption{Best fit parameters inferred from the disk-halo decomposition using all six DM models. }
\begin{adjustbox}{width=1\textwidth}
\centering
\small

\begin{tabular}{ccccccccc}
\hline
\hline
MUSE ID & Model & $V_{\rm vir}$ [km/s] & $\log M_{\rm vir}$ [$M_\odot$] & $c_{\rm vir}$ & $r_s$ [kpc] & $\rho_s$ [$M_\odot$\,kpc$^{-3}$] & $\log X$ & $\alpha$ (Einasto) \\
\hline
 26 & DC14 &
$103.3^{+4.1}_{-3.8}$ &
$11.3^{+0.05}_{-0.05}$ &
$24.4^{+0.6}_{-2.2}$ &
$2.4^{+0.3}_{-0.1}$ &
$(1.5^{+0.09}_{-0.28}) \times 10^8$ &
$-2.01^{+0.08}_{-0.06}$ & 
\\[3pt]

 6877 & DC14 &
$106^{+24.1}_{-9.5}$ &
$11.5^{+0.27}_{-0.12}$ &
$19.6^{+2.7}_{-13.7}$ &
$3.8^{+15.0}_{-0.7}$ &
$(5.1^{+1.9}_{-4.8}) \times 10^7$ &
$-2.4^{+0.82}_{-0.58}$ &
\\[3pt]

  958 & DC14 &
$73.5^{+59.5}_{-48.0}$ &
$11.2^{+0.76}_{-1.34}$ &
$10.4^{+7.9}_{-4.6}$ &
$6.8^{+11.5}_{-5.3}$ &
$(6.3^{+17.8}_{-4.6}) \times 10^6$ &
$-2.6^{+0.75}_{-0.33}$ &
\\[3pt]

26 & NFW & 
$199.3^{+42.54}_{-38.83}$ & 
$12.2^{+0.25}_{-0.28}$ & 
$5.8^{+2.07}_{-0.83}$ & 
$29.1^{+11.53}_{-10.99}$ & 
$(4.7^{+4.93}_{-1.39}) \times 10^6$ & 
& \\[3pt]

6877 & NFW & 
$203.6^{+30.88}_{-45.68}$ & 
$12.3^{+0.18}_{-0.33}$ & 
$5.6^{+1.66}_{-0.61}$ & 
$40.8^{+16.36}_{-11.12}$ & 
$(2.5^{+2.08}_{-0.58}) \times 10^6$ & 
& \\[3pt]

958 & NFW & 
$28.5^{+49.52}_{-12.18}$ & 
$9.9^{+1.31}_{-0.73}$ & 
$9.5^{+13.56}_{-4.29}$ & 
$4.5^{+16.63}_{-3.33}$ & 
$(4.6^{+38.55}_{-3.51}) \times 10^6$ & 
& \\[3pt]

26 & Dekel-Zhao & 
$105.1^{+5.23}_{-3.62}$ & 
$11.4^{+0.04}_{-0.05}$ & 
$24.6^{+0.40}_{-1.70}$ & 
$0.05^{+0.08}_{-0.01}$ & 
$(8.3^{+6.79}_{-3.19}) \times 10^7$\textsuperscript{*} & 
$-2.3^{+0.13}_{-0.54}$ & \\[3pt]

6877 & Dekel-Zhao & 
$130.2^{+51.2}_{-20.4}$ & 
$11.8^{+0.43}_{-0.22}$ & 
$13.7^{+7.74}_{-3.01}$ & 
$1.0^{+0.74}_{-0.74}$ & 
$(4.5^{+12.43}_{-2.00}) \times 10^7$\textsuperscript{*} & 
$-2.3^{+0.28}_{-0.13}$ & \\[3pt]

958 & Dekel-Zhao & 
$50.2^{+62.3}_{-28.0}$ & 
$10.7^{+1.05}_{-0.45}$ & 
$9.3^{+12.46}_{-4.15}$ & 
$2.21^{+2.17}_{-2.21}$ & 
$(3.92^{+251}_{-25.4}) \times 10^7$\textsuperscript{*} & 
$-2.56^{+0.87}_{-0.45}$ & \\[3pt]

26 & Burkert & 
$90.3^{+5.57}_{-4.82}$ & 
$11.2^{+0.08}_{-0.07}$ & 
$24.8^{+0.14}_{-0.57}$ & 
$3.1^{+0.21}_{-0.17}$ & 
$(1.5^{+0.237}_{-0.091}) \times 10^8$ & 
& \\[3pt]

6877 & Burkert & 
$95.53^{+7.56}_{-5.95}$ & 
$11.36^{+0.10}_{-0.08}$ & 
$24.55^{+0.44}_{-1.26}$ & 
$4.39^{+0.53}_{-0.33}$ & 
$(8.78^{+0.41}_{-1.13}) \times 10^7$ & 
& \\[3pt]

958 & Burkert & 
$152.58^{+139.21}_{-99.76}$ & 
$12.10^{+0.84}_{-1.38}$ & 
$12.60^{+4.47}_{-2.95}$ & 
$18.86^{+21.47}_{-13.59}$ & 
$(8.57^{+9.87}_{-4.13}) \times 10^6$ & 
& \\[3pt]

26 & coreNFW & 
$263.4^{+33.23}_{-47.78}$ & 
$12.5^{+0.15}_{-0.26}$ & 
$5.6^{+1.39}_{-0.56}$ & 
$40.4^{+8.94}_{-13.76}$ & 
$(4.3^{+0.92}_{-0.29}) \times 10^6$ & 
& \\[3pt]

6877 & coreNFW & 
$246.2^{+29.08}_{-46.45}$ & 
$12.6^{+0.15}_{-0.27}$ & 
$5.4^{+1.18}_{-0.39}$ & 
$51.3^{+10.18}_{-17.06}$ & 
$(2.27^{+0.13}_{-0.36}) \times 10^6$ & 
& \\[3pt]

958 & coreNFW & 
$34.4^{+30.14}_{-16.30}$ & 
$10.2^{+0.82}_{-0.84}$ & 
$8.3^{+11.98}_{-3.19}$ & 
$6.3^{+11.92}_{-4.87}$ & 
$(3.38^{+27.78}_{-2.29}) \times 10^6$ & 
& \\[3pt]

26 & Einasto & 
$52.7^{+3.53}_{-2.17}$ & 
$10.5^{+0.08}_{-0.05}$ & 
$16.9^{+0.62}_{-1.12}$ & 
$2.7^{+0.24}_{-0.16}$ & 
$(9.5^{+1.49}_{-2.10}) \times 10^7$ & 
& 
$1.8^{+0.45}_{-0.16}$ \\[3pt]

6877 & Einasto & 
$62.3^{+35.08}_{-7.98}$ & 
$10.8^{+0.58}_{-0.18}$ & 
$12.9^{+0.71}_{-1.59}$ & 
$5.4^{+4.33}_{-0.86}$ & 
$(2.00^{+1.25}_{-0.77}) \times 10^7$ & 
& 
$1.3^{+0.68}_{-0.66}$ \\[3pt]

958 & Einasto & 
$152.6^{+135.09}_{-112.05}$ & 
$12.1^{+0.83}_{-1.73}$ & 
$7.2^{+2.12}_{-1.86}$ & 
$33.2^{+37.13}_{-26.05}$ & 
$(1.89^{+2.40}_{-1.44}) \times 10^6$ & 
& 
$1.24^{+0.80}_{-0.72}$ \\[3pt]

\hline
\end{tabular}
\end{adjustbox}
\tablefoot{\textsuperscript{*}For the Dekel-Zhao model, $\rho_s$ corresponds to the density at 150 pc due to, sometimes,  negative $\gamma$ values (see \citealt{fre} for a discussion), which result in unphysical $\rho_{\rm s}$.}
\label{dmbestfit}
\end{table*}

\end{appendix}

\end{document}